\documentclass[aps,pra,twocolumn,amsfonts,superscriptaddress,showpacs]{revtex4-1}

\usepackage{amsmath,amssymb,mathrsfs}
\usepackage{latexsym}
\usepackage{graphicx}
\usepackage[T1]{fontenc}
\usepackage{bbold}
\usepackage{epstopdf}
\usepackage{graphicx,epstopdf,color}
\usepackage{amsfonts}
\usepackage{amsmath,amssymb,mathrsfs}
\usepackage{subfigure}
\usepackage{hyperref}
\usepackage{xspace}
\usepackage{bm}

\newcommand{\ket}[1]{\left|#1\right\rangle}
\newcommand{\bra}[1]{\left\langle#1\right|}

\allowdisplaybreaks[1]
\begin{document}

\title{Anisotropic Quantum Spin Hall Effect, Spin-Orbital Textures and Mott Transition}

\author{Tianhan Liu}
\affiliation{Laboratoire de Physique Th\'eorique et Hautes \'Energies (LPTHE), Universit\'e Pierre et Marie Curie - Paris 6, 75252 Paris, France}
\affiliation{Centre de Physique Th\'eorique (CPHT), \'Ecole Polytechnique, CNRS, 91128 Palaiseau C\' edex, France}
\author{Beno\^{\i}t Dou\c{c}ot}
\affiliation{Laboratoire de Physique Th\'eorique et Hautes \'Energies (LPTHE), Universit\'e Pierre et Marie Curie - Paris 6, 75252 Paris, France}
\author{Karyn Le Hur}
\affiliation{Centre de Physique Th\'eorique (CPHT), \'Ecole Polytechnique, CNRS, 91128 Palaiseau C\' edex, France}

\begin{abstract}
We investigate the interplay between topological effects and Mott physics in two dimensions on a graphene-like lattice, via a tight-binding model containing an anisotropic spin-orbit coupling on the next-nearest-neighbour links and the Hubbard interaction. We thoroughly analyze the  resulting phases, namely a topological band insulator phase or anisotropic quantum Spin Hall phase until moderate interactions, a N\'eel and Spiral phase at large interactions in the Mott regime, as well as the formation of a spin-orbital texture in the bulk at the Mott transition. The emergent magnetic orders at large interactions  are analyzed through a spin wave analysis and mathematical arguments.  At weak interactions, by analogy with the Kane-Mele model, the system is described through a $\mathbb{Z}_2$ topological invariant. In addition, we describe how  the anisotropic spin-orbit coupling already produces an exotic spin texture at the edges. The physics at the Mott transition is described in terms of a $U(1)$ slave rotor theory. Taking into account gauge fluctuations around the mean-field saddle point solution, we show how the spin texture now proliferates into the bulk above the Mott critical point. The latter emerges from the response of the spinons under the insertion of monopoles and this becomes more pronounced as the spin-orbit coupling becomes prevalent. We discuss implications of our predictions for thin films of the iridate compound Na$_2$IrO$_3$ and also graphene-like systems.
\end{abstract}

\maketitle

\section{Introduction}

Studies of topological phases have lately been a main topic in condensed matter physics \cite{Laughlin_FQHE,Haldane1983,Wen,ReadMoore,Leboeuf,Kitaev,Nayak,DoucotIoffe,ti-reviews,Majoranas,spinl,Misguich,Mong}. The topological index has been discussed by Thouless, Kohmoto, Nightingale, and den Nijs (TKNN) in the integer quantum Hall system \cite{TKNN}, in which they found that the topological index TKNN number is related to the Hall conductance. More specifically, the Hall conductance can be related to the first Chern class of a U(1) principal fiber bundle on a torus.  The seminal work by Haldane in 1988 \cite{Haldane} proposed a model on graphene with ``artificial'' gauge fields breaking time-reversal symmetry. This is referred to in the literature as the quantum anomalous Hall effect. The concept of a band insulator preserving time-reversal symmetry with a non-trivial topological invariant was generalized by Kane and Mele in the context of quantum spin Hall physics (QSH) in two dimensions induced by spin-orbit coupling \cite{kane-mele}. In particular, the $\mathbb{Z}_2$ topological invariant is related to time-reversal symmetry \cite{fu-kane,moore-07prb121306}. The quantum spin Hall effect (QSH) which is robust in the presence of disorder \cite{disorder} and weak interactions \cite{wu-06prl106401,ti+int,pesin-10np376,RachelLeHur,Kallin,Hohenadler,Wu} has been observed experimentally in two-dimensional HgTe systems \cite{koenig-07s766,bernevig-06s1757}. A three-dimensional analogue has also been observed in various materials \cite{Hsieh_1,Hsieh_2,Chen,Zhang_H,BenoitKamran,Brune,Crauste}. The concept of non-interacting topological insulator has been thence founded theoretically \cite{kane-mele,bernevig-06s1757,koenig-07s766} and experimentally \cite{Hsieh_1,Hsieh_2,Chen,Zhang_H}, and the investigation of the subsequent helical edge transport has also been clarified \cite{wu-06prl106401,Schmidt,Adroguer,IonKaryn}. Interaction effects could eventually substitute the spin-orbit coupling and stabilize a topological band insulating state of matter \cite{Raghu}. The recent progress on the implementation of topological phases in artificial and tunable systems such as neutral atoms \cite{atoms} and photon systems \cite{photons,photons1,AlexKaryn} should also be underlined. In particular, Floquet-type topological insulators can be engineered through time-dependent perturbations \cite{photons,Floquet} and periodic alternating magnetic fields \cite{Taillefumier}.

Strong interactions in the context of topological phases can result in a plethora of interesting phenomena. For example, when the system is partially filled as in the fractional quantum Hall effect, strong interactions would play the role of changing the statistics of the electrons \cite{Laughlin_FQHE,Jain,Su,Tsui}, and contribute to the establishment of the topological order in the system \cite{Wen}. Similar fractional states of matter have been predicted to occur in the physics of Chern insulators \cite{FCI}. Strong spin-orbit interactions in the case of Ir-based transitional-metal oxides could lead to a spin liquid phase \cite{Krempa} with a topological invariant beyond the Mott transition, which is referred to as the topological Mott insulator \cite{pesin-10np376,RachelLeHur,Krempa0,Ruegg}. Other exotic phases such as the Weyl semi-metal and axion insulator may emerge as a result of interactions \cite{Krempa,Ari}. A chiral spin liquid has been potentially detected in three-dimensional iridates \cite{Nakatsuji}. A strong Hubbard interaction with the interplay of spin-orbit coupling also triggers a Mott transition with the appearance of magnetic order in the XY plane in the Kane-Mele-Hubbard model \cite{RachelLeHur}.

The physics of iridates incorporates the electron-electron interaction and spin-orbit interaction \cite{Takagi1,BJKim,Takagi2,Martins,Alaska}. In relation with topological phases, the investigation in the iridate family has aroused both theoretical \cite{pesin-10np376,Ruegg,Krempa,Shitade,Chaloupka,Jackeli_Khaliullin,Jiang,Natalia,Maria,RachelThomale,Kargarian,Bhatt,You,Vishwanath,Kimichi_2,Roser} and experimental interests \cite{Gegenwart1,Gegenwart2,Damascelli,Gretarsson,Takagi} in particular due to the possible realisation of the Kitaev exactly solved anyon model \cite{Kitaev}. Concerning the iridate compound Na$_2$IrO$_3$, the Heisenberg-Kitaev model on the honeycomb lattice and its variants provides a relatively good (even though not complete \cite{Valenti}) description of this compound in the Mott phase \cite{Trebst} and its phase diagram has been investigated numerically \cite{Jiang}.  Models with nearest-neighbour Heisenberg-Kitaev coupling on the honeycomb lattice \cite{Jackeli} are believed to be a quite proper description of strong-correlated members in the iridate family. A zig-zag order has been identified experimentally \cite{Liu}. It is relevant to underline that deep in the Mott phase, in principle, the emergent magnetic ordering for large spin-orbit couplings depend on the details of the model Hamiltonian \cite{RachelThomale,Chaloupka,Singh}. Recently, Lithium-based two-dimensional iridates have also been investigated experimentally \cite{Cao}.Two-dimensional iridates are also believed to be a good host of Quantum Spin Hall physics, as emphasized in Ref. \onlinecite{Shitade} through a next-nearest-neighbour anisotropic spin-orbit coupling model. The recent realization of thin films of Na$_2$IrO$_3$ \cite{Jenderka} tend to favor the occurrence of a two-dimensional (anisotropic) Quantum Spin Hall phase. In this paper, we adopt the point of view of Refs. \onlinecite{Ruegg,RachelThomale} and thoroughly analyze the emergent phase diagram as a function of the anisotropic spin-orbit coupling and the Hubbard interaction. In particular, the interplay of the Quantum Spin Hall physics and the Heisenberg-Kitaev magnetic model is yet to be studied, and the frustration effects about the magnetism is yet to be clarified. In contrast to the $S_z$ conserving models \cite{kane-mele,RachelLeHur,Hohenadler,Wu}, here we observe a full breakdown of the spin-rotation symmetry and some additional degree of magnetic frustration. Another relevant question to study concerns the role of the anisotropy on the  $\mathbb{Z}_2$ quantum Quantum Spin Hall state. Throughout this paper, we restrict ourselves to a model with an on-site Hubbard interaction on the honeycomb lattice and next-nearest-neighbor anisotropic spin-orbit coupling.

The paper is organized as follows. Firstly, we introduce the model and discuss the resulting band structure as well as the formation of a $\mathbb{Z}_2$ quantum Spin Hall phase in the presence of an anisotropic spin-orbit coupling.  We address a relatively simple case of zigzag edges with $x$ type links parallel to the boundary and show the dependence of the edge spin transport on the boundary. In particular, we illustrate how the two counter-propagating helical edge states, protected by the topological $\mathbb{Z}_2$ invariant, yield a spin polarization which depends on the relative strength of the spin-orbit coupling. In Sec. II, following Refs. \onlinecite{RachelThomale,Kargarian,Lhuillier}, we present a detailed investigation of the magnetism in which we highlight the frustration effects induced by the interplay between the nearest-neighbor spin exchange $J_1$ and the ``effective'' next nearest neighbor $J_2$ spin coupling. Such a frustration effect manifests itself already at the classical level and further when quantum fluctuations are taken into account. In Sec. III, we investigate the emergent Mott physics stemming from the Hubbard interaction especially the intermediate interacting regime through a U(1) slave-rotor approach \cite{Florens,Paramekanti,LeeLee,Miniatura} which has been applied to include topological effects \cite{pesin-10np376,RachelLeHur}. The emergent spin texture formation in the bulk above the critical Mott point due to the gauge fluctuations is clarified and its possible connection to the Spiral order at large interactions is addressed. The Mott transition is embodied by the disappearance of the helical edge states due to the breaking of time-reversal symmetry. Appendices are devoted to technical details.

\begin{figure}[h]
\includegraphics[width=1\linewidth]{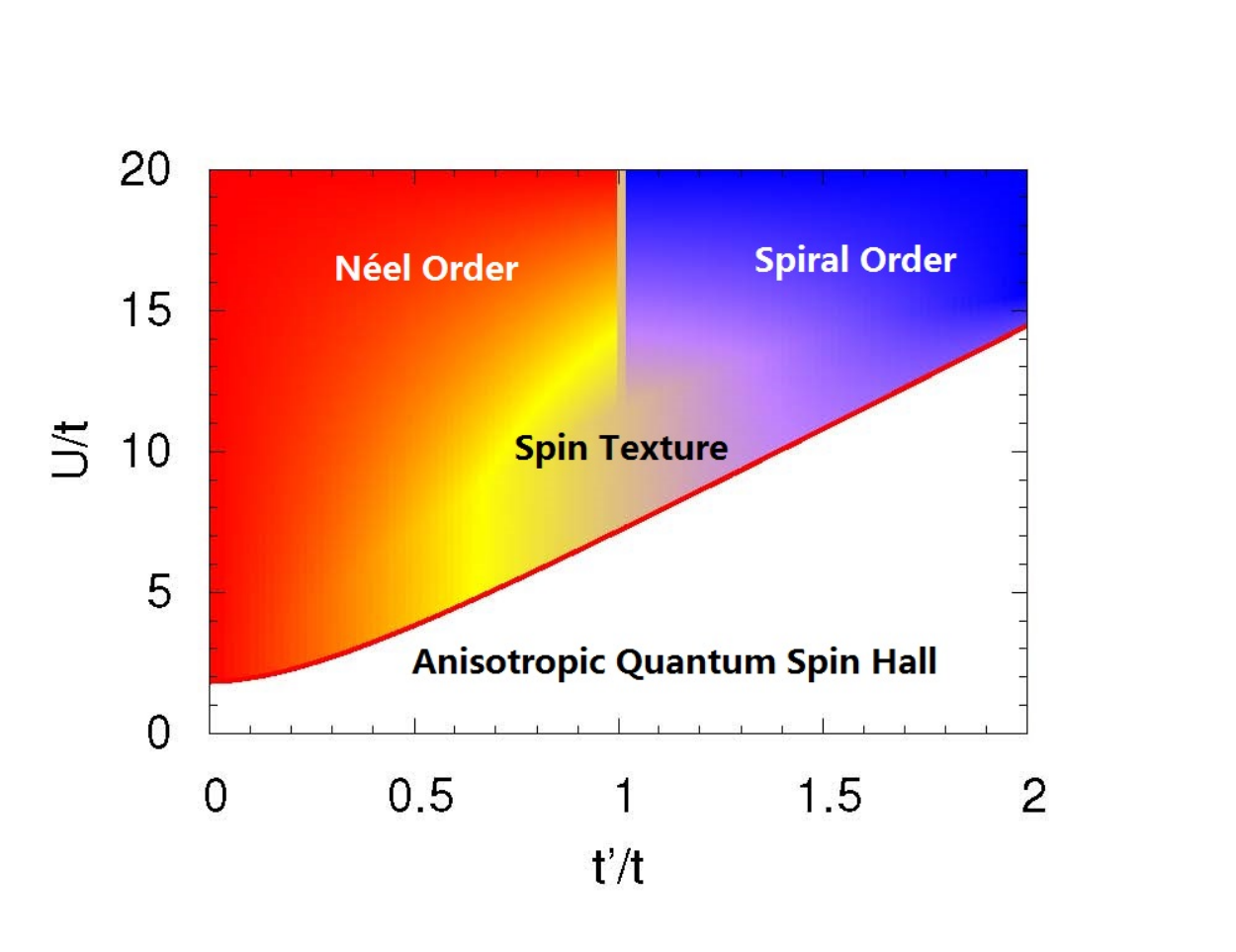}
\caption{(color online) Our Phase diagram. When $U<U_c$ (red line), the system is in the class of a $\mathbb{Z}_2$ two-dimensional topological band insulator. The edge modes are embodied by a peculiar spin texture as a result of the anisotropic spin-orbit coupling. We then refer to this phase as Anisotropic Quantum Spin Hall (AQSH) phase. Above the Mott critical point $U_c$, the spin texture now progressively develops into the bulk when increasing the spin-orbit coupling strength. At large interactions U, we identify two magnetic phases, the N\'eel and the Spiral phase.} \label{fig:mott_transition}
\end{figure}

\subsection{Model and Brief Summary of Results}

\begin{figure}[ht]
\includegraphics[width=0.85\linewidth]{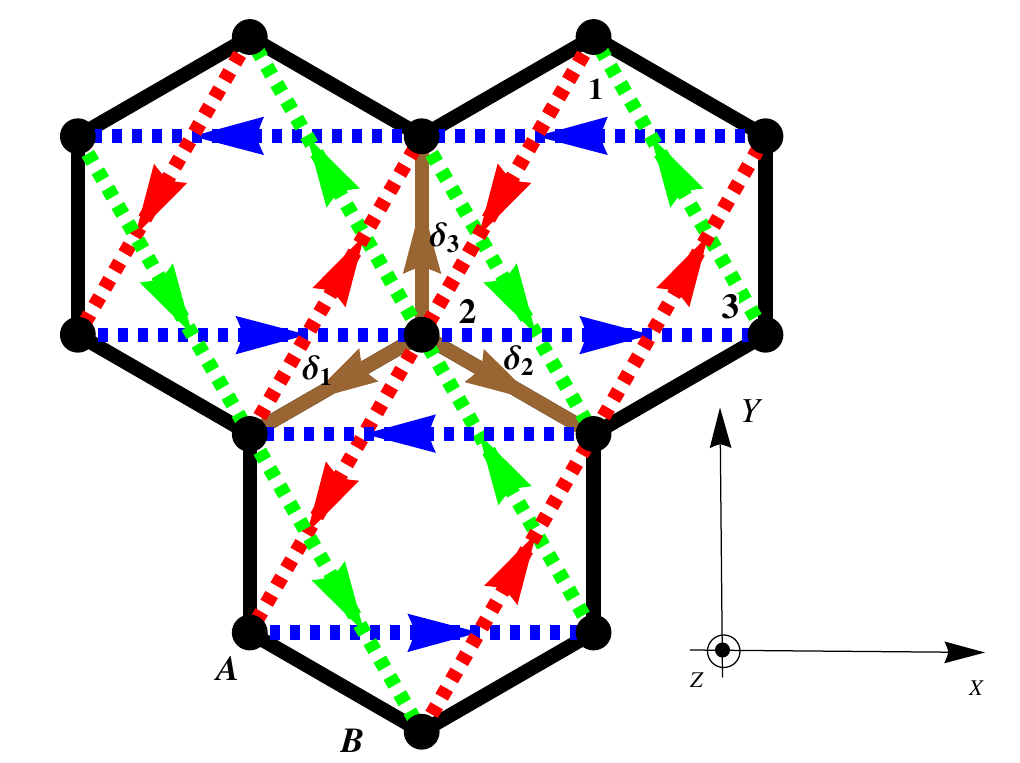}
\caption{(color online) Illustration of the tight-binding model on the honeycomb lattice with complex next-nearest-neighbor spin-orbit couplings entailing hopping of $it'\sigma_x$ on the $x$ red link, $it'\sigma_y$ on the $y$ green link, and $it'\sigma_z$ on the blue $z$ link, in which $\sigma_w$, $w=x,y,z$ is the Pauli matrix acting on the space of spins. The anisotropic spin-orbit coupling makes the spin no longer a conserved quantity in the system.} \label{fig:lattice}
\end{figure}

Hereafter, combining theoretical and numerical procedures, our primary goal is to carefully address the phase diagram summarized in Fig. \ref{fig:mott_transition} of the quite generic tight-binding model at half-filling on the honeycomb lattice with an Hubbard on-site interaction and next-nearest-neighbor anisotropic spin-orbit coupling. The physics of this model is potentially related to the correlated iridate compound Na$_2$IrO$_3$ \cite{Shitade} and possibly to other materials with spin-orbit coupling. The Hamiltonian consists of electrons hopping between nearest-neighbor sites with a strength $t$ similar to graphene and hopping between next-nearest-neighbor sites with a complex and anisotropic strength of $it'\sigma_x$, $it'\sigma_y$ and $it'\sigma_z$ in the counter-clockwise direction as in Fig. \ref{fig:lattice}.  For any finite $t'$, the sign in front of $t$ is not important. This model has been previously studied in the context of Quantum Spin Hall physics and magnetism \cite{Ruegg,Shitade,RachelThomale}. We add an on-site Hubbard interaction in order to describe the iridate family of strongly correlated materials.

The (sodium-iridate) model Hamiltonian is written as
\begin{eqnarray}\label{Shitade_Hubbard}
\begin{split}
H_{0}&=\sum_{\boldsymbol{<i,j>}}t c_{i\sigma}^{\dagger}c_{j\sigma}+\sum_{\boldsymbol{\ll i,j \gg}}i t'\sigma_{\sigma\sigma'}^w c_{i\sigma}^{\dagger}c_{j\sigma'}\\
H &= H_{0}+H_I\\
H_I &=\sum_i Un_{i\uparrow}n_{i\downarrow},
\end{split}
\end{eqnarray}
where $\left<i,j\right>$ denotes a sum over the nearest neighbor and $\ll i,j\gg$ denotes a sum over the next-nearest-neighbors, and $\sigma_{\sigma\sigma'}^w$ is a Pauli matrix with $w=x$ on the x link painted in red, $w=y$ on the y link painted in green and $w=z$ on the z link painted in blue as in Fig. \ref{fig:lattice}. To be precise, the hopping strengths of electrons on the next-nearest-neighbor links are denoted $it'\sigma_x$ on the red link $it'\sigma_y$ on the green link and $it'\sigma_z$ on the z link. Here, the electrons travel in a counterclockwise orientation. The second nearest-neighbor hopping strengths pick a minus sign if electrons travel in the clockwise orientation.

For the sake of clarity, in this Sec. I A, we present a brief summary of the results that will be shown subsequently. In the weak interaction limit, this model lies in the phase of a two-dimensional topological band insulator, in which the chemical potentiel lies between the valence and conduction bands, but edge states still exist and are protected by the $\mathbb{Z}_2$ topological invariant \cite{fu-kane} which can be generalized for interacting systems \cite{KrempaKim,GurarieWessel,Budich,Wang,Wang_SCZhang}.

\begin{figure}[htb]
\includegraphics[width=1.0\linewidth]{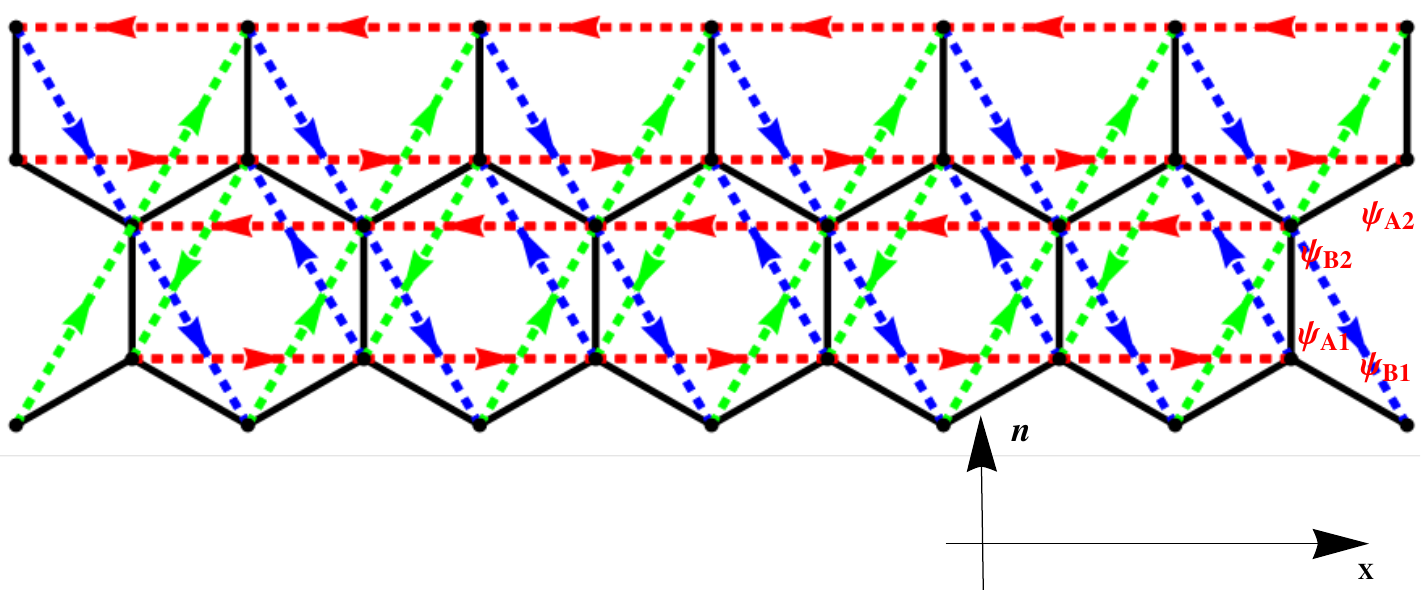}
\includegraphics[scale=0.55]{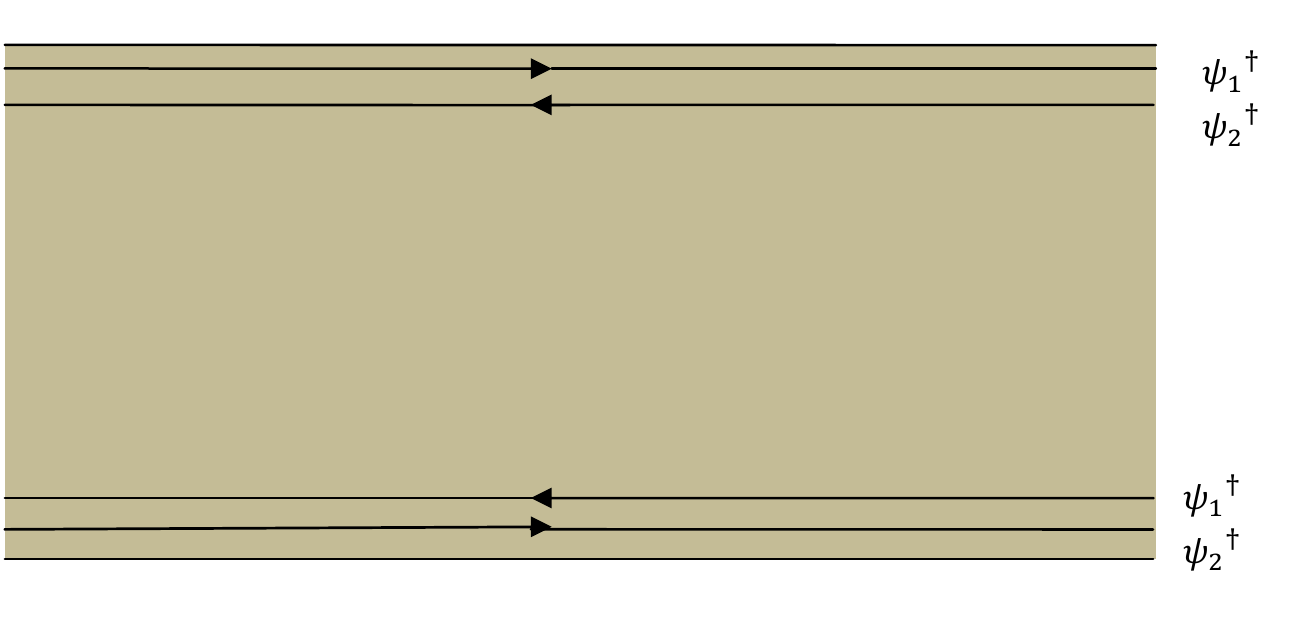}
\caption{(color online) Upper panel: The lower edge of the semi-infinite system with edges parallel to the x-type links. The system consists of layers of one-dimensional chains coupled together, and the edge mode decays exponentially when moving into the bulk. Lower panel: the chiral edge transport corresponding to the boundary configuration. Two helical edge modes with opposite spin polarization counter-propagate on the boundary of the system.}
\label{fig.edge_transport}
\end{figure}

\begin{figure}[htb]
\includegraphics[scale=0.6]{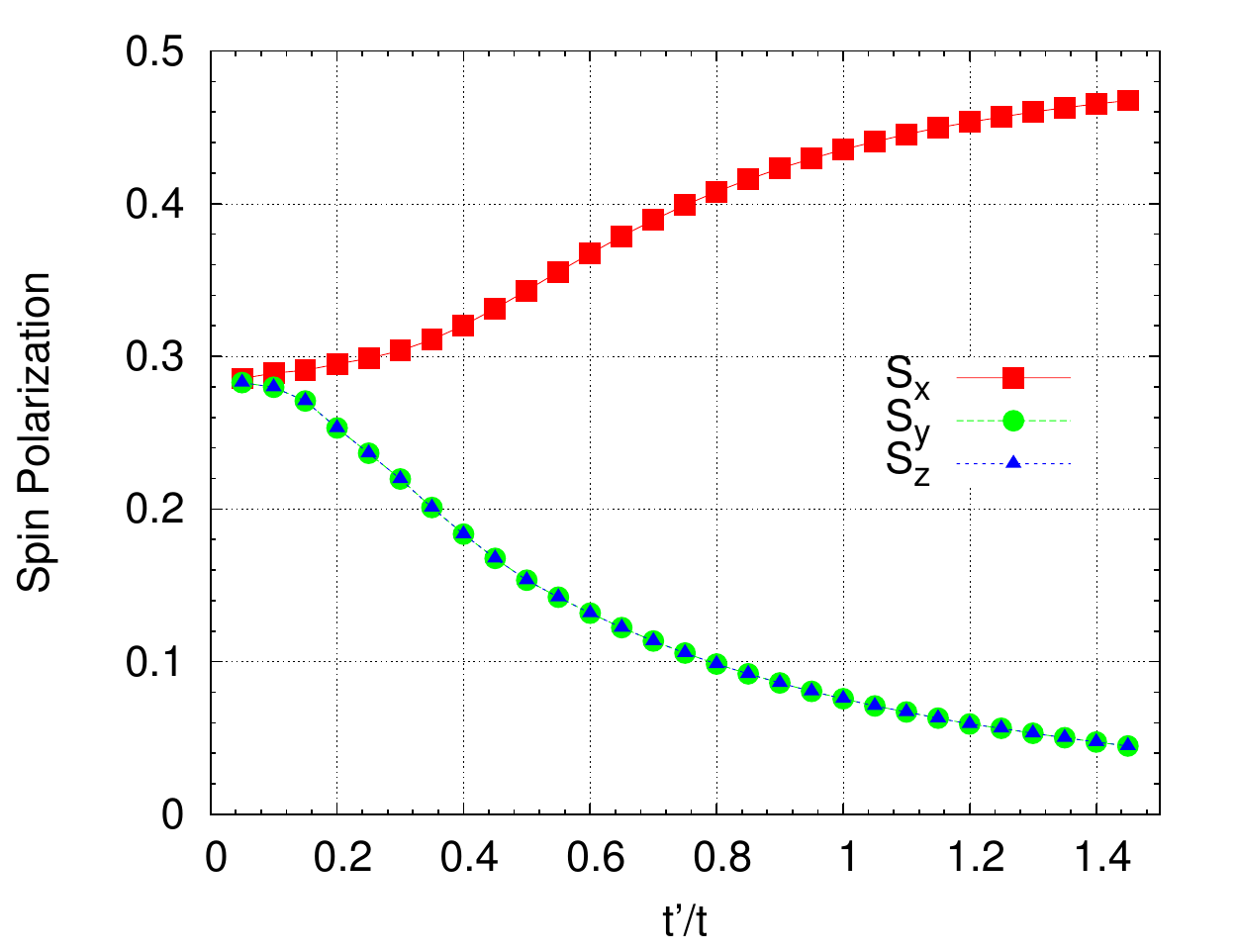}
\caption{(color online) In the weak interaction regime, the anisotropic spin-orbit model lies in the phase of a topological band insulator on a cylinder, in which only the wave-vector $k_x$ is a good quantum number. The two counter-propagating helical edge states protected by the $\mathbb{Z}_2$ topological invariant of the system have spin polarizations which explicitly depend on the ratio $t'/t$.  Here, we show the spin polarization components of the edge state with a wave-vector $k_x$ where $S_x$ is maximum, as a function of $t'/t$ (see, for example, Fig. 7). $S_x$ prevails over $S_y$ and $S_z$ at large $t'/t$.}\label{fig.edge_spin}
\end{figure}

 In this Anisotropic Quantum Spin Hall (AQSH) phase, spin is not conserved and spin current is not a well-defined quantity because of the anisotropic spin-orbit coupling while the edge spin physics depends highly on the ratio $t'/t$. To illustrate this point, we have studied the edge transport in the case of zigzag boundaries as in Fig. \ref{fig.edge_transport} applying the transfer matrix method summarized in Appendix \ref{Transfer_matrix} and numerical diagonalization of the system on a cylinder in Sec. \ref{spin_transport}. On the two edges of the system, we identify two counter-propagating helical spin states with opposite polarizations as a reminiscence of the Kane-Mele model \cite{kane-mele}. As shown in Fig. \ref{fig.edge_spin}, when $t'/t$ is small, the spin polarization has equal components in the $x$, $y$ and $z$ directions, and when $t'/t$ is large, one spin polarization component dominates  and this dominant spin polarization coincides with the type of next-nearest-neighbour links parallel to the boundary (see  Fig. \ref{fig.edge_spin}), which implies that helical edge states point in $x$ ($y$, $z$) direction if the two edges are parallel to the $x$ ($y$ and $z$) type link, respectively.  At a mean-field level, the interaction adds an effective chemical potential and the AQSH phase is robust as long as the chemical potential does not touch the conduction (valence) band \cite{RachelLeHur}.

When the Hubbard on-site interaction and the spin-orbit coupling become large enough, the topological band insulator phase will be affected by Mott physics. In particular, the helical edge states will fade away \cite{Wu} without invoking the closing of the single-particle excitation gap in the bulk. We resort to the U(1) slave-rotor representation \cite{Florens,Paramekanti,LeeLee} to investigate the occurrence of Mott physics in the system instead of the slave-spin representation  \cite{Biermann,Nand} which in the present model can accomodate extra vison excitations \cite{Ruegg}.

\begin{figure}[h]
\includegraphics[scale=0.6]{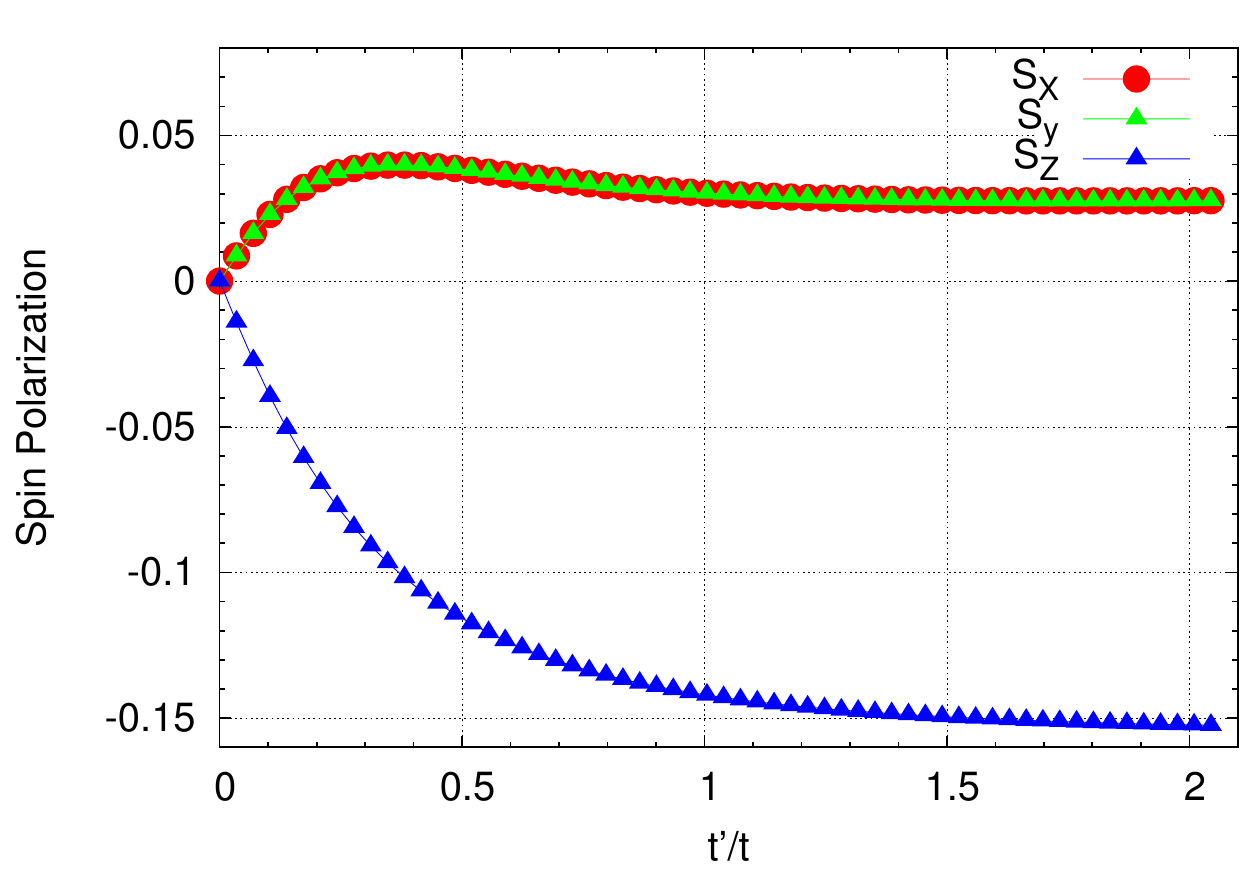}
\caption{(color online) Spin texture in the intermediate U regime induced by the fluctuating gauge field within the U(1) slave-rotor theory. The spin polarization on site $1$ in Fig. \ref{fig:lattice} as a function of $t'/t$. When $t'/t\ll 1$,  the subordinate spin polarization is in the same order as the dominant spin polarization $S_x,S_y\approx -0.6 S_z$ (see Fig. 14). When $t'/t> 1$ the subordinate spin polarization becomes (much) smaller in front of the dominant polarization: $S_x,S_y\approx -0.2 S_z$. The spin texture above the Mott quantum critical point seems to evolve very gradually. Site $1$ is facing the z type links in the system and it acquires a dominant z spin component. The spin texture on other different sites carries a symmetry which is a combination of a $2\pi/3$ rotation around the core of the fluctuating flux and a spin permutation, a symmetry inherent to this anisotropy model.} \label{fig:spin_texture_diagram}
\end{figure}

Beyond the Mott critical point, electrons are fractionalized into chargeons and spinons, and the chargeons are localized whereas the emergent fluctuating gauge field in our theory will induce the spinons to form a spin texture around the fluctuating flux. This spin texture could be then interpreted as a precursory effect of the formation of magnetic order above the Mott critical point. The spinon response to the gauge fluctuation (insertion of monopoles) will be explicitly computed in Sec. \ref{spin_texture_formation}. We also notice that the spin texture is very sensitive to the strength of the spin-orbit coupling $t'/t$, as summarized in Fig. \ref{fig:spin_texture_diagram}, which is  much analogous to the edge spin physics of the AQSH phase.

The anisotropy is embodied by the fact that the dominant spin polarization on a given site coincides with the type of spin-orbit coupling on the next-nearest-neighbour link that it is confronted with when facing the core of the inserted flux (site $1$ with $z$, site $2$ with $y$ and site $3$ with $x$).  If a fluctuating flux is inserted into the center of the plaquette with sites $1$,$2$ and $3$, we focus on the spin texture on a z type site $1$ as in Fig. \ref{fig:lattice}; when $t'/t\ll 1$ the spin projections satisfy $S_x,S_y\approx-0.6S_z$ while for $t'/ t>1$ then $S_x,S_y\approx -0.2 S_z$, as exemplified in Fig. \ref{fig:spin_texture_diagram}. Thanks to the inherent symmetry of a combination of a $2\pi/3$ rotation and spin permutation of the anisotropic spin-orbit coupling model, the spin texture preserves this symmetry of the rotation around the core of the inserted flux and spin permutation. Namely, the symmetry operator $U=R(\frac{2\pi}{3})\sigma$ commutes with the Hamiltonian. $R(\frac{2\pi}{3})$ is a $2\pi/3$ rotation around the flux core: $R(\frac{2\pi}{3})\vec{r}_1=\vec{r}_2$, $R(\frac{2\pi}{3})\vec{r}_2=\vec{r}_3$ and $R(\frac{2\pi}{3})\vec{r}_3=\vec{r}_1$, in which $\vec{r}_1,\vec{r}_2,\vec{r}_3$ are the coordinates of the sites $1$, $2$, $3$ indicated in Fig. \ref{fig:lattice}. The permutation $\sigma$ gives $\sigma(S_z)=S_y$, $\sigma(S_y)=S_x$, and $\sigma(S_x)=S_z$. Under the inversion symmetry with respect to the localized flux the spin polarization is reversed. The prevalent spin texture(s) developing by increasing the ratio $t'/t$ in the intermediate interaction regime can be related to the edge transport in the AQSH phase, by applying an analogy of the Laughlin's U(1) charge pump argument \cite{laughlin} of U(1) flux insertion onto the cylinder. Here the spin texture formation is rather associated with the spin pump under the fluctuating fluxes above Mott critical point.

The formation of spin texture in the bulk above the Mott critical point breaks time-reversal symmetry resulting in the disappearance of the edge modes. The Mott transition is manifested by the peculiar magnetism driven by the spin-orbit coupling and interactions as well as the destruction of edge transport.

\begin{figure}[b]
\includegraphics[scale=0.3]{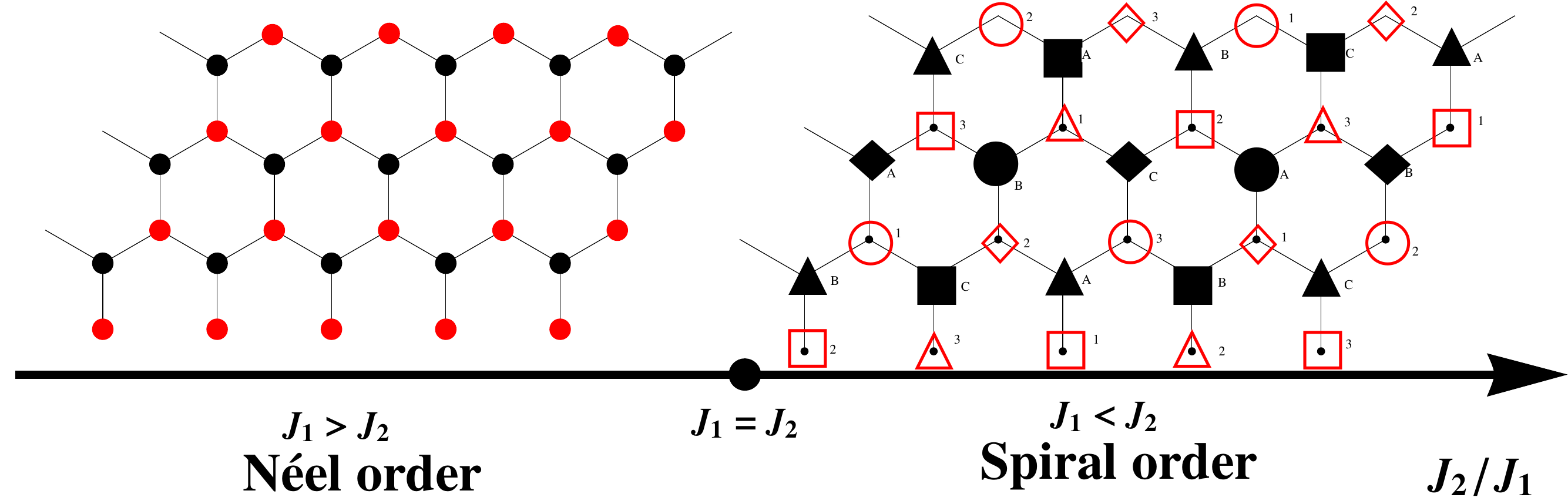}
\caption{(color online) The magnetic phase diagram for the tight-binding model with anisotropic spin-orbit coupling on the honeycomb lattice in the limit of infinite $U$ described by Eq. \ref{J_1_J_2_magnetism}. The $J_1-J_2$ model is highly frustrated because of the hexagonal geometry and the anisotropy of the $J_2$ coupling.  We identify the bipartite N\'eel phase at $J_1>J_2$, the Spiral phase with $24$ sublattices at $J_1<J_2$ and both phases are frustrated either at the classical or the quantum level.} \label{fig:phase_diagram}
\end{figure}

Another relevant result found in this paper concerns the type of magnetism at large interactions. The super-exchange magnetism is investigated in the strong coupling limit as shown in Fig. \ref{fig:phase_diagram}. The nearest-neighbor hopping and the next-nearest-neighbor anisotropic spin-orbit coupling now mimic the $J_1$ \& $J_2$ model with $J_2$ being the Heisenberg-Kitaev coupling.  The magnetic phase diagram of this $J_1$ \& $J_2$ model will be studied using a combination of spin-wave theory and mathematical arguments, then complementing the previous analysis of Ref. \cite{RachelThomale} obtained using a fermionic functional Renormalization Group approach and exact diagonalization \cite{Kargarian}. More specifically, we find a N\'eel order when $J_1>J_2$ and a two-copy locked non-colinear Spiral order with $24$ sublattices when $J_1<J_2$. Both magnetic phases are highly frustrated due to the anisotropic spin-orbit coupling and the geometry of the lattice.

For the N\'eel phase, the $J_2$ coupling frustrates the magnetic order at a quantum level.  The N\'eel order parameter is fixed in the $x$, $y$ and $z$ direction since the zero-point energy would be higher in other directions. As a result of the anisotropy, the Goldstone mode in this frustrated N\'eel magnetic order develops a gap.

For the Spiral phase, the Heisenberg-Kitaev coupling and the triangular geometry tend to imply an enlarged unit cell with four patterns spiraling forward along one direction on the two triangular lattices; see Fig. \ref{fig:phase_diagram}. This conclusion is in agreement with Ref. \onlinecite{RachelThomale} (however, we diagree on the ordering wave-vectors associated with the Spiral phase). The spiral order can be viewed as a $120^{\circ}$ N\'eel order with four patterns giving a $12$ sublattice magnetic order on each triangular sublattice, and the nearest-neighbor anti-ferromagnetic coupling then locks the two transformed $120^{\circ}$ N\'eel orders. Analogously for $J_2>J_1$, switching on the $J_1$ term lifts the massive ground state degeneracy of the Spiral phase found for $J_1=0$.

\subsection{The Anisotropic Quantum Spin Hall Phase}\label{spin_transport}

\begin{figure}[htb]
\includegraphics[scale=0.5]{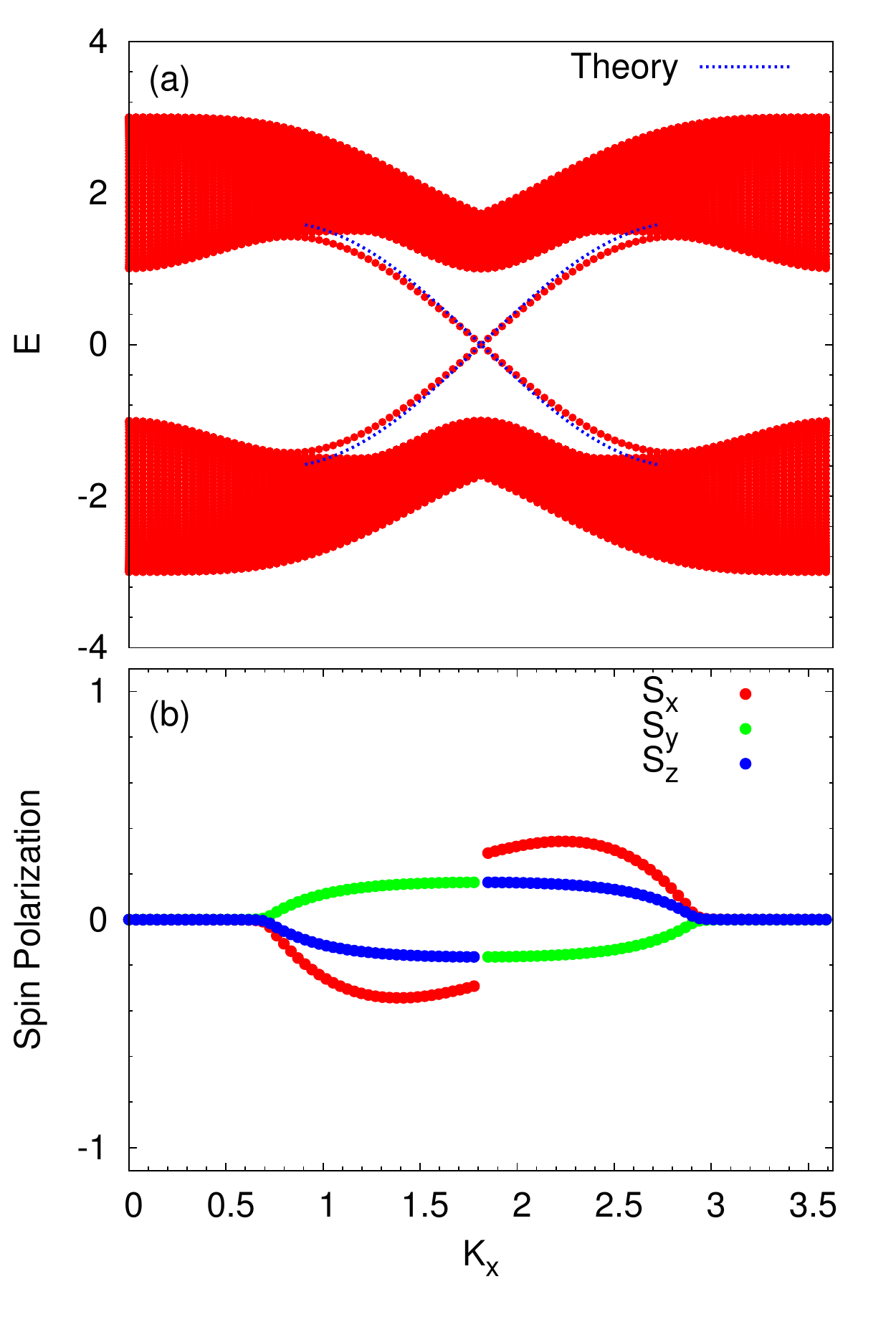}
\caption{(color online) The edge states of the anisotropic spin-orbit coupling model with zigzag boundary and $x$ links parallel to the boundary. (a): Spectrum of a system with anisotropic spin-orbit coupling on a cylinder at $t'/t=0.5$ obtained from numerical diagonalization of 70 layers of a one-dimensional system described by the Schr\"{o}dinger equation \ref{edge_layers}. The non-trivial $\mathbb{Z}_2$ topological invariants ensures an helical edge states with opposite spin polarization according to Kramers theorem. The energy dispersion obtained analytically using transfer matrix in Appendix \ref{Transfer_matrix} fits well the numerics. (b): The different components of the spin polarization measured on the lower edge of the state with the lowest positive energy in the spectrum as a function of momentum obtained from diagonalization of the system. We observe that states with opposite Fermi velocities on both sides of $k_x=\frac{\pi}{\sqrt{3}}$ have opposite spin polarizations, thus implying helical spin transport on the edge and the dominant spin component corresponds to the type of links parallel to the boundary.}\label{fig:shitade_cylinder}
\end{figure}

Here, we flesh out the theoretical investigations of the $\mathbb{Z}_2$ two-dimensional Quantum Spin Hall phase, in the presence of such a spin-orbit anisotropy \cite{Shitade}. Neglecting the Hubbard interaction in the first place, we diagonalize the tight binding model $H_{\text{0}}$ by Fourier transformation:
\begin{equation}
H_{\text{0}}=\sum_{\vec{k}} \Psi_{\vec{k}}^\dagger h(\vec{k}) \Psi_{\vec{k}}
\end{equation}
 in which the wave function in the momentum representation exhibits four components $\Psi_{\vec{k}}^\dagger=(a_{\vec{k}\uparrow}^\dagger, b_{\vec{k}\uparrow}^{\dagger},a_{\vec{k}\downarrow}^\dagger,b_{\vec{k}\downarrow}^\dagger)$ and the two sublattices of the honeycomb (A and B) give rise to the corresponding electron creation operators $a^\dagger$ and $b^{\dagger}$. We then identify
\begin{eqnarray}
h(\vec{k}) &=& (\tau_x \Re e +\tau_y \Im m)g(\vec{k}) \\ \nonumber
&+& (m_x\sigma_x+m_y\sigma_y+m_z\sigma_z)\tau_z,
\end{eqnarray}
where $\tau_{x}$, $\tau_{y}$ and $\tau_{z}$ are Pauli matrices acting on the sublattice isospin A \& B while $\sigma_x$, $\sigma_y$ and $\sigma_z$ are Pauli matrices acting on the spin space $\uparrow$ and $\downarrow$.

For convenience, we have introduced the notations
\begin{equation}
g(\vec{k})=\sum_{\boldsymbol i} t e^{i\vec{k}\cdot \vec{\delta}_i}
\end{equation}
and
\begin{eqnarray}
m_x &=&2t'\sin(\vec{k}\cdot\vec{R}_x) \\ \nonumber
 m_y&=&2t'\sin(\vec{k}\cdot\vec{R}_y) \\ \nonumber
 m_z &=&2t'\sin(\vec{k}\cdot\vec{R}_z).
 \end{eqnarray}
  Here, $\vec{\delta}_1=(-\frac{\sqrt{3}}{2},-\frac{1}{2})a$, $\vec{\delta}_2=(\frac{\sqrt{3}}{2},-\frac{1}{2})a$ and $\vec{\delta}_3=(0,1)a$ refer to vectors connecting the nearest neighbours (see Fig. 1), while $\vec{R}_x=(-\frac{\sqrt{3}}{2},-\frac{3}{2})a$, $\vec{R}_y=(-\frac{\sqrt{3}}{2},\frac{3}{2})a$ and $\vec{R}_z=(\sqrt{3},0)a$ represent vectors connecting next nearest neighboring sites. Moreover, $a$ is the length of a bond on a given hexagon and we set it equal to $1$ in the rest of the article for convenience.

The Hamiltonian represents a two band system with two energy levels:
\begin{eqnarray}
E(\vec{k})&=& \pm E_0(\vec{k}) \\ \nonumber
&=&\pm\sqrt{m_x^2(\vec{k})+m_y^2(\vec{k})+m_z^2(\vec{k})+|g(\vec{k})|^2}.
\end{eqnarray}
The system is an insulator with a gap $\Delta(k)=2E_0(k)$.

Each band is doubly degenerate and it is convenient to introduce the band projectors associated to the upper and lower band $P_{\pm}$ respectively such that $2P_{\pm}$ is equal to
\begin{equation}
\left[1\pm\left(\frac{\tau_x\Re e g}{E_0}+\frac{\tau_y\Im m g}{E_0}+\frac{\tau_z}{E_0}(m_x\sigma_x+m_y\sigma_y+m_z\sigma_z)\right)\right].
\end{equation}

The non-trivial topology is encoded in the $\mathbb{Z}_2$ invariant \cite{fu-kane} namely the product of the time-reversal polarization for the four time-reversal and inversion symmetric points:
\begin{equation}
 (-1)^{\nu}=\prod_{i=1}^4 \gamma_i=-1;
 \end{equation}
here, we have defined $\gamma_i=-\hbox{sgn}(\Re e g(\Gamma_i))$ and $\Gamma_i=(0,0);(0,\frac{2\pi}{3});(\frac{\pm\pi}{\sqrt{3}},\frac{2\pi}{3})$. The $\mathbb{Z}_2$ topological invariant depicts a twist of the rank 2 ground-state wave function in the first Brillouin zone.

As a result of the non-conservation of the spin in the system, the spin polarization of the helical edge states is more sophisticated than in the Kane-Mele model. To thoroughly analyze this point, we consider a system with two zigzag boundaries as layers of one-dimensional chains coupled together as illustrated in Fig. \ref{fig.edge_transport}. The Schr\"{o}dinger equation of such a system takes the form:
\begin{widetext}
\begin{eqnarray}\label{edge_layers}
&&\left(\begin{array}{cc}
-it'(e^{-i\frac{\sqrt{3}}{2}k_x}\sigma_z-e^{i\frac{\sqrt{3}}{2}k_x}\sigma_y) & -t  \\
0 & it'(e^{-i\frac{\sqrt{3}}{2}k_x}\sigma_z-e^{i\frac{\sqrt{3}}{2}k_x}\sigma_y)
\end{array}
\right)\left(\begin{array}{c} \psi_A^{n+1} \\ \psi_B^{n+1} \end{array} \right) \\
&+&
\left(\begin{array}{cc}
E+2t'\sin\sqrt{3}k_x\sigma_x & -2t\cos\frac{\sqrt{3}}{2}k_x  \\
-2t\cos\frac{\sqrt{3}}{2}k_x & E-2t'\sin\sqrt{3}k_x\sigma_x
\end{array}
\right)\left(\begin{array}{c} \psi_A^{n} \\ \psi_B^{n} \end{array} \right)\nonumber\\
&+&\left(\begin{array}{cc}
it'(e^{i\frac{\sqrt{3}}{2}k_x}\sigma_z-e^{-i\frac{\sqrt{3}}{2}k_x}\sigma_y) & 0  \\
-t & -it'(e^{i\frac{\sqrt{3}}{2}k_x}\sigma_z-e^{-i\frac{\sqrt{3}}{2}k_x}\sigma_y)
\end{array}
\right)\left(\begin{array}{c} \psi_A^{n-1} \\ \psi_B^{n-1} \end{array} \right)=0.
\end{eqnarray}
\end{widetext}
We then perform a numerical diagonalization of such a system with 70 layers of one-dimensional chains (see Fig. \ref{fig:shitade_cylinder}) and a purely analytical transfer matrix approach is developed in Appendix \ref{Transfer_matrix}. We address a system with boundaries parallel to the $x$-type links and the resulting spin polarization depends on how the system is cut and on the ratio $t'/t$. We observe that there are two edge modes crossing the gap connecting the upper and lower bands according to the results obtained from the numerical diagonalization presented in Fig. \ref{fig:shitade_cylinder} (upper panel).

We have studied the spin polarization of the lowest positive energy state by measuring its spin polarization on the boundary: spin have opposite components respectively at $k_x>\frac{\pi}{\sqrt{3}}$ and $k_x<\frac{\pi}{\sqrt{3}}$; since the Fermi velocity in these two intervals separated by $k_x=\frac{\pi}{\sqrt{3}}$ are opposite as well, this implies two counter-propagating states with opposite spin polarization. The energy dispersion obtained analytically in Appendix A fits well the edge states plotted in the spectrum in Fig. \ref{fig:shitade_cylinder}.  As a result, we have two counter-propagating states with linear energy dispersion in the spectrum on both upper and lower edges: the state with one polarization propagating to the left (right) on the lower (upper) edge and the state with the opposite polarization propagating to the right (left) on the lower (upper) edge as in Fig. \ref{fig.edge_transport}. The time-reversal symmetry forbids the (elastic) backscattering allowing for helical edge spin transport.

\begin{figure}[htb]
\includegraphics[scale=0.35]{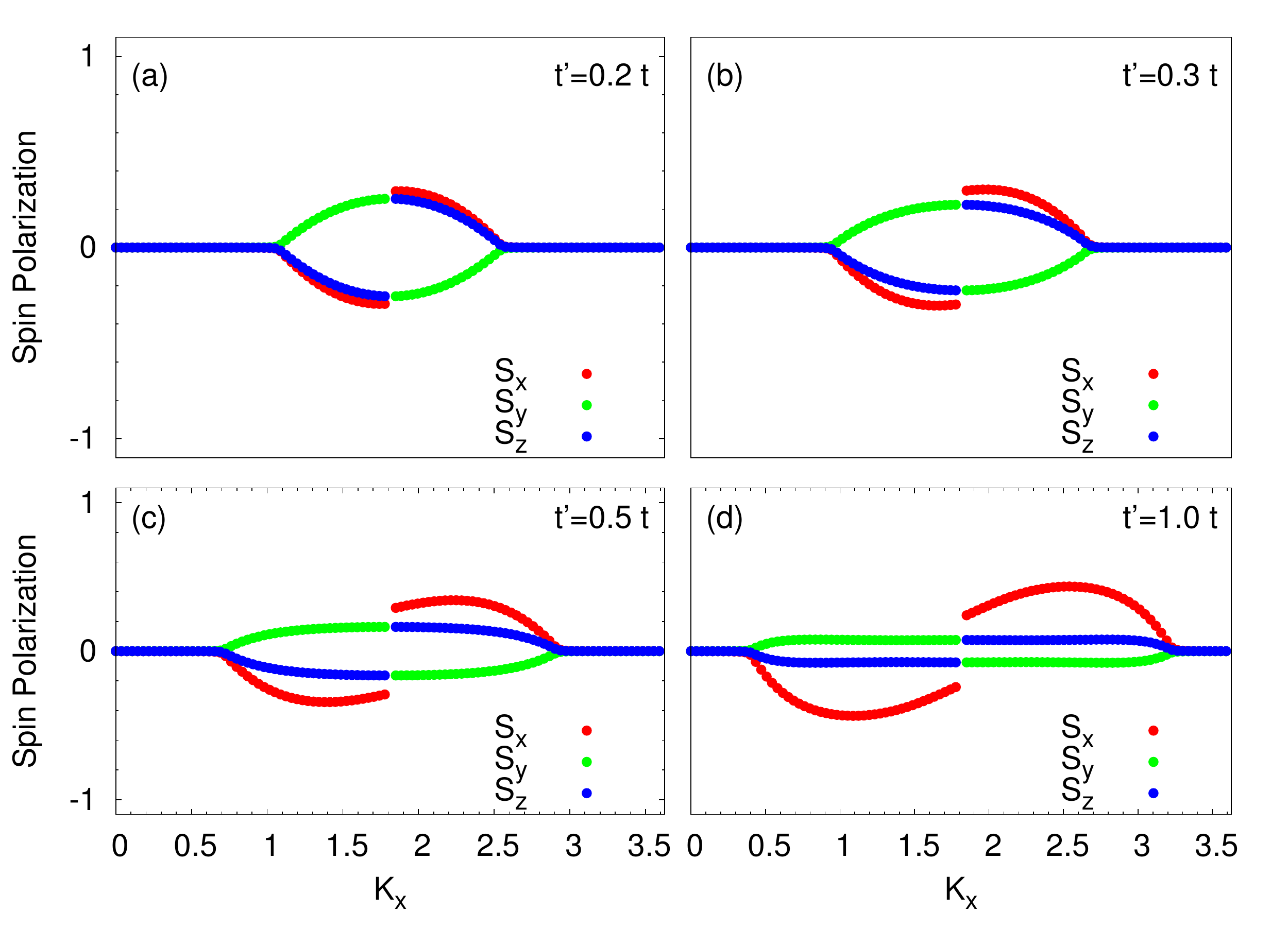}
\caption{(color online) Spin polarization of the lowest positive energy state  for the exactly diagonalized Hamiltonian on a cylinder with x links parallel to the boundary (see Fig. 3). $k_x$ refers to the wavevector along the boundary. Spin polarization at the edge for (a) $t'=0.2t$, (b) $t'=0.3t$, (c) $t'=0.5t$, (d) $t'=1.0t$. The $x$ component becomes dominant when $t'/t$ increases. The spin polarization at the momentum $k_x$ with the maximal dominant component is shown in Fig. \ref{fig.edge_spin} as a function of $t'/t$.}
\label{fig.edge_transport_spin}
\end{figure}

Consequently, the effective Hamiltonian on the lower edge can be described as a helical Luttinger liquid with two types of wave functions $\ket{\Psi_1}$, $\ket{\Psi_2}$ with opposite spin polarizations (see Fig. 3). The spin polarization of the two helical states, which varies as a function of $t'/t$, is studied using exact diagonalization of the system on a cylinder. As shown in Fig. \ref{fig.edge_transport_spin} lower panel, when $t'\ll t$ the helical states have equal components in all spin polarizations; when $t'/t$ increases the helical states have a x component gradually dominating the spin polarization.

At a general level, one can show either using a mean-field type argument or by invoking the U(1) slave-rotor theory \cite{RachelLeHur}, that such a Quantum Spin Hall phase is robust towards finite to moderate interactions. The notion of topological invariants can also been extended for an interacting system \cite{KrempaKim,GurarieWessel,Budich}. In Sec. III, we shall study in more details the emergence of the Mott transition resulting in the
disappearance of the helical edge modes.

\section{Magnetism}\label{magnetism}

In this Section, we investigate the magnetism emerging in the limit of ``infinite'' interactions, the possible magnetic orders and the phase transition(s) between these phases. This analysis complements the recent analysis performed via a fermionic functional renormalization group method \cite{RachelThomale} and via exact diagonalization \cite{Kargarian}. Since the electron-hole excitations in this limit would cost an energy proportional to $U$, electrons are subject to virtual tunneling  processes in which they exchange their positions while leaving the filling unchanged. The induced super-exchange magnetism is a second-order process in $H_0$:

\begin{equation}\label{J_1_J_2_magnetism}
H_{J_1J_2}=J_1\sum_{\boldsymbol{<i,j>}}\vec{S}_i\cdot\vec{S}_j+J_2\sum_{\boldsymbol{\ll i,j\gg}}(S_i^wS_j^w-S_i^uS_j^u-S_i^vS_j^v)
\end{equation}
where $J_2=2t'^2/U$ and $J_1=2t^2/U$.

The term with $J_2$ indicates a next nearest-neighbor link in $w$ spin polarization, with $w=x,y,z$ on respectively red, green and blue links in Fig. \ref{fig:lattice}, $u$ and $v$ are other spin polarizations than $w$. The coexistence of first and second neighbor couplings, the anisotropy in the next-nearest-neighbour coupling as well as the lattice geometry implies frustrated magnetism under which different scenarios like enlarged unit cells, disappearance of Goldstone modes and reduction of possible classical ground states would be concerned.

When evaluating the classical energy of the magnetic order, we identify two magnetic phases: the N\'eel order at $J_1>J_2$ and the two copies of locked Spiral order on the two triangular sublattices at $J_1<J_2$ with the critical point $J_1=J_2$ as in Fig. \ref{fig:phase_diagram}. We also performed a spin wave analysis based on the classical magnetic order. Analytical and numerical investigations of the magnetism at all $J_1/J_2$ ratios are presented below, for completeness. We recover the existence of a quantum phase transition at $J_1\approx J_2$. It shall be noted that $J_1-J_2$ (also including $J_3$) spin models have been studied in various contexts \cite{J1J2,spinl,Bishop,Lhuillier}.

\subsection{N\' eel Phase for $J_1>J_2$}

The magnetic phase in the $J_1\gg J_2$ regime is the well-known bipartite N\'eel order on the bipartite honeycomb lattice: $\vec{S}_A=-\vec{S}_B$ and the classical energy of this state per site is $E_{\text{N\'eel}}=-\frac{3J_1}{2}\vec{S}^2-J_2\vec{S}^2$.

In the absence of next-nearest-neighbor frustration, there exists a Goldstone mode underlying the whole original continuous spin symmetry $SU(2)$ on the unit sphere. At the level of this N\'eel order, we carried out a semiclassical spin wave analysis in order to compute the quantum corrections to the energy of the N\'eel state. The anisotropy in the $J_2$ coupling lifts the degeneracy between the different possible orientations of the N\' eel order parameter.

We begin by writing the Holstein-Primakoff representation of the spin in the $z$ polarization, then we rotate the $z$ quantization axis by the Euler rotation matrix in order to describe quantum fluctuations in all the spontaneously broken symmetry cases: we rotate the z axis first around the y axis by an angle $\theta$ then around z axis by an angle $\phi$, resulting in

\begin{equation}
R(\phi,\theta)=R_z(\phi)R_y(\theta)=\left(\begin{array}{ccc} \cos\theta & -\sin\theta & 0 \\ \cos\phi\sin\theta & \cos\phi\cos\theta & -\sin\phi \\ \sin\phi\sin\theta & \sin\phi\cos\theta & \cos\phi \end{array}\right)
\end{equation}

\begin{equation}\label{quantification_axis}
\left(\begin{array}{c}S_{A0}^z \\ S_{A0}^x \\ S_{A0}^y\end{array}\right)=\left(\begin{array}{c}
            S-a^{\dagger}a \\
            \frac{\sqrt{2S}}{2}(a^{\dagger}+a)\\
            \frac{\sqrt{2S}}{2i}(a-a^{\dagger})
            \end{array}\right),
\end{equation}
\begin{equation}
\left(\begin{array}{c}S_{B0}^z \\ S_{B0}^x \\ S_{B0}^y\end{array}\right)=\left(\begin{array}{c}
            -S+b^{\dagger}b \\
            \frac{\sqrt{2S}}{2}(b^{\dagger}+b)\\
            \frac{\sqrt{2S}}{2i}(b^{\dagger}-b)
            \end{array}\right),
\end{equation}

\begin{equation}
\left(\begin{array}{c}S_{A}^z \\ S_{A}^x \\ S_{A}^y\end{array}\right)=R(\phi,\theta)\left(\begin{array}{c}S_{A0}^z \\ S_{A0}^x \\ S_{A0}^y\end{array}\right),
\end{equation}
\begin{equation}
\left(\begin{array}{c}S_{B}^z \\ S_{B}^x \\ S_{B}^y\end{array}\right)=R(\phi,\theta)\left(\begin{array}{c}S_{B0}^z \\ S_{B0}^x \\ S_{B0}^y\end{array}\right).
\end{equation}

We insert the above semiclassical spin representation back into Eq. \ref{J_1_J_2_magnetism}, then we will obtain the Bogoliubov-De Gennes type effective Hamiltonian describing the quantum fluctuation about the N\'eel state:
\begin{equation}
H=\sum_{\vec{q}}\Phi_{\vec{q}}^{\dagger} H_{\vec{q}} \Phi_{\vec{q}}-\frac{J_1}{2}NS^2z-J_2NS^2,
\end{equation}
 where $\Phi_{\vec{q}}^{\dagger}=(a_{\vec{q}},b_{-\vec{q}}^{\dagger},a_{-\vec{q}}^{\dagger},b_{\vec{q}})$, $z=3$ is the coordinate number, $N$ the number of sites, and we define

\begin{widetext}
\begin{equation}
H_q=\left(\begin{array}{cccc} \gamma_z & \gamma_{\vec{q}}^{\star} & \gamma_{xy}^{\star} & 0 \\ \gamma_{\vec{q}} & \gamma_z & 0 & \gamma_{xy}^{\star} \\ \gamma_{xy} & 0 & \gamma_z & \gamma_{\vec{q}}^{\star} \\ 0 & \gamma_{xy} & \gamma_{\vec{q}} & \gamma_z \end{array}\right),
\end{equation}

\begin{eqnarray}
\begin{split}
\gamma_{\vec{q}}&=J_1S\sum_i\exp(i\vec{q}\cdot\vec{\delta}_i)\\
\gamma_z&=3J_1S+2J_2S-2J_2S[\cos^2\phi\sin^2\theta\cos(\vec{q}\cdot\vec{R}_x)+\sin^2\phi\sin^2\theta\cos(\vec{q}\cdot\vec{R}_y)+\sin^2\theta\cos(\vec{q}\cdot\vec{R}_z)]\nonumber\\
\gamma_{xy}&=J_2S[\exp(i\vec{q}\cdot\vec{R}_z)\sin^2\theta+\exp(i\vec{q}\cdot\vec{R}_x)(\cos^2\phi\cos^2\theta\-\sin^2\phi-i\sin2\phi\cos\theta) \nonumber\\
&+\exp(i\vec{q}\cdot\vec{R}_y)(\sin^2\phi\cos^2\theta-\cos^2\phi+i\sin2\phi\cos\theta)].
\end{split}
\end{eqnarray}
\end{widetext}

\begin{figure}[t]
\vskip -0.6cm
\includegraphics[scale=0.7]{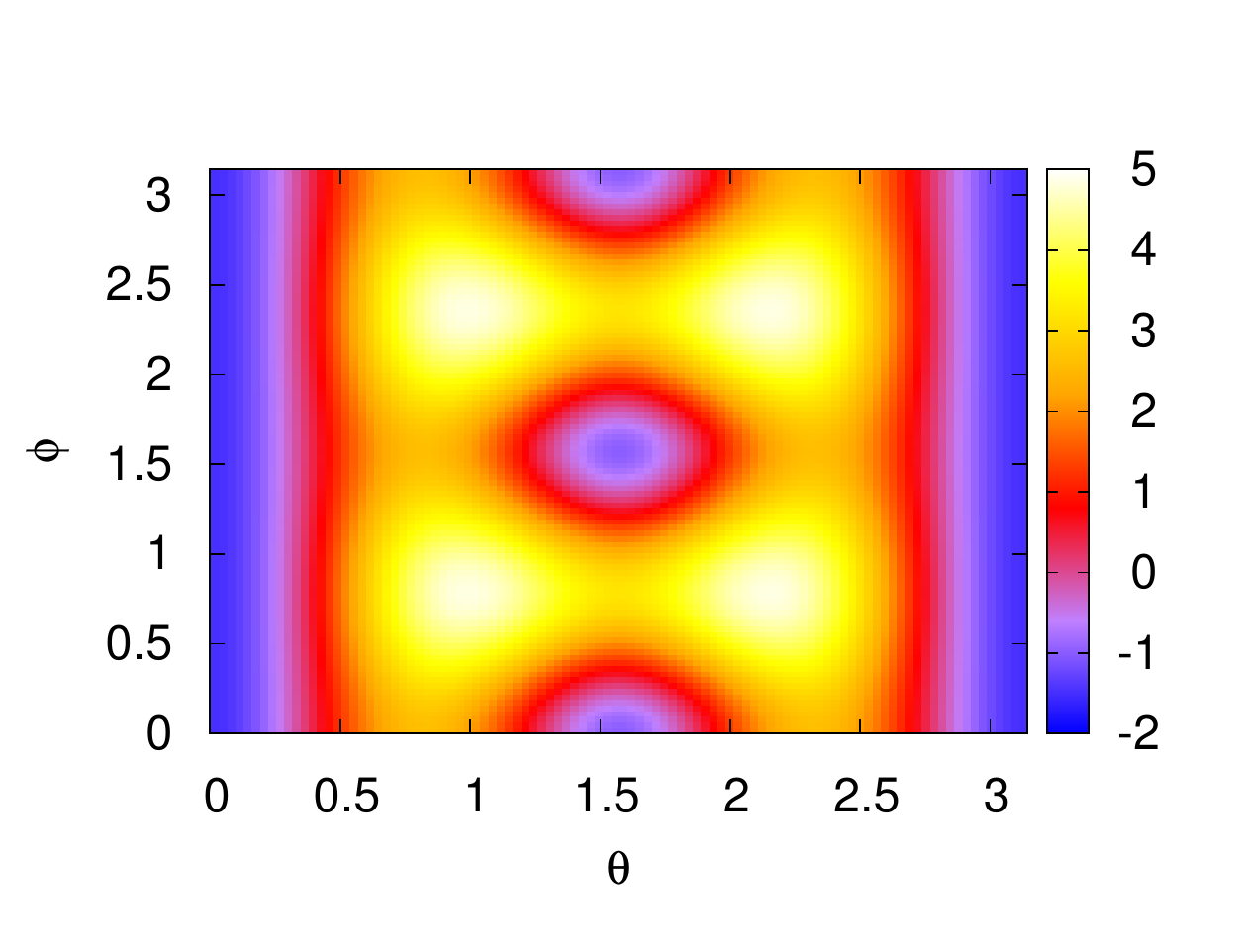}
\caption{(color online) Color topography of the vacuum energy as a function of $\theta$ and $\phi$ in the N\'eel order phase when $J_1>J_2$, in which $\theta$ and $\phi$ indicate the Euler angles describing the quantization axis of the N\'eel order. The minimum of the vacuum energy is taken when the N\' eel order parameter coincides with the x y and z direction for the quantization axis. The next-nearest-neighbor anisotropic coupling reduces the $SU(2)$ symmetry of the vacuum states for a conventional N\'eel order to a discrete symmetry of three possible order parameters of this frustrated N\'eel order. } \label{fg:zero_point_energy}
\end{figure}

We apply the Bogoliubov-De Gennes method to diagonalize the Hamiltonian: $\alpha_{\vec{q}}=u_1a_{\vec{q}}+v_1b_{-\vec{q}}^{\dagger}+u_2a_{-\vec{q}}^{\dagger}+v_2b_{\vec{q}}$, and $[\alpha_{\vec{q}}, H]=\omega_{\vec{q}}\alpha_{\vec{q}}$, then we will obtain the excitation energies for the spin wave and the corresponding wave function $\alpha_{i\vec{q}}; i=1,2,3,4$. Thereafter, we have diagonalized the Hamiltonian with four energy levels:

\begin{equation}
\omega_{i\vec{q}}=\pm\sqrt{\gamma_z^2-(|\gamma_{\vec{q}}|\pm|\gamma_{xy}|)^2} \quad i=1,2,3,4
\end{equation}

\begin{equation}
H=\sum_{\vec{q}}\omega_{\vec{q}}(\alpha_{1\vec{q}}^{\dagger}\alpha_{1\vec{q}}+\alpha_{2\vec{q}}\alpha_{2\vec{q}}^{\dagger}+\alpha_{3\vec{q}}^{\dagger}\alpha_{3\vec{q}}+\alpha_{4\vec{q}}\alpha_{4\vec{q}}^{\dagger}).
\end{equation}
By putting the Hamiltonian in `time order' (commuting $\alpha_{2\vec{q}}\alpha_{2\vec{q}}^{\dagger}$ and $\alpha_{4\vec{q}}\alpha_{4\vec{q}}^{\dagger}$), we obtain the energy of the vacuum:

\begin{equation}
E_0=2\sum_{\vec{q}}\omega_{\vec{q}}=\sum_{\vec{q}}2\sqrt{\gamma_z^2-(|\gamma_{\vec{q}}|+|\gamma_{xy}|)^2}.
\end{equation}

Noticing that the vacuum energy depends on the two Euler angles $\theta$ and $\phi$,  the vacuum quantum fluctuations shall choose an angle that minimizes $E_0$. Numerically, we find that the minimal vacuum energy is taken when the quantization axis coincides with the x y and z axis (see Fig. 9). The Goldstone mode is no longer soft in this case, because when we shift from one spontaneously broken symmetry vacuum to another, the variation of the vacuum energy makes this `transversal' mode energetic, thus destroying the Goldstone mode. Conclusively, the spin wave analysis infers that the N\'eel phase in the limit of $J_1>J_2$  loses its Goldstone mode due to the anisotropy, and that the zero-point vacuum fluctuations select only N\'eel orders pointing along the x, y and z directions.

\subsection{Non-Colinear Spiral Phase for $J_1<J_2$}

Next, we focus on the Spiral phase of $J_1<J_2$.

\begin{figure}[t]
\includegraphics[scale=0.38]{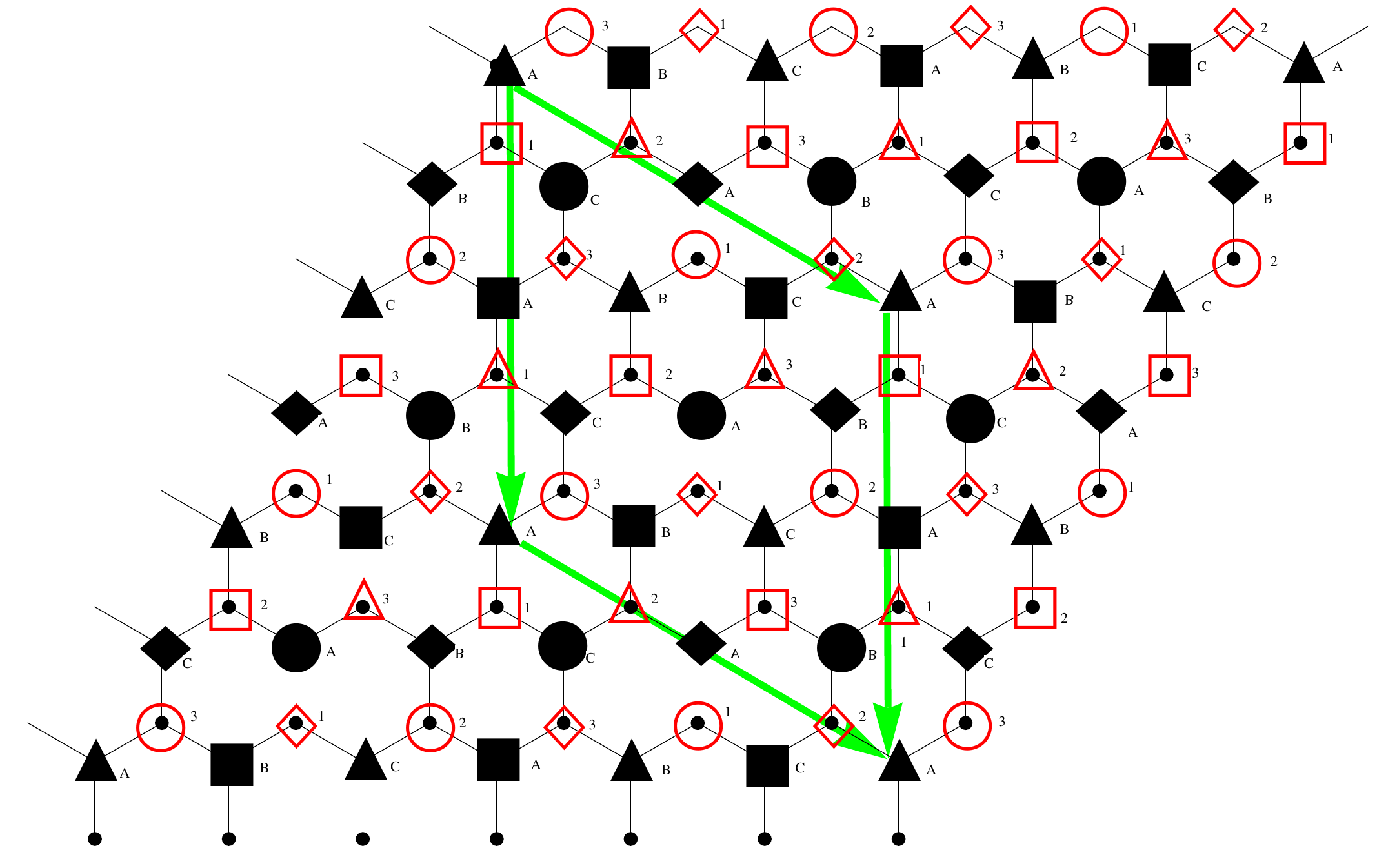}
\includegraphics[scale=0.3]{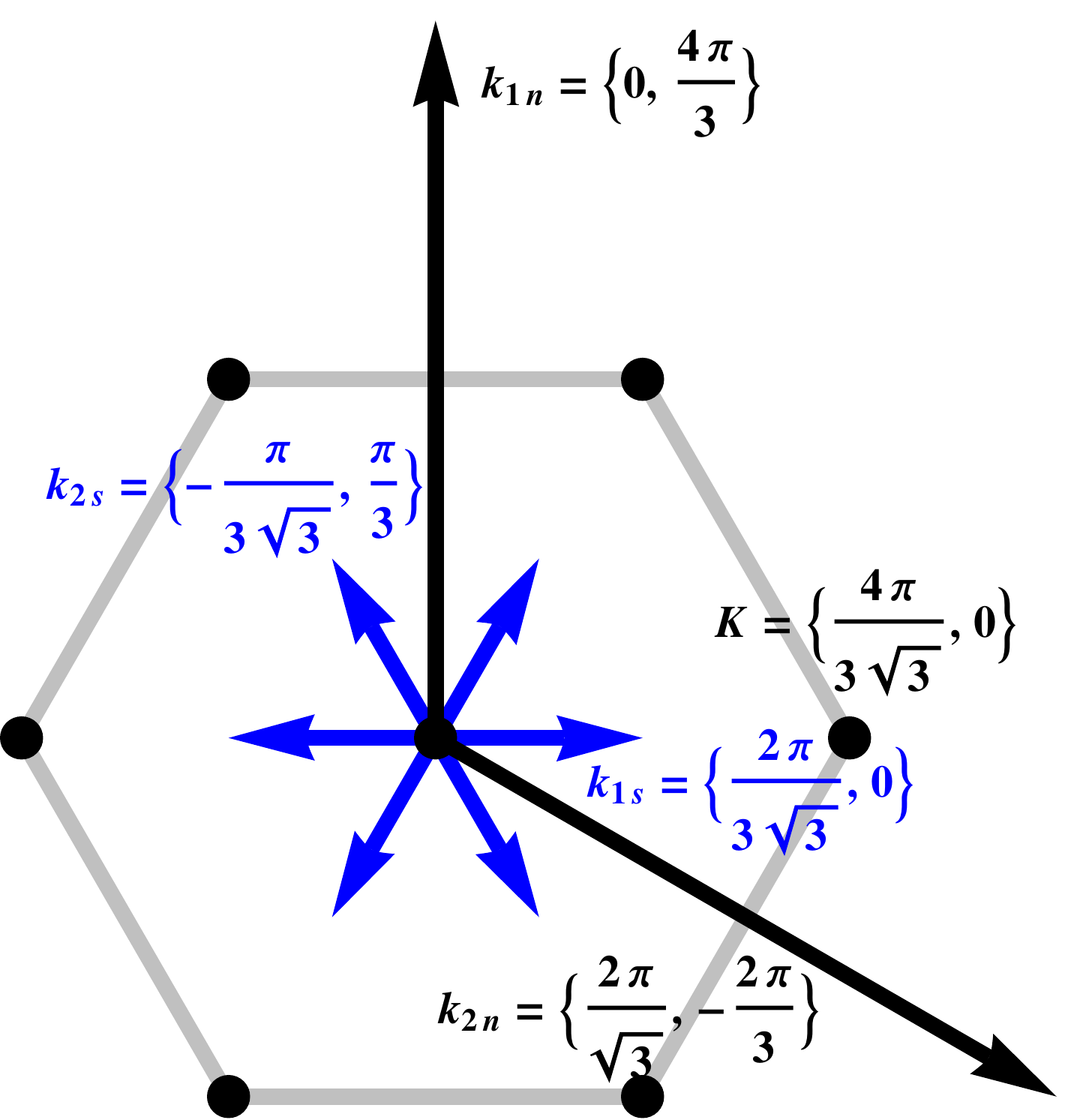}
\caption{(color online) The global transformation brings the $J_2$ anisotropic magnetic model to an anti-ferromagnetic spin model on triangular lattices with four patterns:
$\square\diamondsuit\bigcirc\bigtriangleup$ with black patterns on one sublattice and red patterns on the other. The nearest-neighbor $J_1$ Heisenberg coupling locks the angles between two copies of spiral orders and fixing the relative arrangement of the 4 patterns between the two sublattices as shown in the figure. The sites on which we studied the nearest-neighbor Heisenberg coupling namely the local fields $\vec{h}_{\bigtriangleup}^1$, $\vec{h}_{\square}^2$ and $\vec{h}_{\bigcirc}^3$ in Eqs. \ref{triangle_field}, 26 and 27 are painted in red color with their number indicating the sublattice for the transformed $120^{\circ}$ N\'eel order. In Green, we depict the 12 sublattices (sites) on each triangular sublattice with 4 patterns. We also represent the wave-vectors associated with the Spiral phase (in blue) and with the N\' eel phase (in black). The grey hexagon connects the Dirac points.}
\label{fig.4_sublattices}
\end{figure}

If we only take into account the $J_2$ magnetic coupling, following Ref. \cite{RachelThomale}, we can apply a global transformation to bring the spin model in Eq. \ref{J_1_J_2_magnetism} into an $SU(2)$ anti-ferromagnetic Heisenberg model on the triangular sublattices by introducing 4 patterns, namely:
$H_{J_2}=J_2\sum \vec{\widetilde{S_i}}\cdot\vec{\widetilde{S_j}}$ where $S_i^l=\epsilon_i^l\cdot\widetilde{S_i^l}$ and $l=x,y,z$
such that the global transformation obeys the following condition:
\begin{eqnarray}
&\epsilon_i^z\epsilon_j^z=1 \quad \epsilon_i^y\epsilon_j^y=-1 \quad \epsilon_i^x\epsilon_j^x=-1 \nonumber\\
&\epsilon_j^z\epsilon_k^z=-1 \quad \epsilon_j^y\epsilon_k^y=-1 \quad \epsilon_j^x\epsilon_k^x=1\nonumber\\
&\epsilon_k^z\epsilon_i^z=-1 \quad \epsilon_k^y\epsilon_i^y=1 \quad \epsilon_k^x\epsilon_i^x=-1,
\end{eqnarray}
where $\epsilon_l^w=\pm1 \quad(l=i,j,k; \quad w=x,y,z)$. We can thus find out the four solutions of $\left(\begin{array}{c}
            \epsilon^x_{i,j,k} \\
            \epsilon^y_{i,j,k} \\
            \epsilon^z_{i,j,k}
          \end{array}\right)$:

\begin{equation}
\left(\begin{array}{c}
            \epsilon^x_{i} \\
            \epsilon^y_{i} \\
            \epsilon^z_{i}
            \end{array}\right)=\left(\begin{array}{c}
            1 \\
            1 \\
            1
            \end{array}\right)\square, \left(\begin{array}{c}
            -1 \\
            -1 \\
            1
            \end{array}\right) \bigtriangleup,\left(\begin{array}{c}
            -1 \\
            1 \\
            -1
            \end{array}\right)\bigcirc,\left(\begin{array}{c}
            1 \\
            -1 \\
            -1
            \end{array}\right)\diamondsuit.
\end{equation}

The x, y, z links are all transformed into Heisenberg anti-ferromagnetic links after the global transformation but the introduced four patterns are paved to every sites. Then, the classical ground state on the triangular lattices is obviously the coplanar $120^{\circ}$ N\'eel order for the transformed anti-ferromagnetic Heisenberg model, consisting of 3 sublattices ($A$, $B$, $C$ or $1$, $2$, $3$ in Fig. 10). The magnetic order will be a Spiral order with $12$ sublattices on each triangular sublattice with 4 patterns
$\square\diamondsuit\bigcirc\bigtriangleup$ paved according to the following constraints:
\begin{itemize}
\item X-link: $\square\bigcirc$ or $\diamondsuit\bigtriangleup$
\item Y-link: $\square\diamondsuit$ or $\bigtriangleup\bigcirc$
\item Z-link: $\square\bigtriangleup$ or $\bigcirc\diamondsuit$.
\end{itemize}

\begin{figure}[t]
\includegraphics[scale=0.45]{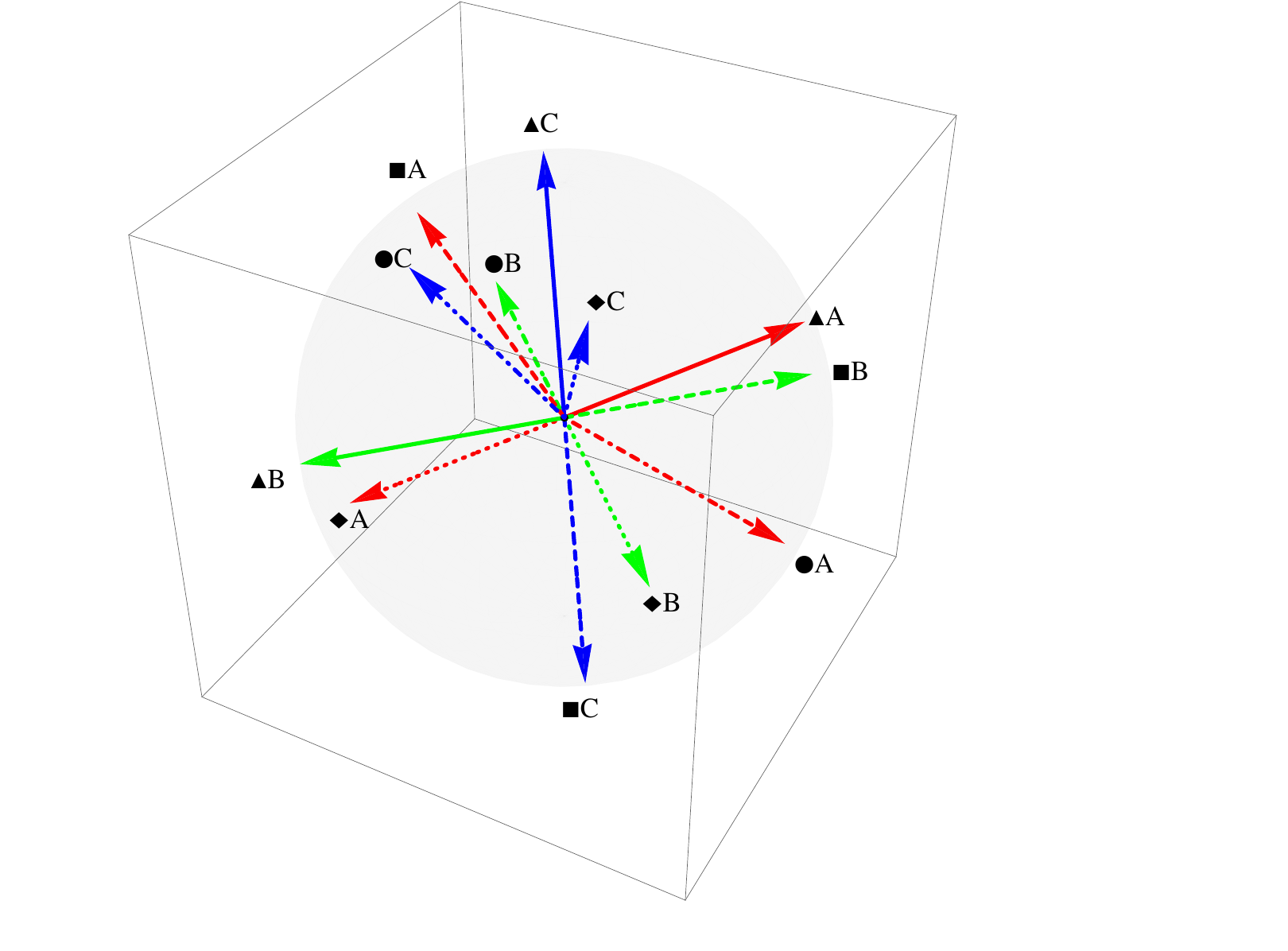}
\caption{(color online) Here, we represent the orientations of the 12  black sites in the Green unit cell of the Spiral phase.}
\label{3D}
\end{figure}

The magnetic order is spiral in that the 4 patterns $\square\diamondsuit\bigcirc\bigtriangleup$ and the 3 spins of the $120^{\circ}$ N\'eel order are alternating when moving in one direction on the lattice. It is important to underline that in the Spiral phase, the spin order is non-colinear. See Fig. \ref{3D}. 
In the absence of the $J_1$ coupling, the two copies of spiral order can rotate with respect to each other freely and the relative arrangement of the 4 patterns can be arbitrary between the two sublattices.At the classical level, the $J_1$ anti-ferromagnetic coupling shall impose the choice of the 4 pattern paving on the alternative sublattice once the 4 pattern paving is fixed in one triangular sublattice as in Fig. \ref{fig.4_sublattices}. Meanwhile, if we consider the $J_1$ coupling in terms of the 3 spins of the
$120^{\circ}$ N\'eel order after the global transformation, the two copies of the transformed $120^{\circ}$ spins would be mutually locked reducing the degree of freedom of the angle between the two copies of the transformed $120^{\circ}$ N\'eel order. The spiral order likewise the $120^{\circ}$ N\'eel state has another degree of freedom, namely the direction of the N\' eel order parameter. The energy of the $J_1$ coupling would depend on the latter. The minimization with regard to this degree of freedom would still reduce the possible choices of the N\' eel order parameter. We shall clarify the minimization of $J_1$ coupling energy with regard to these factors.

The $120^{\circ}$ N\'eel state imposes that spins on the three vertices A, B and C of a triangle $\vec{\widetilde{S}}_A+\vec{\widetilde{S}}_B+\vec{\widetilde{S}}_C=0$. The $J_1$ Heisenberg coupling is equivalent to a local magnetic field produced by the three nearest neighbour spins of the alternative copy of the spiral order on the other copy of the $4$ sublattice spiral order:
\begin{equation}\label{magnetism_4_sublattices}
H_{J_1}=J_1\sum_{\boldsymbol i}(\vec{h}_{i\bigtriangleup}\cdot\vec{S}_{i\bigtriangleup}+\vec{h}_{i\square}\cdot\vec{S}_{i\square}+\vec{h}_{i\diamondsuit}\cdot\vec{S}_{i\diamondsuit}+\vec{h}_{i\bigcirc}\cdot\vec{S}_{i\bigcirc}),
\end{equation}
where the sum is carried out in terms of $4$ sublattices. Considering the $3$-sublattices N\'eel order, we have to sum over $12$ sites in order to get the classical energy of the $J_1$ coupling. However, when writing down the local magnetic field stemming from the $J_1$ coupling, we found that the $4$ patterns could be simplified, and that we need only to sum over the $3$ sublattices of the $120^{\circ}$ order.

The effective local  magnetic fields due to the nearest-neighbor Heisenberg coupling on the sites with numbers painted in red in Fig. \ref{fig.4_sublattices} are:
\begin{equation}\label{triangle_field}
\vec{h}_{\bigtriangleup}^1=D_{\square}\vec{\widetilde{S}}_A+D_{\bigcirc}\vec{\widetilde{S}}_B+D_{\diamondsuit}\vec{\widetilde{S}}_C=2D_{\bigtriangleup}\left(\begin{array}{c}
            \widetilde{S}_{B}^x \\
            \widetilde{S}_{C}^y \\
            \widetilde{S}_{A}^z
          \end{array}\right)
\end{equation}
\begin{equation}\label{square_field}
\vec{h}_{\square}^2=D_{\bigtriangleup}\vec{\widetilde{S}}_B+D_{\bigcirc}\vec{\widetilde{S}}_A+D_{\diamondsuit}\vec{\widetilde{S}}_C=2D_{\square}\left(\begin{array}{c}
            \widetilde{S}_{C}^x \\
            \widetilde{S}_{A}^y \\
            \widetilde{S}_{B}^z
            \end{array}\right)
\end{equation}

\begin{equation}\label{circle_field}
\vec{h}_{\bigcirc}^3=D_{\diamondsuit}\vec{\widetilde{S}}_C+D_{\square}\vec{\widetilde{S}}_B+D_{\bigtriangleup}\vec{\widetilde{S}}_A=2D_{\bigcirc}\left(\begin{array}{c}
            \widetilde{S}_{A}^x \\
            \widetilde{S}_{B}^y \\
            \widetilde{S}_{C}^z
            \end{array}\right)
\end{equation}
where $D_{\square}=\left(\begin{array}{ccc}
1 & 0 & 0 \\
0 & 1 & 0 \\
0 & 0 & 1
\end{array}
\right)$, $
D_{\triangle}=\left(\begin{array}{ccc}
-1 & 0 & 0 \\
0 & -1 & 0 \\
0 & 0 & 1
\end{array}
\right)$, $
D_{\bigcirc}=\left(\begin{array}{ccc}
-1 & 0 & 0 \\
0 & 1 & 0 \\
0 & 0 & -1
\end{array}
\right)$, $
D_{\diamondsuit}=\left(\begin{array}{ccc}
1 & 0 & 0 \\
0 & -1 & 0 \\
0 & 0 & -1
\end{array}
\right).$

We found that the sum of the $J_1$ coupling on these three sites is independent of the $4$ patterns:
\begin{eqnarray}\label{4_sublattice_simplification}
\begin{split}
&J_1(\vec{h}_{1\bigtriangleup}\cdot\vec{S}_{1\bigtriangleup}+\vec{h}_{2\square}\cdot\vec{S}_{2\square}+\vec{h}_{3\diamondsuit}\cdot\vec{S}_{3\diamondsuit})\\
=&J_1(D_{\bigtriangleup}\vec{\widetilde{h}}_1D_{\bigtriangleup}\vec{\widetilde{S}}_1+D_{\square}\vec{\widetilde{h}}_2D_{\square}\vec{\widetilde{S}}_2+D_{\diamondsuit}\vec{\widetilde{h}}_3D_{\diamondsuit}\vec{\widetilde{S}}_3)\\
=&J_1(\vec{\widetilde{h}}_1\cdot\vec{\widetilde{S}}_1+\vec{\widetilde{h}}_2\cdot\vec{\widetilde{S}}_2+\vec{\widetilde{h}}_3\cdot\vec{\widetilde{S}}_3),
\end{split}
\end{eqnarray}
in which $\vec{\widetilde{h}}_1=2\left(\begin{array}{c}
            \widetilde{S}_{B}^x \\
            \widetilde{S}_{C}^y \\
            \widetilde{S}_{A}^z
          \end{array}\right)$, $\vec{\widetilde{h}}_2=2\left(\begin{array}{c}
            \widetilde{S}_{C}^x \\
            \widetilde{S}_{A}^y \\
            \widetilde{S}_{B}^z
            \end{array}\right)$ and $\vec{\widetilde{h}}_3=2\left(\begin{array}{c}
            \widetilde{S}_{A}^x \\
            \widetilde{S}_{B}^y \\
            \widetilde{S}_{C}^z
            \end{array}\right)$.
As a result, the $J_1$ coupling turns into:
\begin{equation}\label{J_2_simplified}
H_{J_1}=J_1\sum_i\vec{\widetilde{h}}_i\cdot\vec{\widetilde{S}}_i,
\end{equation}
in which the sum is carried over the $3$ sublattices of the $120^{\circ}$ N\'eel order and the transformed local magnetic field $\vec{\widetilde{h}}=D_{X}\vec{h}_{X}$ ($X=\square\diamondsuit\bigcirc\bigtriangleup$) is independent of the choice of the 4 patterns.

We observe one property of these local fields that would allow us to simplify the analysis in terms of the choice of the N\' eel order parameter of the $120^{\circ}$ transformed N\'eel order and the relative angle between the two copies of spiral order:
\begin{equation}\label{120_Neel_on_alternative_lattice}
\vec{\widetilde{h_1}}+\vec{\widetilde{h_2}}+\vec{\widetilde{h_3}}=0.
\end{equation}
Then, the minimization of energy of the $J_1$ nearest neighbor coupling in equation \ref{J_2_simplified} can be fulfilled by the use of
Cauchy-Schwarz Inequality:
\begin{widetext}
\begin{equation}\label{Cauchy_Schwartz}
\begin{split}
\vec{\widetilde{h}}_1\cdot\vec{\widetilde{S}}_1+\vec{\widetilde{h}}_2\cdot\vec{\widetilde{S}}_2+\vec{\widetilde{h}}_3\cdot\vec{\widetilde{S}}_3\ge-(||\vec{\widetilde{h}}_1||\cdot||\vec{\widetilde{S}}_1||+||\vec{\widetilde{h}}_2||\cdot||\vec{\widetilde{S}}_2||+||\vec{\widetilde{h}}_3||\cdot||\vec{\widetilde{S}}_3||)\ge-\sqrt{3(||\vec{\widetilde{h}}_1||^2+||\vec{\widetilde{h}}_2||^2+||\vec{\widetilde{h}}_3||^2)}=-6.
\end{split}
\end{equation}
\end{widetext}

The two equalities in Eq. \ref{Cauchy_Schwartz} are taken simultaneously when the norms of the three local magnetic fields on the other copy of the triangular sublattice are equal as in Eq. \ref{equality}:
\begin{eqnarray}\label{equality}
\begin{split}
&||\vec{\widetilde{h}}_1||=||\vec{\widetilde{h}}_2||=||\vec{\widetilde{h}}_3||\\
&\vec{\widetilde{S}}_1=-\frac{1}{2}\vec{\widetilde{h}}_1 \quad \vec{\widetilde{S}}_1=-\frac{1}{2}\vec{\widetilde{h}}_2 \quad \vec{\widetilde{S}}_3=-\frac{1}{2}\vec{\widetilde{h}}_3.
\end{split}
\end{eqnarray}

Since all the spins are prone to align in the opposite direction to the local magnetic field to lower the energy of the ground state, the equality of norms of the three magnetic field on the alternative triangular sublattice coincidentally implies as well: $\vec{\widetilde{S}}_1+\vec{\widetilde{S}}_2+\vec{\widetilde{S}}_3=0$, in other words the $120^{\circ}$ N\'eel state for $\vec{\widetilde{S}}$ on the alternative sublattice.

Accordingly, the spiral order for $\vec{S}$ on the alternative sublattice is favored when the energy of the nearest-neighbor Heisenberg coupling is minimized, and the latter locks the angle between the two copies of spiral order of the ground state obtained from further analysis of Eq. \ref{equality}. The fixing procedure of the relative arrangement between the two sublattices is presented in Fig. \ref{fig.4_sublattices}. We will further study Eq. \ref{equality} to find out how the choice of the N\' eel order parameter for the $120^{\circ}$ N\'eel order and the angle between the two copies of spiral order are constrained for the minimization of the classical energy.

Eqs. \ref{equality} impose extra restrictions on the three $120^{\circ}$ N\'eel vectors, and these supplementary restrictions to spins will reduce the $SU(2)$ continuous symmetry for quantization axis choice to a smaller group:
\begin{equation}\label{conditions}
\begin{cases}
\widetilde{S}_B^{x2}+\widetilde{S}_C^{y2}+\widetilde{S}_A^{z2}=\widetilde{S}_A^{x2}+\widetilde{S}_B^{y2}+\widetilde{S}_C^{z2}=\widetilde{S}_C^{x2}+\widetilde{S}_A^{y2}+\widetilde{S}_B^{z2}=1\\
||\vec{\widetilde{S}}_A||=||\vec{\widetilde{S}}_B||=||\vec{\widetilde{S}}_C||=1\\
\vec{\widetilde{S}}_A+\vec{\widetilde{S}}_B+\vec{\widetilde{S}}_C=0.
\end{cases}
\end{equation}
The last two equations in Eqs. \ref{conditions} is implied by the construction of three arbitrary vectors in the space with $120^{\circ}$ between each other by means of Olinde-Rodrigue formula:
\begin{equation}
\begin{cases}
\vec{\widetilde{S}}_A=\cos(\alpha)\vec{u}+\sin(\alpha)\vec{v}\\
\vec{\widetilde{S}}_B=\cos(\alpha+\frac{2\pi}{3})\vec{u}+\sin(\alpha+\frac{2\pi}{3})\vec{v}\\
\vec{\widetilde{S}}_C=\cos(\alpha-\frac{2\pi}{3})\vec{u}+\sin(\alpha-\frac{2\pi}{3})\vec{v}
\end{cases}
\end{equation}
The two vectors $\vec{u}$ and $\vec{v}$ indicate the plane in which the N\' eel order parameter of the black sublattice lives:
$\vec{u}=\left(\begin{array}{c} \cos\theta \\ \sin\theta \\ 0 \end{array} \right) \quad \vec{v}=\left(\begin{array}{c} -\sin\phi\sin\theta \\ \sin\phi\cos\theta \\ \cos\phi \end{array} \right)$, then $\vec{n}$ is the normal vector to the plane defined by $(\widetilde{S}_A,\widetilde{S}_B,\widetilde{S}_C)$:
\begin{equation}
\vec{n}=\vec{u}\wedge\vec{v}=\left(\begin{array}{c} \sin\theta\cos\phi \\ -\cos\theta\cos\phi \\ \sin\phi \end{array} \right).
\end{equation}
$\vec{n}$ plays the role of rotation axis of the three vectors composing the $120^{\circ}$ N\'eel state.

Resolution of Eqs. \ref{conditions} gives a group of solutions for $\theta$ and $\phi$, therefore a family of rotation axes of the $120^{\circ}$ N\'eel state on the unitary sphere. More precisely, we obtain the equations:
\begin{widetext}
\begin{equation}
\begin{cases}\label{condition_angles}
(\cos\alpha\cos\theta-\sin\alpha\sin\phi\sin\theta)^2+(\cos(\alpha+\frac{2\pi}{3})\sin\theta+\sin(\alpha+\frac{2\pi}{3})\sin\phi\cos\theta)^2+(\sin(\alpha-\frac{2\pi}{3})\cos\phi)^2=1\\
(\cos(\alpha-\frac{2\pi}{3})\cos\theta-\sin(\alpha-\frac{2\pi}{3})\sin\phi\sin\theta)^2+(\cos\alpha\sin\theta+\sin\alpha\sin\phi\cos\theta)^2+(\sin(\alpha+\frac{2\pi}{3})\cos\phi)^2=1.
\end{cases}
\end{equation}
\end{widetext}
The numerical solution gives that the rotational axis for the $120^{\circ}$ N\'eel order (namely $\vec{n}$) takes the form(s):
\begin{equation}
\frac{1}{\sqrt{3}}\left(\begin{array}{c} -1 \\ -1 \\ -1 \end{array} \right), \frac{1}{\sqrt{3}}\left(\begin{array}{c} \, 1 \\ -1 \\ \,1 \end{array} \right), \frac{1}{\sqrt{3}}\left(\begin{array}{c} \, 1 \\ \, 1 \\ -1 \end{array} \right), \frac{1}{\sqrt{3}}\left(\begin{array}{c} -1 \\ 1 \\ 1 \end{array} \right),
\end{equation}
and the three $120^{\circ}$ spins can rotate freely around these axes ($\alpha$ can take any value). Within our choice of notations, the solution $\vec{n}$ is equivalent to $-\vec{n}$ because they describe the same `plane' of solutions for the spins. We should notice that there exists a spin permutation symmetry $\sigma$ for the group formed by these axes and this symmetry is a reminiscence of symmetry group preserved by the original model:
\begin{equation}
\vec{n}'=\sigma\vec{n}=\left(\begin{array}{ccc}
0 & 0 & 1 \\
1 & 0 & 0 \\
0 & 1 & 0
\end{array}
\right)\vec{n}.
\end{equation}

By analogy with the case $J_1>J_2$ phase, the vacuum quantum fluctuations would depend on the rotational degrees of freedom $\alpha$ and the vacuum energy minimization would reduce the group of symmetry for ground state from a continuous rotational group to a discrete group similar to the $J_1>J_2$ phase. The spin wave analysis, however, is not pursued here owing to its complexity, but we can infer the absence of gapless Goldstone modes due to the quantum fluctuations in the presence of the anisotropic magnetic frustration.

\subsection{Phase Transition at $J_1=J_2$}

The classical energy per site for the N\'eel state is $E_{\text{N\'eel}}=-\frac{3J_1}{2}S^2-J_2S^2$, and the classical energy per site for spiral order is $E_{\text{Spiral}}=-\frac{3J_2}{2}S^2-J_1S^2$. Apparently, a quantum phase transition would occur in varying the ratio of $J_1/J_2$ and a first-order phase transition at the critical point $J_1=J_2$ where the $E_{\text{N\'eel}}=E_{\text{Spiral}}$.

The phase transition can be visualized by studying the deformation of the transformed $120^{\circ}$ N\'eel order from the spiral phase. The deformation of the copies of the $120^{\circ}$ N\'eel order can be manifested by the following expressions:

\begin{eqnarray}\label{deformed_triangle}
\begin{split}
&\vec{\widetilde{S}}_A+\vec{\widetilde{S}}_B+\vec{\widetilde{S}}_C=\vec{\epsilon}\\
&\vec{\widetilde{S}}_1+\vec{\widetilde{S}}_2+\vec{\widetilde{S}}_3=\vec{\eta},
\end{split}
\end{eqnarray}
in which $\vec{\epsilon}$ and $\vec{\eta}$ are vectors describing deformations of the three spins on respectively the two triangular sublattices. The $J_2$ coupling is
\begin{eqnarray}
\begin{split}
H_{J_2}=&J_2\sum (\vec{\widetilde{S}}_1\cdot \vec{\widetilde{S}}_2+\vec{\widetilde{S}}_2\cdot \vec{\widetilde{S}}_3+\vec{\widetilde{S}}_3\cdot\vec{\widetilde{S}}_1)\\
=&\frac{1}{2}J_2\sum [(\vec{\widetilde{S}}_1+\vec{\widetilde{S}}_2+\vec{\widetilde{S}}_3)^2-3||\vec{\widetilde{S}}||^2]\\
=&\frac{1}{2}J_2\sum (\vec{\eta}^2-3||\vec{\widetilde{S}}||^2),
\end{split}
\end{eqnarray}
in which the sum is carried out over all the triangles of the sublattice. Then the energy variation of the $J_2$ coupling would be:
\begin{equation}
\Delta E_{J_2}=\frac{1}{2}J_2\sum( \vec{\epsilon}^2+\vec{\eta}^2).
\end{equation}

For the $J_1$ coupling, we can proceed with the similar analysis as Eqs. \ref{triangle_field},\ref{circle_field} and \ref{square_field}:
\begin{equation}
\vec{h}_{\bigtriangleup}^1=\left(\begin{array}{c}
            \widetilde{S}_{A}^x-\widetilde{S}_{B}^x+\widetilde{S}_{C}^x \\
            \widetilde{S}_{A}^y+\widetilde{S}_{B}^y-\widetilde{S}_{C}^y \\
            \widetilde{S}_{A}^z-\widetilde{S}_{B}^z-\widetilde{S}_{C}^z
          \end{array}\right)=D_{\bigtriangleup}(\vec{\widetilde{h}}_1-\vec{\epsilon}),
\end{equation}

\begin{equation}\label{square_field}
\vec{h}_{\square}^2=\left(\begin{array}{c}
            -\widetilde{S}_{A}^x-\widetilde{S}_{B}^x+\widetilde{S}_{C}^x \\
            \widetilde{S}_{A}^y-\widetilde{S}_{B}^y-\widetilde{S}_{C}^y \\
            -\widetilde{S}_{A}^z+\widetilde{S}_{B}^z-\widetilde{S}_{C}^z
          \end{array}\right)=D_{\square}(\vec{\widetilde{h}}_2-\vec{\epsilon}),
\end{equation}

\begin{equation}\label{circle_field}
\vec{h}_{\bigcirc}^3=\left(\begin{array}{c}
            -\widetilde{S}_{A}^x+\widetilde{S}_{B}^x+\widetilde{S}_{C}^x \\
            -\widetilde{S}_{A}^y+\widetilde{S}_{B}^y-\widetilde{S}_{C}^y \\
            \widetilde{S}_{A}^z+\widetilde{S}_{B}^z-\widetilde{S}_{C}^z
          \end{array}\right)=D_{\bigcirc}(\vec{\widetilde{h}}_3-\vec{\epsilon}).
\end{equation}
We also have:
\begin{equation}
\vec{\widetilde{h}}_1 +\vec{\widetilde{h}}_2+\vec{\widetilde{h}}_3 = 2\vec{\epsilon}.
\end{equation}
We can pursue the same procedure as in Eq. \ref{4_sublattice_simplification} to get rid of the sum over $4$ patterns and obtain the $J_1$ coupling:
\begin{eqnarray}\label{J_1_variation}
\begin{split}
H_{J_1}&=J_1\sum((\vec{\widetilde{h}}_1-\vec{\epsilon})\cdot\vec{\widetilde{S}}_1+(\vec{\widetilde{h}}_2-\vec{\epsilon})\cdot\vec{\widetilde{S}}_2+(\vec{\widetilde{h}}_3-\vec{\epsilon})\cdot\vec{\widetilde{S}}_3)\\
&=J_1\sum(\vec{\widetilde{h}}_1\cdot\vec{\widetilde{S}}_1+\vec{\widetilde{h}}_2\cdot\vec{\widetilde{S}}_2+\vec{\widetilde{h}}_3\cdot\vec{\widetilde{S}}_3-\vec{\epsilon}\cdot\vec{\eta}).
\end{split}
\end{eqnarray}
The conditions in Eq. \ref{equality} are satisfied for both the N\'eel and Spiral orders, then the first three terms in Eq. \ref{J_1_variation} is a constant. Thereafter we could obtain an expression of the energy variation per site as a function of $\vec{\epsilon}$ and $\vec{\eta}$:

\begin{equation}\label{energy_variation_deformation}
\Delta E_{\text{Spiral}}=\frac{1}{36}[J_2(\vec{\epsilon}^2+\vec{\eta}^2)+2J_1\vec{\epsilon}\cdot\vec{\eta}].
\end{equation}
The energy variation of the deformed $120^{\circ}$ N\'eel triangle is a positive semi-definite form of $\vec{\epsilon}$ and $\vec{\eta}$ when $J_1<J_2$ on the one hand, the minimal energy variation $\Delta E_{\text{Spiral}}=0$ is obtained when $\vec{\epsilon}=\vec{\eta}=0$; when $J_1>J_2$ on the other hand, Eq. \ref{energy_variation_deformation} is no more a positive semi-definite form, the energy variation due to the deformation is capable of lowering the classical energy, and the minimal energy is reached when $\vec{\epsilon}=-\vec{\eta}$ and $||\vec{\epsilon}||=||\vec{\eta}||=3||\vec{S}||$. Note that here $\vec{\epsilon}$ is large and we don't have a small deformation. This implies that the spins on the two sublattices are oriented in opposite directions and spins on the same sublattice point in a unanimous direction, or the bipartite N\'eel order. We remark also $\Delta E_{\text{Spiral}}=E_{\text{Spiral}}-E_{\text{N\'eel}}$, which signifies that the deformation energy of the $120^{\circ}$ triangle exactly lowers the energy of the spiral magnetic order to that of N\'eel order when $J_1>J_2$.

Consequently, the magnetic order at all $J_1/J_2$ ratios is the bipartite N\'eel order when $J_1>J_2$ and the two copies of locked Spiral order when $J_1<J_2$. This approach rather suggests the emergence of a quantum critical point when $J_1=J_2$. In both phases, Goldstone modes are absent because of the vacuum quantum fluctuation selection.

\section{Mott transition}

To address the Mott transition (characterized here for example by the disappearance of the helical edge modes), as mentioned earlier in the text, we shall use the U(1) slave-rotor theory method \cite{Florens, Paramekanti}. A physical electron can be viewed as a spin and a charge (chargeon) glued together. At the Mott critical point $U_c$ depicted by a red line in Fig. 1,  spin and charge become disentangled and charge is localized in a Mott state.  It is perhaps important to stress that the single-electron gap does not close at the Mott transition and that the single-electron Green's function should now reveal a two-peak structure above the Mott gap. A gauge field will however emerge in this spin-charge separation physics describing the confining force between the charge and the spin above the Mott transition. The nature of this confining force might determine whether above the Mott critical point the system is in a spin liquid phase \cite{Ruegg} or already in a magnetically ordered phase. This question is beyond the scope of this work and we shall only describe how the pseudo spin-orbital texture will develop from the edges into the bulk at the Mott transition. First, in the anisotropic spin-orbit coupling model with an Hubbard on-site interaction defined in Eq. \ref{Shitade_Hubbard}, the AQSH phase will disappear when the on-site Hubbard interaction will exceed a certain critical value $U_c$, that needs to be determined.

The Mott transition is characterized by the acquisition of a gap for the chargeon then localizing the charge particle. The critical value $U_c$ of the Mott transition as a function of the anisotropic spin-orbit coupling-Hubbard model will be proved in this Section to be exactly the same as for the Kane-Mele-Hubbard model \cite{RachelLeHur}: the chargeon effective Hamiltonian in the spin-charge fractionalized representation is the same as in the Kane-Mele-Hubbard model after doing the mean-field approximation. However, spinons that will be subject to the strong gauge field fluctuations behave distinctly for the anisotropic spin-orbit coupling model. By attaching a gauge field \cite{polyakov,IoffeLarkin,HerbutSubir,pepin} to the chargeon to describe the residual degrees of freedom in the phase of the localized chargeon, we will establish a gauge theory that will incorporate the apparition and proliferation of monopoles \cite{polyakov}. The monopoles will affect the spinons by insertion of fluxes, and the spinons respond to these fluxes by forming spin textures around the inserted flux. The gauge fluctuations in this anisotropic spin-orbit coupling model with on-site Hubbard interaction triggers anisotropic spin textures while the spin texture would be homogeneous in the $XY$ plane in the Kane-Mele Hubbard model above the Mott critical point \cite{RachelLeHur,Lee}.

The U(1) slave-particle representation \cite{Florens,Paramekanti} consists in cracking the physical electron down to the fermionic spinon particle for the spin and the bosonic chargeon particle for the charge. On each site of the system, there could be $4$ electron states: $\ket{\phi}$, $\ket{\uparrow}$, $\ket{\downarrow}$ and $\ket{\uparrow\downarrow}$, and different representation of slave particle uses different description of these $4$ electron states. Two representations are currently applied to describe the Mott transition, namely the U(1) slave-rotor representation \cite{Florens} and the $\mathbb{Z}_2$ slave-spin representation \cite{Biermann}, \cite{RueggSigrist}. In the U(1) formulation, the `superfluid' phase of the rotors is characterized by an ordered rotor meaning the coherence of the wave function over the whole system. The `Mott' phase, in which electrons are rather localized on lattice sites (rather than in $k$-space), is characterized by disordered rotors implying the loss of coherence of the wave function. The phase transition is described by the gap acquisition of the rotors and the disappearance of the quasiparticle poles in the electronic Green's function.

In contrast, in the $\mathbb{Z}_2$ slave-spin representation, the `superfluid' phase is represented by ordered Ising spins of the quantum Ising model in a transverse field, and the Mott phase is embodied by disordered Ising spins. The main difference between the two representations lies in the gauge fluctuations: the $\mathbb{Z}_2$ effective gauge field predicts a phase with exotic vison excitations \cite{SenthilFisher1,SenthilFisher2,Ruegg,SubirZ2}, while the U(1) Maxwellian gauge theory only implies magnetic monopoles and is also widely used in the context of studies of Hubbard models \cite{Florens,Paramekanti,LeeLee,HermeleRanLeeWen}. We choose here the U(1) rotor representation to study the Mott transition \cite{Florens,Paramekanti}.

\subsection{Mott Transition in U(1) Slave Rotor Theory}

The U(1) slave-rotor representation \cite{Florens,Paramekanti} consists of labelling the $4$ state Hilbert space by angular momentum: $\ket{\uparrow}_e=\ket{\uparrow}_s\ket{0}_{\theta}$, $\ket{\downarrow}_e=\ket{\downarrow}_s\ket{0}_{\theta}$, $\ket{\uparrow\downarrow}_e=\ket{\uparrow\downarrow}_s\ket{1}_{\theta}$ and $\ket{\phi}_e=\ket{\phi}_s\ket{-1}_{\theta}$. The creation of a physical electron is the creation of a spin in the spinon Hilbert space accompanied by raising the angular momentum in the rotor Hilbert space, while the measure of the number of electron is the measure of the angular momentum:
\begin{equation}
c_{\sigma}^{\dagger}=f_{\sigma}^{\dagger}e^{i\theta}\quad c_{\sigma}=f_{\sigma}e^{-i\theta},
\end{equation}
in which $f_{\sigma}^{\dagger}$ is a spinon creation operator with spin $\sigma$, and $e^{i\theta}$ is an angular momentum raising operator. The Hubbard interaction Hamiltonian turns into $H_I=\sum_i \frac{U}{2}(n_i-1)^2=\sum_i \frac{U}{2}L_i^2$, in which we used the fact that we consider the case of half filling.

Hence, following the lines of thoughts of the Kane-Mele-Hubbard model \cite{RachelLeHur}, we can write the Hamiltonian in the U(1) slave rotor representation as:
\begin{eqnarray}
H_{\text{rotor}}&=&\sum_i\frac{U}{2}{L}_i^2+\sum_{\left<i,j\right>}\sum_{\sigma}tf_{i\sigma}^{\dagger}f_{j\sigma}e^{i\theta_{i}-i\theta_j}\nonumber\\&+&\sum_{\ll i,j \gg}\sum_{\sigma,\sigma'}i t'f_{i\sigma}^{\dagger}f_{j\sigma'}\sigma_{\sigma\sigma'}^we^{i\theta_{i}-i\theta_j}
\end{eqnarray}
When applying the rotor formalism, we enlarge the Hilbert space, therefore an extra constraint needs to be imposed:
\begin{equation}\label{rotor_constraint}
L_i=\sum_{\sigma}\left[f_{i\sigma}^{\dagger}f_{i\sigma}-\frac{1}{2}\right].
\end{equation}

In the Hamiltonian formalism, we can replace $e^{i\theta_{i}-i\theta_j}$ and $f_{i\sigma}^{\dagger}f_{j\sigma}$ by their mean-field ansatz, and separate the spinon and chargeon. By working out the ground state mean value of these replaced observables, we obtain the self-consistent equations to solve, or specifically:
\begin{equation}\label{spinon_hamiltonian}
H_f=\sum_{<i,j>}tQ_ff_{i\sigma}^{\dagger}f_{j\sigma}+\sum_{\ll i,j\gg}it'\widetilde{Q}_f\sigma_{\sigma\sigma'}^wf_{i\sigma}^{\dagger}f_{j\sigma'}
\end{equation}
\begin{equation}
H_{\theta}=\sum_{<i,j>}tQ_x\cos(\theta_i-\theta_j)+\sum_{\ll i,j \gg}t'\widetilde{Q}_x\cos(\theta_i-\theta_j)+\frac{U}{2}{L}_i^2
\end{equation}
\begin{equation}
\begin{cases}
\left<e^{i\theta_{i}-i\theta_j}\right>_{\left<i,j\right>}=Q_f \quad \left<e^{i\theta_{i}-i\theta_j}\right>_{\ll i,j \gg}=\widetilde{Q}_f \\
\left<f_{i\sigma}^{\dagger}f_{j\sigma}\right>_{\left<i,j\right>}=Q_x \quad \left<i\sigma_{\sigma\sigma'}^wf_{i\sigma}^{\dagger}f_{j\sigma'}\right>_{\ll i,j \gg}=\widetilde{Q}_x.
\end{cases}
\end{equation}
We can obtain an effective rotor Hamiltonian by making use of the mean field ansatz and solving  Eqs. \ref{spinon_hamiltonian}, which is the anisotropic spin-orbit coupling model itself.
\begin{equation}
H_{\theta}=-\sum_{<i,j>}K\cos(\theta_i-\theta_j)-\sum_{\ll i,j \gg}G\cos(\theta_i-\theta_j)+\sum_i\frac{U}{2}{L}_i^2,
\end{equation}
where
\begin{eqnarray}
K &=&\sum_{\vec{k}}\frac{|Q_fg(\vec{k})|^2}{\widetilde{E}_0(\vec{k})} \\ \nonumber
G &=& \sum_{\vec{k}}\sum_w\frac{(2\widetilde{Q}_ft'\sin( \vec{k}.\vec{R}_w))^2}{\widetilde{E}_0(\vec{k})}
\end{eqnarray}
and
\begin{equation}
\widetilde{E}_0(\vec{k})=\sqrt{|Q_fg(\vec{k})|^2+\sum_w(2\widetilde{Q}_ft'\sin(\vec{ k}\cdot\vec{R}_w))^2}.
\end{equation}
We recall that $w=x,y,z$. We observe that the effective rotor Hamiltonian is a non-frustrated $XY$ model with first and second neighbours on the honeycomb lattice. By resorting to the one-site mean-field approximation as in Ref. \cite{RachelLeHur}, we can identify the critical interaction:
\begin{eqnarray}
\begin{split}
&\left<\cos\theta\right>=-\frac{2K}{U} \\
&U_c=\frac{4}{N_{\Lambda}}\sum_{\vec{k}}|g(\vec{k})|
\end{split}
\end{eqnarray}
in which $N_{\Lambda}$ denotes the number of unit cells.

\begin{widetext}

In order to do the main field approximation in a more explicit way, we pursue here the Lagrangian formalism of which we can carry out the saddle-point approximation more easily in the path integral formulation. We keep the same notation for mean-field ansatz but they can take different values in the Lagrangian formalism from those in the Hamiltonian formalism.

The Hubbard interaction $\frac{U}{2}L_i^2$ in the rotor representation is a kinetic term, and the constraint in Eq. \ref{rotor_constraint} is now imposed through the addition of the Lagrangian multiplier $\sum_ih_i\sum_{\sigma}(f_{i\sigma}^{\dagger}f_{i\sigma}-L_i-\frac{1}{2})$ to the Lagrangian. By using $i\partial_{\tau}\theta_i=\frac{\partial H}{\partial L_i}$, we then obtain the whole action:
\begin{eqnarray}
S &=&\int_0^{\beta}d\tau \left[\sum_i(iL_i\partial_{\tau}\theta_i+f_{i\sigma}^{\dagger}\partial_{\tau}f_{i\sigma})+\mathcal{H}\right]\\
&=&\int_0^{\beta}d\tau \biggl[\sum_{i,\sigma}f_{i\sigma}^{\dagger}(\partial_{\tau}+h_i)f_{i\sigma}+\frac{1}{2U}\sum_i(\partial_{\tau}\theta_i+ih_i)^2+\sum_i \left(\frac{h_i^2}{2U}-h_i\right)\nonumber\\
&+&t\sum_{<i,j>,\sigma}f_{i\sigma}^{\dagger}f_{j\sigma}e^{i\theta_{i}-i\theta_j}+h.c.+it'\sum_{\ll i,j \gg}\sum_{\sigma\sigma'}\sigma_{\sigma\sigma'}^wf_{i\sigma}^{\dagger}f_{j\sigma'}e^{i\theta_{i}-i\theta_j}\biggl],
\end{eqnarray}
\end{widetext}
where
\begin{equation}
\mathcal{H}=H_{\text{rotor}}+\sum_i\sum_{\sigma}\left[f_{i\sigma}^{\dagger}f_{i\sigma}-L_i-\frac{1}{2}\right]
\end{equation}
such that the constraint in Eq. \ref{rotor_constraint} is imposed at a mean-field level.

To solve the rotor model in the above action, firstly we shall replace the rotor $e^{i\theta_i}$ by a $O(2)$ complex bosonic field $X_i$ and treat the constraint $|X_i|^2=1$ by a mean field self-consistent equation (and formally treat the fermion $f^{\dagger}$ as a complex field $f^{\star}$). The Lagrangian for the rotors then takes the form:
\begin{widetext}
\begin{eqnarray}
{\cal{L}}_x&=&\sum_{\vec{k}}-g(\vec{k})Q_xX_{\vec{k}}^{a\star}X_{\vec{k}}^b-g(\vec{k})^{\star}Q_xX_{\vec{k}}^aX_{\vec{k}}^{b\star}+\tilde{Q}_x\sum_kt'g_2(\vec{k})(X_{\vec{k}}^{a\star}X_{\vec{k}}^a+X_{\vec{k}}^{b\star}X_{\vec{k}}^b)+\sum_{\vec{k}}\rho X_{\vec{k}}^{\star}X_{\vec{k}}\nonumber\\
&=&\sum_{\vec{k}}-|g|Q_xX_{\vec{k}}^{l\star}X_{\vec{k}}^l+|g|Q_xX_{\vec{k}}^{u\star}X_{\vec{k}}^u+\sum_{\vec{k}} \tilde{Q}_xt'g_2(\vec{k})(X_{\vec{k}}^{l\star}X_{\vec{k}}^l+X_{\vec{k}}^{u\star}X_{\vec{k}}^u)+\sum_{\vec{k}}\rho X_{\vec{k}}^{\star}X_{\vec{k}},
\end{eqnarray}
\end{widetext}
in which $g_2(\vec{k})=\cos\left(\vec{k}\cdot\vec{R}_x\right)+\cos\left(\vec{k}\cdot\vec{R}_y\right)+\cos\left(\vec{k}\cdot\vec{R}_z\right)$. We can derive the Green function for the quantum rotor:
\begin{equation}
G_x=\frac{1}{\frac{\nu_n^2}{U}+\rho+\xi_{\vec{k}}},
\end{equation}
in which $\xi_{\vec{k}}=-Q_x|g(\vec{k})|+t'\tilde{Q}_xg_2(\vec{k})$, and $\nu_n$ is the Mastubara frequency.

Then we can use the self-consistent equation of the saddle-point for the rotor field $\left<|X_i|^2\right>=\sum_{\vec{k}}\frac{1}{G_x}=1$ to determine the critical value of U:

\begin{equation}
1=\frac{U}{N_{\Lambda}}\sum_{\vec{k}}\frac{1}{\sqrt{\Delta_g^2+4U(\xi_{\vec{k}}-\min_{\vec{k}}(\xi_{\vec{k}}))}},
\end{equation}
where $\Delta_g=2\sqrt{U(\rho+\min_k(\xi_{\vec{k}}))}$ describes the gap acquired by the rotors which turns to be zero at the critical point due to the condensation of the rotors resulting in an extra constraint on the Lagrangian multiplier $\rho$.

The rotor gap formally becomes non-zero in the Mott insulating phase. Then we get the expression of the critical value $U_c$ when $\Delta_g=0$:
\begin{equation}
U_c=\left[\frac{1}{2N_{\Lambda}}\sum_{\vec{k}}\frac{1}{\sqrt{\xi_{\vec{k}}-\min_{\vec{k}}(\xi_{\vec{k}}})}\right]^{-2}.
\end{equation}
We can numerically evaluate the Mott transition versus the spin-orbit coupling $t'$ and it turns out that $U_c$ increases monotonously when increasing $t'$, substantiating the spin-induced induced Mott transition (see Fig. \ref{fig:mott_transition}). The critical line determining the Mott transition is identical  to that in
the Kane-Mele-Hubbard model \cite{RachelLeHur}.

\subsection{Gauge Fluctuations above the Mott Transition}

When we break the physical electron down to the fermionic spinon and the bosonic rotor, then emerges a $U(1)$ gauge symmetry:
\begin{equation}
f_i^{\dagger}\rightarrow f_i^{\dagger}e^{i\phi_i}, e^{i\theta_i}\rightarrow e^{i\theta_i-i\phi_i}
\end{equation}
 that binds the chargeon and spinon together. In the Mott phase, the rotors become disordered, and the local phase of the rotors fluctuates considerably. We describe this local gauge fluctuations by attaching a field strength $\mathcal{A}^c$ simultaneously to the spinon and chargeons. Then we can integrate out the rotors to get an effective action of the fluctuating gauge field \cite{LeeLee,RachelLeHur} and describe the effects of the fluctuating gauge field on the spinons. The response of the spinons to the fluctuating gauge field clarifies the emergence of a peculiar spin texture in the bulk.

We first apply the Hubbard-Stratonovich transformation to decouple the rotor field and the spinon field by using the complex Gaussian integral equality: $\int d\overline{z}dz \exp(-|z|^2+uz+w\overline{z})=\exp(uw)$ in which $z$ and $\overline{z}$ are the auxiliary field and $\left<z\right>=w$ and $\left<\overline{z}\right>=u$ are the saddle point. For the anisotropic spin-orbit coupling model with an on-site Hubbard interaction, then this results in the effective Lagrangian:
\begin{widetext}
\begin{eqnarray}
\cal{L} &=& \sum_if_{i\sigma}^{\dagger}(\partial_{\tau}+h_i)f_{i\sigma}+\frac{1}{2U}(\partial_{\tau}\theta_i+ih_i)^2\nonumber\\
&+&\sum_{<i,j>}(-t|\eta_{ij}|^2-t|\eta_{ji}|^2+tf_{i\sigma}^{\dagger}f_{j\sigma}\eta_{ij}+tf_{j\sigma}^{\dagger}f_{i\sigma}\eta_{ji}+te^{i(\theta_i-\theta_j)}\eta_{ij}^{\star}+te^{i(\theta_j-\theta_i)}\eta_{ji}^{\star})\nonumber\\&+&\sum_{\ll i,j \gg}(-t'|\zeta_{ij}|^2-t'|\zeta_{ji}|^2+it'f_{i\sigma}^{\dagger}f_{j\sigma'}\sigma_{\sigma\sigma'}^w\zeta_{ij}+it'f_{j\sigma}^{\dagger}f_{i\sigma'}\sigma_{\sigma\sigma'}^w\zeta_{ji}+t'e^{i(\theta_i-\theta_j)}\zeta_{ij}^{\star}+t'e^{i(\theta_j-\theta_i)}\zeta_{ji}^{\star}),
\end{eqnarray}
\end{widetext}
where at the level of the saddle point solution $\eta_{ij}=\left<e^{i(\theta_i-\theta_j)}\right>_{\left<i,j\right>}$, $\eta_{ij}^{\star}=\left<f_{i\sigma}^{\dagger}f_{j\sigma}\right>_{\left<i,j\right>}$, $\zeta_{ij}=\left<e^{i(\theta_i-\theta_j)}\right>_{\ll i,j \gg}$ and $\zeta_{ij}^{\star}=\left<if_{i\sigma}^{\dagger}f_{j\sigma'}\sigma_{\sigma\sigma'}^w\right>_{\ll i,j \gg}$ respectively on the nearest-neighbor and next-nearest neighbor links, (similar relations of saddle points hold for $\eta_{ji}$, $\eta_{ji}^{\star}$, $\zeta_{ji}$ and $\zeta_{ji}^{\star}$) and it is worth noticing that $\eta_{ij}\neq\eta_{ji}$ and $\zeta_{ij}\neq\zeta_{ji}$.

In the rotor ordered phase, $\eta_{ij}=\eta_{ji}^{\star}$ (same with $\zeta_{ij}$ on the next-nearest-neighbours) and the gauge fluctuation is suppressed, while in the rotor disordered phase $\eta_{ij}$ and $\eta_{ji}$ become independent, and this can be described by attaching a field strength $\mathcal{A}_{ij}^c$ to the behavior of the link variables $\eta_{ij}$ and $\zeta_{ij}$ and the strong fluctuations of the gauge field elucidates the difference for the link variable in the two phases for the rotors:

\begin{equation}
\begin{cases}
\zeta_{ij}\rightarrow\zeta_{ij}e^{i\mathcal{A}_{ij}^c} \quad \zeta_{ij}^{\star}\rightarrow\zeta_{ij}^{\star}e^{-i\mathcal{A}_{ij}^c}\\
\eta_{ij}\rightarrow\eta_{ij}e^{i\mathcal{A}_{ij}^c} \quad \eta_{ij}^{\star}\rightarrow\eta_{ij}^{\star}e^{-i\mathcal{A}_{ij}^c}
\end{cases}
\end{equation}

We explicitly introduce a temporal gauge field $\mathcal{A}_i^{\tau c}$ at site $i$ in the action.
We then obtain the spinon and rotor Lagrangians:
\begin{eqnarray}\label{spinon_lagrangian}
&\mathcal{L}_f &=\sum_i\sum_{\sigma}f_{i\sigma}^{\dagger}(\partial_{\tau}-i\mathcal{A}_i^{\tau c}+h_i)f_{i\sigma}+t\sum_{<i,j>}f_{i\sigma}^{\dagger}f_{j\sigma}\eta_{ij}e^{i\mathcal{A}_{ij}^c}\nonumber\\
&+& it'\sum_{\ll i,j \gg}\zeta_{ij}f_{i\sigma}^{\dagger}f_{j\sigma'}\sigma_{\sigma\sigma'}^we^{i\mathcal{A}_{ij}^c}
\end{eqnarray}
\begin{eqnarray}\label{rotor_lagrangian}
&\mathcal{L}_{\theta} &=\sum_i\frac{(\partial_{\tau}\theta_i-\mathcal{A}_i^{\tau c}-ih_i)^2}{2U}+t\sum_{<i,j>}e^{i(\theta_i-\theta_j-\mathcal{A}_{ij}^c)}\eta_{ij}^{\star}\nonumber\\
&+&t'\sum_{\ll i,j \gg}\zeta_{ij}^{\star}e^{i(\theta_i-\theta_j-\mathcal{A}_{ij}^c)}.
\end{eqnarray}
Integrating out the rotor $e^{i\theta}$, we get a Maxwellian gauge theory with coupling constants depending on the rotor gap with $\Delta_g$ indicating  the magnitude of the gauge fluctuations:
\begin{eqnarray}
\mathcal{L}_{\mathcal{A}^c} &=& \sum_{\triangle}\left(\frac{t' |\zeta_{ij}|}{\Delta_g}\right)^3\cos(\nabla\times\mathcal{A}^c)  \\ \nonumber
&+&\frac{1}{2U\Delta_g}(\partial_{\tau}\mathcal{A}_i^{\tau c}-\partial_{x}\mathcal{A}_i^{\tau c})^2 ,
\end{eqnarray}
where the sum is carried out on all the triangle plaquettes; $\Delta_g$ is the rotor gap and $\nabla\times\mathcal{A}^c=\mathcal{A}^c_{ij}+\mathcal{A}^c_{jk}+\mathcal{A}^c_{ki}$ on one triangle ($i,j,k$ are the three vertices of the triangle) and $\partial_x\mathcal{A}_i^{\tau c}=\mathcal{A}_i^{\tau c}-\mathcal{A}_j^{\tau c}$. In the rotor ordered phase, the rotor gap $\Delta_g=0$, so it costs an infinite energy to insert any magnetic flux into the system, namely the gauge field barely fluctuates. In contrast,  in the rotor disordered phase, the rotor gap $\Delta_g$ becomes finite making the insertion of the magnetic flux possible. Because the gauge field is compact, the insertion of a $2\pi$ flux (monopoles in $2+1$ dimensions \cite{polyakov}) $\nabla\times\mathcal{A}^c=2\pi$ leaves the Maxwellian gauge action invariant, which means magnetic fluxes can be inserted adiabatically into the system without any cost of energy. This implies the proliferation of the monopoles in the space: the monopole correlation function in the space $\left<m^{\star}(\vec{r})m(\vec{0})\right>$ is a constant, in which $m^{\star}(\vec{r})$ creates a $2\pi$ flux at $\vec{r}$.

We now address the spinon response to this adiabatic insertion of monopoles.

\subsection{Spin Texture upon Insertion of Flux} \label{spin_texture_formation}

The spinon Lagrangian under gauge fluctuation is the anisotropic spin-orbit coupling model in Eq. \ref{Shitade_Hubbard} upon insertion of flux of $2\pi$ brought by the rotor gauge field $\mathcal{A}_{ij}^c$.

The spin-orbit coupling implies the spin Hall physics elucidated in Sec. \ref{spin_transport}. In the context of Gedanken experiment by Laughlin \cite{laughlin}, the insertion of a $U(1)$ flux leads to an edge charge transport. In the context of spin Hall effect, the insertion of a U(1) flux implies a contour in the first Brillouin zone enclosing the time reversal points; this contour denotes an exchange of Kramer pairs, therefore a $\mathbb{Z}_2$ spin pump \cite{fu-kane_2}. In other words, flux insertion triggers spin transport on the edge. The spinon of the anisotropic spin-orbit coupling system is a similar system, and flux insertion should incur spin transport in the system (see Fig. 12). Some spin `charge' would be transported near the monopole core as the `edge' of the system. However, several difficulties are encountered in the anisotropic spin-orbit coupling model, in contrast to the Kane-Mele-Hubbard model \cite{RachelLeHur}: 1. the non conservation of spin number and the not-defined spin current would make the Kubo formalism inapplicable here; 2. the insertion of the magnetic flux is local instead of onto the whole system as in the case of Laughlin U(1) pump and the $\mathbb{Z}_2$ spin pump.

In order to study the spin behavior around the fluctuating gauge field, here we apply the perturbation theory, and quantitatively describe how local spin observables on a given site are affected when a $2\pi$ magnetic flux of the gauge field $\mathcal{A}^c$ is adiabatically inserted into the spinon system. We describe the adiabatic insertion of a magnetic flux of the chargeon gauge field by making the gauge field dependent on time $\mathcal{A}_{ij}^c(\tau)=\mathcal{A}_{ij}^ce^{\eta \tau}$, $\eta>0$ in the time interval of $\tau\in]-\infty,0]$ and the gauge field with a field strength $\mathcal{A}_{ij}^c$ is inserted adiabatically within this time interval. Considering the exceptional anisotropic properties of the system, we shall investigate the lattice gauge field on each link around the flux, presuming that gauge fields on different links $X$, $Y$ or $Z$ might have different influences on the spin polarization at a given site that we measure. We expect that a certain spin texture might appear around the inserted flux due to the spin-Hall nature of the system, which is the main subject here.

We get back to the Hamiltonian formalism and apply the perturbation method. The observable we measure is
\begin{equation}
S_M^{\alpha}=f_{MJ\sigma}^{\dagger}f_{MJ\sigma'}\sigma_{\sigma\sigma'}^{\alpha},
\end{equation}
 in which $\alpha$ is the spin polarization, $\vec{R}_M$ is the site at which we measure the spin and $J=A,B$ is the sublattice isospin of the corresponding site. Resorting to the time evolution operator, we can express the spin polarization variation under the flux insertion perturbation $\delta\mathcal{H}=(\mathcal{H}_S-\mathcal{H}_S^0)$ in which $\mathcal{H}_S$ is the spinon Hamiltonian after the gauge insertion and $\mathcal{H}_S^0$ is the original spinon Hamiltonian:
\begin{eqnarray}
\delta S_M^{\alpha} &=& e^{\int_{-\infty}^0i\delta\mathcal{H}d\tau}S_M^{\alpha}e^{-\int_{-\infty}^0i\delta\mathcal{H}d\tau}-S_M^{\alpha} \\ \nonumber
&=&\left[i\int_{-\infty}^0\delta\mathcal{H}d\tau,S_M^{\alpha}\right].
\end{eqnarray}

\begin{widetext}
The original and the perturbed spinon Hamiltonians are explicitly given by:
\begin{eqnarray}\label{spinon_perturbation}
\begin{split}
\mathcal{H}_{S}^0(\tau)&=\sum_{\left<i,j\right>}tQ_ff_{i\sigma}^{\dagger}(\tau)f_{j\sigma}(\tau)+it'\sum_{\ll i,j \gg}\tilde{Q}_ff_{i\sigma}^{\dagger}(\tau)f_{j\sigma'}(\tau)\sigma_{\sigma\sigma'}^w\\
\mathcal{H}_{S}(\tau)&=\sum_{\left<i,j\right>}tQ_ff_{i\sigma}^{\dagger}(\tau)f_{j\sigma}(\tau)e^{i\mathcal{A}_{ij}^c}+it'\sum_{\ll i,j \gg}\tilde{Q}_ff_{i\sigma}^{\dagger}(\tau)f_{j\sigma'}(\tau)\sigma_{\sigma\sigma'}^we^{i\mathcal{A}_{ij}^c}
\end{split}
\end{eqnarray}
such that $\delta\mathcal{H}(\tau)=\mathcal{H}_S-\mathcal{H}_S^0$ becomes equal to:
\begin{eqnarray}\label{delta_H}
\begin{split}
&\approx\sum_{\vec{k},\vec{k}',\vec{q}}\sum_{\sigma\sigma'}f_{I\sigma}^{\dagger}(\vec{k},\tau)f_{I'\sigma'}(\vec{k} ',\tau)\biggl(\sum_{\substack{\left<\vec{r}_i,\vec{r}_j\right>\\  \vec{\rho}=\vec{r}_i-\vec{r}_j}}itQ_f(\tau^x_{II'}\Re e+\tau^y_{II'}\Im m)\mathbb{1}_{\sigma\sigma'}\mathcal{A}_{\rho}^c(\vec{q})\exp(-i\vec{k}\cdot\vec{r}_i+i\vec{k}'\cdot\vec{r}_j+i\vec{q}\cdot\vec{R}_i)\\
&-t'\tilde{Q}_f\sum_{\substack{\ll r_i,r_j \gg\\ \vec{\rho}_w=\vec{r}_i-\vec{r}_j}}\tau^z_{II'}\sigma_{\sigma\sigma'}^w\mathcal{A}_{\rho_w}^c(\vec{q})\left[\exp(-i\vec{k}\cdot\vec{r}_i+i\vec{k}'\cdot\vec{r}_j+i\vec{q}\cdot\vec{R}_i)+\exp(-i\vec{k}\cdot\vec{r}_j+i\vec{k}'\cdot\vec{r}_i+i\vec{q}\cdot\vec{R}_i)\right]\biggl) \quad \tau\in]-\infty,0],
\end{split}
\end{eqnarray}

\begin{figure}[h]
\includegraphics[width=0.8\linewidth]{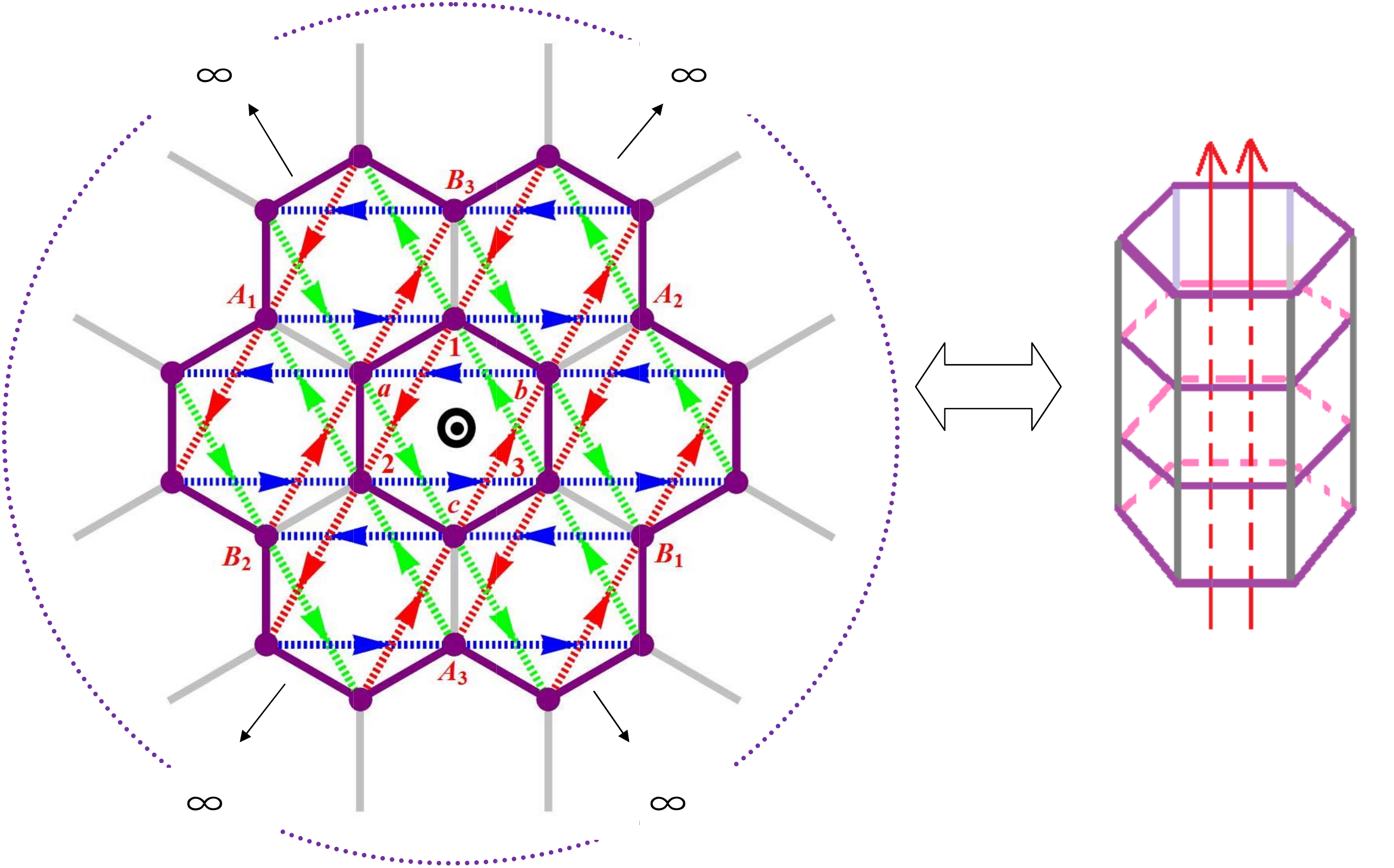}
\caption{(color online) The anisotropic spin texture developing into the bulk above the Mott critical point $U_c$ as a function of $t'/t$ could be associated with the spin physics on the edge by invoking the $U(1)$ pump argument to a system on a cylinder by Laughlin \cite{laughlin}: the centered plaquette with flux inserted could be viewed as one edge and the infinity of the system as another. The different sites in Table \ref{spin_texture_value} are labeled in the figure. The spin physics of insertion of flux could be mapped to the edge spin transport on a cylinder under the insertion of flux and the emergent spin texture could be viewed as `spin charge'. When the gauge field fluctuations insert monopoles (flux in $2+1$ dimensions) into the system, the $U(1)$ spin pump would induce a spin texture around the `edges', namely the core of the monopoles. The spin texture as a spin response summed over all momenta shared similar configurations as the spin transport in Sec. \ref{spin_transport}: when $t'/t\ll 1$ the three components are comparable while $t'/t\gg 1$ one dominant component of spin polarization will appear. The dominant spin polarization polarization depends on the type of links intersected by the line connecting the measured site and the monopole core. It resembles the dependence of the dominant spin polarization component on the types of links to which the boundary is parallel in the context of edge spin physics in the AQSH phase.} \label{fig:gauge_field_configuration}
\end{figure}

\end{widetext}
in which $\vec{r}_i=\vec{R}_i+\vec{r}_I$ and $\vec{r}_j=\vec{R}_i+\vec{r}_J$ are coordinates on which spinon excitations due to the gauge field is considered. $\vec{r}_I$ and $\vec{r}_J$ are vectors connecting the center of the studied plaquette and the corresponding sites indicated in Eq. \ref{premier_deuxieme_voisin}; see Fig. 13. In Eq. \ref{delta_H}, we add up by hand the two terms of hopping on the next-nearest-neighbour links in order to avoid the ambiguity of $\pm i$ when electrons hop along or against the link orientation. Though the lattice is translational invariant, the gauge field is not, rendering the problem of Fourier transformation more sophisticated. We apply the Fourier transformation to derive the spinon response:
\begin{equation}
f_{iI\sigma}(\tau)=\frac{1}{\sqrt{N}}\sum_{\vec{k},\tau}e^{i\vec{k}\cdot\vec{r}_i}f_{I\sigma}(\vec{k},\tau),
\end{equation}
knowing that the spinon system has also two gapped bands as a reminiscence of AQSH phase. The energy of the bands of the spinon system and the band projectors are given explicitly by:

\begin{equation}
\epsilon_{\vec{k}}=\sqrt{|Q_fg(\vec{k})|^2+(\tilde{Q}_f)^2(m_x^2(\vec{k})+m_y^2(\vec{k})+m_z^2(\vec{k}))}
\end{equation}
\begin{eqnarray}
\begin{split}
P_{\pm}({\vec{k}})_{II'\sigma\sigma'}&=\frac{1}{2}[1\pm\frac{1}{\epsilon_{\vec{k}}}[Q_fg(\vec{k})(\tau^x_{II'}\Re e+\tau^y_{II'}\Im m)\mathbb{1}_{\sigma\sigma'}
\\
&+\tilde{Q}_f\tau^z_{II'}(\sigma^x_{\sigma\sigma'}m_x+\sigma^y_{\sigma\sigma'}m_y+\sigma^z_{\sigma\sigma'}m_z)]].
\end{split}
\end{eqnarray}

\begin{figure}[t]
\includegraphics[width=0.42\linewidth]{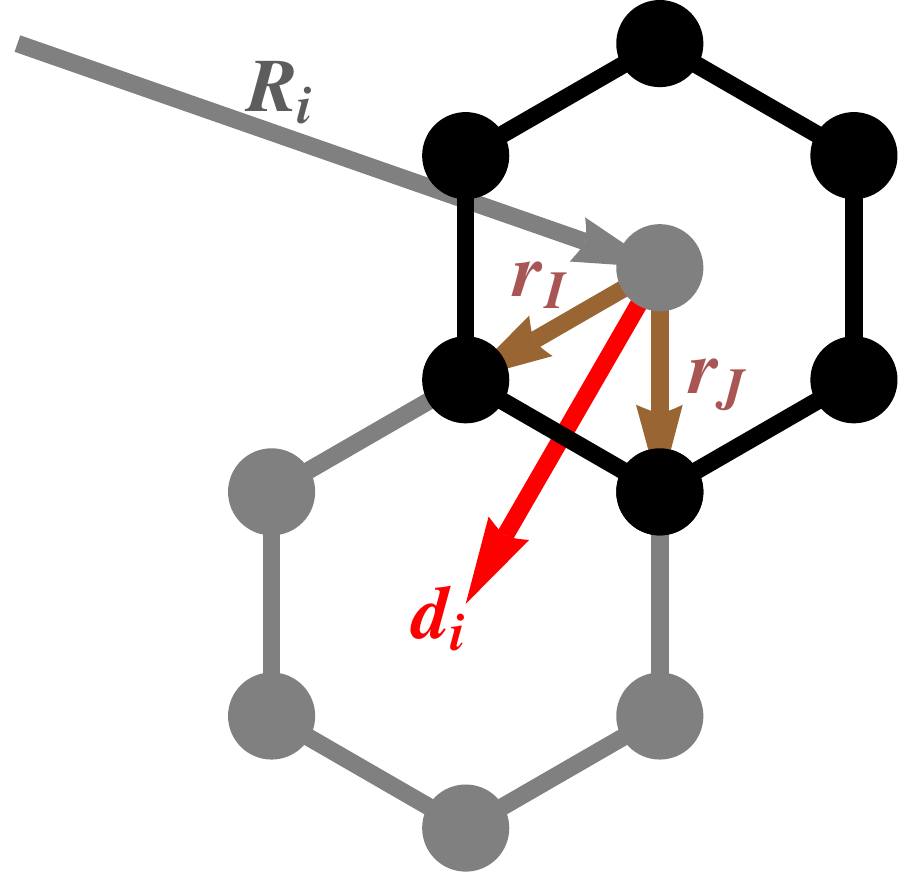}
\includegraphics[width=0.43\linewidth]{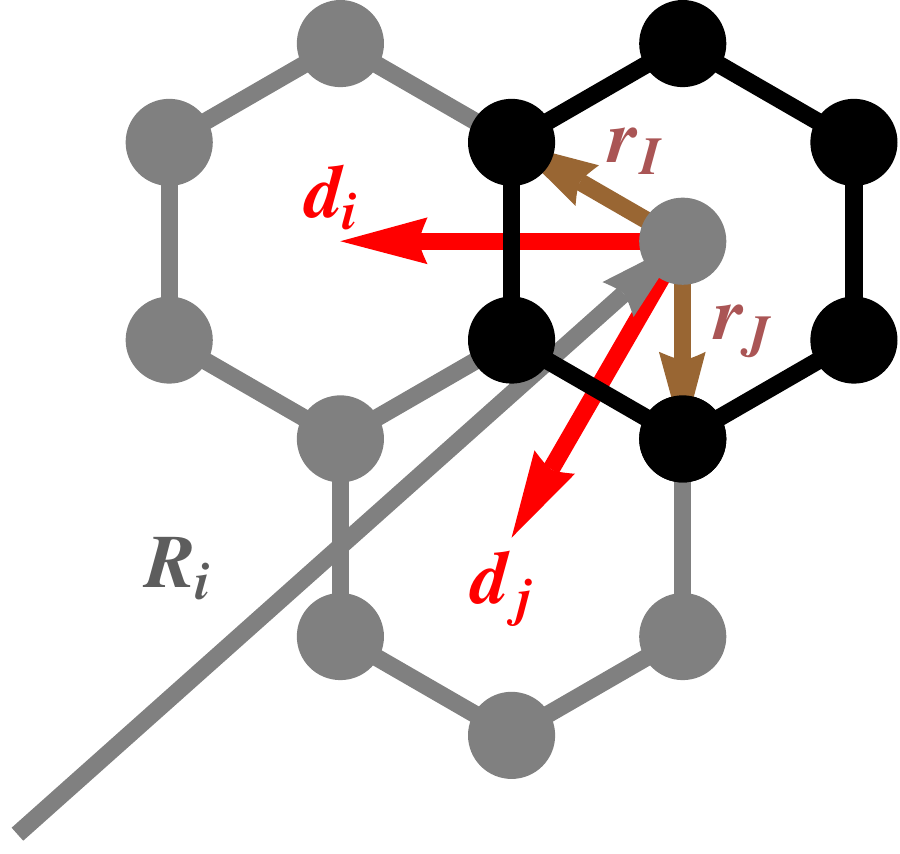}
\caption{(color online) The configuration of $\vec{r}_I$, $\vec{r}_J$ and $\vec{R}_i$ related to Eq. \ref{delta_H}, in which $\vec{r}_I$ $\vec{r}_J$ are vectors connecting the plaquette centers to its vertices and $|\vec{r}_I-\vec{r}_J|$ denotes the first neighbour link; $\vec{R}_i$ gives the coordinates of the studied plaquette. We have to pay special attention to coordinates in Eq. \ref{delta_H}: $\vec{r}_i=\vec{r}_I+\vec{R}_i$ and $\vec{r}_j=\vec{r}_J+\vec{R}_i$. The sum over the coordinates of the studied plaquette at $\vec{R}_i$ in Eq. (\ref{delta_H}) shall induce the momentum conservation $\vec{k}-\vec{k}'=\vec{q}$ of the spinon excitations under the monopole insertion. The vectors $\vec{d}_I$ and $\vec{d}_J$ are vectors connecting the plaquettes for the configuration of gauge fields on the honeycomb lattice; see Appendix \ref{lattice_gauge_configuration}.} \label{fig:premier_second_voisin}
\end{figure}

The configuration of the lattice gauge field is explained in Appendix \ref{lattice_gauge_configuration} using the loop variable method and $\mathcal{A}^c(\vec{q})$ is the Fourier transformed form of the lattice gauge field. The idea of loop variable construction is to write the gauge field on a given link as the difference of the loop variables on the two juxtaposing plaquettes of the link so that $\nabla\cdot\mathcal{A}^c=0$ is automatically satisfied. If $\phi_{\vec{R}_i}$ and $\phi_{\vec{R}_i+\vec{d}_j}$ are two loop variables on the plaquettes centered at $\vec{R}_i$ and $\vec{R}_i+\vec{d}_j$, then the gauge field along the link vector $\vec{\rho}$ juxtaposed by these two neighbouring plaquettes would be: $\mathcal{A}^c_{\rho}=\phi_{\vec{R}_i}-\phi_{\vec{R}_i+\vec{d}_j}$ and the link vector $\vec{\rho}$ is in the counterclockwise orientation with regard to the center plaquette at $\vec{R}_i$.

If we write  Eq. (\ref{delta_H}) in a matrix form, as in Eq. (\ref{perturbed_spinon_hamiltonian_matrix}), then the spin variation $\delta S_M^{\alpha}$ can be written in a concise form. The commutator of the four fermions in Eq. \ref{spin_variation} generates the band projectors indicating the excitations of particle-hole pairs, and the spinon response is proportional to the flux inserted. The sum over the center plaquette coordinates $\vec{R}_i$ in Eq. (\ref{delta_H}) imposes the momentum conservation of $\vec{k}-\vec{k}'=\vec{q}$ which means that the gauge fluctuations excite particle-hole pairs with the momentum exchange $\vec{q}$ equal to the momentum of the fluctuating gauge field.

It is convenient to introduce the notations:

\begin{equation}\label{perturbed_spinon_hamiltonian_matrix}
\delta\mathcal{H}(\tau)=\frac{1}{N}\sum_{\vec{k},\vec{k}'}\sum_{\sigma\sigma'}f_{I\sigma}^{\dagger}(\vec{k},\tau)f_{I'\sigma'}(\vec{k}',\tau)\delta\mathcal{H}^S_{II'\sigma\sigma'}(\tau).
\end{equation}
Then, in Eq. \ref{spin_variation}, the spin polarization variation is written as a trace over spin space of the matrix product measured on the Hilbert space of
the sublattices $\ket{J}$:
\begin{widetext}
\begin{eqnarray}\label{spin_variation}
\begin{split}
\delta S_M^{\alpha}&=\frac{1}{N}\lim_{\eta\rightarrow 0}\sum_{\substack{\vec{k}_1,\vec{k}_1'\\\vec{k}_2,\vec{k}_2'}} \int_{-\infty}^0 d\tau
\left[f_{J\sigma}^{\dagger}(\vec{k}_1,\tau)f_{J\sigma'}(\vec{k}_1',\tau)\sigma_{\sigma\sigma'}^{\alpha}e^{-i(\vec{k}_1-\vec{k}_1')\vec{R}_M},f_{I\tilde{\sigma}}^{\dagger}(\vec{k}_2)f_{I'\tilde{\sigma'}}(\vec{k}_2')\frac{\delta\mathcal{H}_{II'\tilde{\sigma}\tilde{\sigma'}}^S(\tau)}{\epsilon_{\vec{k}_2}-\epsilon_{\vec{k}_2'}-i\eta}\right]\\
&=\frac{1}{N}\lim_{\eta\rightarrow 0}\sum_{\substack{\vec{k},\vec{k}'}}\hbox{Tr}_{\sigma}(\bra{J}[\frac{P_{-}(\vec{k}')\delta\mathcal{H}^sP_+(\vec{k})\sigma^{\alpha}}{\epsilon_{\vec{k}}+\epsilon_{{\vec{k}}'}-i\eta}\exp(-i(\vec{k}-\vec{k}').\vec{R}_M)+\frac{P_{-}(\vec{k})\delta\mathcal{H}^sP_+(\vec{k}')\sigma^{\alpha}}{\epsilon_{\vec{k}'}+\epsilon_{\vec{k}}-i\eta}\exp(i(\vec{k}-\vec{k}').\vec{R}_M)]\ket{J}).
\end{split}
\end{eqnarray}
\end{widetext}

The evaluation of the quantity in Eq. (\ref{spin_variation}) is not so simple because of the integral over the whole first Brillouin zone and therefore we have done this numerically. The anisotropy is manifested by the spin texture dependence on the site $\vec{R}_M$ on which we measure the spin. A table of numerical results of $\delta S_M^{\alpha}$ is listed; see Table I.

\begin{table}
  \centering
  \begin{tabular}{|c|c c c c c c|}
  \hline
  site  & 1 & 2 & 3 & a & b & c \\
  \hline
  $S_x$ & 0.0302 & 0.0302 & -0.142 & 0.142 & -0.0302 & -0.0302 \\
  $S_y$ & 0.0302 & -0.142 & 0.0302 & -0.0302 & 0.142 & -0.0302 \\
  $S_z$ & -0.142 & 0.0302 & 0.0302 & -0.0302 & -0.0302 & 0.142 \\
  \hline
  site  & A1 & B1 & A2 & B2 & A3 & B3 \\
  \hline
  $S_x$ & 0.0314 & -0.0314 & 0.0378 & -0.0378 & 0.0378 & -0.0378 \\
  $S_y$ & 0.0378 & -0.0378 & 0.0314 & -0.0314 & 0.0378 & -0.0378 \\
  $S_z$ & 0.0378 & -0.0378 & 0.0378 & -0.0378 & 0.0314 & -0.0314 \\
  \hline

  \end{tabular}\label{spin_texture_value}
  \caption{Spin texture on the plaquette of inserted flux when $t=t'=1$. The row represents the spin polarization in the x, y and z component and the column represents the sites labeled in Fig. \ref{fig:gauge_field_configuration}.}
\end{table}

The spin texture is very localized around the inserted flux, and numerical studies shows that the spin texture becomes $S^{w}\approx1.0\times10^{-3}$ on the sites that are third neighbours to the center O in Fig. \ref{fig:gauge_field_configuration}. Therefore, we focus on sites around the core of the inserted flux.

From Table I, we observe certain symmetries in the spin texture and these symmetries are in fact inherent to the original spinon system in Eqs. \ref{spinon_hamiltonian} and \ref{spinon_lagrangian}. Specifically, the symmetry of a combination of $2\pi/3$ rotation around the core of the inserted flux and spin polarization permutation. We denote the $2\pi/3$ rotation around the core of the inserted monopole as $R(\frac{2\pi}{3})$ under which different sites are connected:

\begin{equation}
\begin{cases}
R(\frac{2\pi}{3})\vec{R}_1=\vec{R}_2;  R(\frac{2\pi}{3})\vec{R}_2=\vec{R}_3; R(\frac{2\pi}{3})\vec{R}_3=\vec{R}_1\\
R(\frac{2\pi}{3})\vec{R}_a=\vec{R}_b;  R(\frac{2\pi}{3})\vec{R}_b=\vec{R}_c; R(\frac{2\pi}{3})\vec{R}_c=\vec{R}_a\\
R(\frac{2\pi}{3})\vec{R}_{A1}=\vec{R}_{A3};  R(\frac{2\pi}{3})\vec{R}_{A3}=\vec{R}_{A2};  R(\frac{2\pi}{3})\vec{R}_{A2}=\vec{R}_{A1}\\
R(\frac{2\pi}{3})\vec{R}_{B1}=\vec{R}_{B3}; R(\frac{2\pi}{3})\vec{R}_{B3}=\vec{R}_{B2};  R(\frac{2\pi}{3})\vec{R}_{B2}=\vec{R}_{B1}
\end{cases}
\end{equation}
The spin polarization permutation $\sigma$ is defined as follows:
\begin{equation}
\begin{cases}
\sigma(S_z)=S_y\\
\sigma(S_y)=S_x\\
\sigma(S_x)=S_z.
\end{cases}
\end{equation}

\begin{figure}[t]
\includegraphics[scale=0.6]{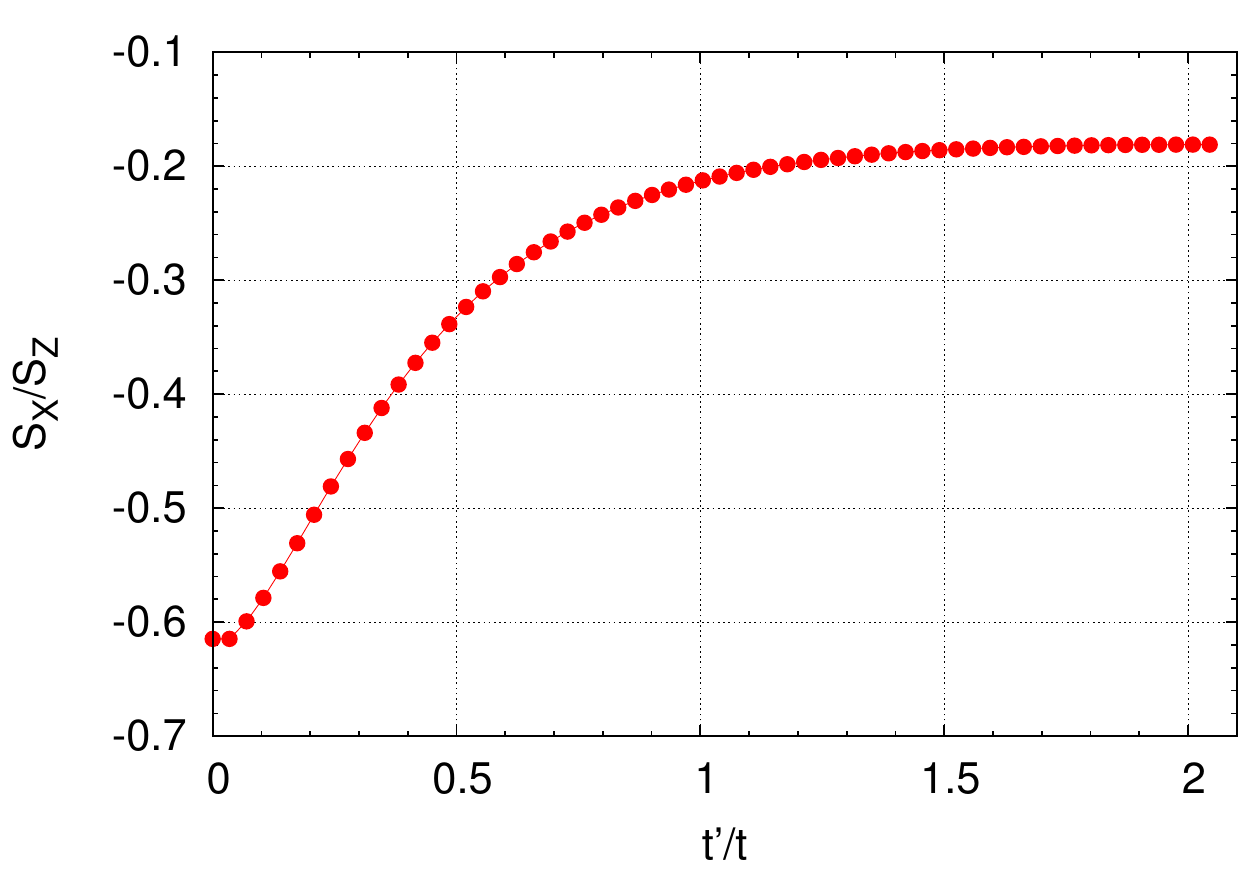}
\caption{(color online) The ratio of $S_x/S_z$ on site 1 in Fig. \ref{fig:gauge_field_configuration} as a function of $t'/t$. This indicates that there are two spin textures when varying $t'/t$: 1. The subordinate spin polarization have an opposite component as the dominant polarization $S_x=S_y=-0.6S_z$ when t' is small compared to t; and 2. $S_z\gg (S_x, S_y)$ when $t'>t$.}\label{fig:spin_texture_ratio}
\end{figure}

If we write the symmetry operator as $U=R(\frac{2\pi}{3})\sigma$, which commutes with the spinon Hamiltonian in Eq. \ref{spinon_hamiltonian}, then the spin texture response on different sites will be related by this symmetry operator. Thus, we confirm the numerical results that $S_1^z=S_2^y=S_3^x$, $S_1^y=S_2^x=S_3^z$ and $S_1^x=S_2^z=S_3^y$, etc. Another symmetry is that spin texture on corresponding sites on different sublattices have opposite signs: $S^w_1=-S_c^w, S^w_2=-S_b^w,S_3^w=-S_a^w$, (w=x,y,z) and identically for the sites A1 \& B1, A2 \& B2, A3 \& B3, etc. This symmetry is also present in the original spinon Hamiltonian in that $i\sigma^w$ is changed into $-i\sigma^w$ for the next-nearest-neighbour hopping on different sublattices.

The anisotropy is manifested by one dominant component of the spin polarization on different types of sites: $S_1^z=S_2^y=S_3^x$ on site $1$, $2$ and $3$. The lines linking these sites and the monopole core intersect respectively the $z$, $y$ and $x$ links, so the dominant spin polarization are $S_1^z=S_2^y=S_3^x$ on site $1$, $2$ and $3$. Accordingly, the dominant spin polarization component on one site corresponds to the type of links intersected by the line linking the monopole core and the site under investigation. The subordinate components and dominant component on each site change differently when $t'/t$ varies, thus generating two different types of spin texture above the Mott critical point as in figure \ref{fig:spin_texture_diagram}. At small $t'/t$, the spin texture tends to zero (proportional to $t'$) because the appearance of spin textures is due to the effective spin-orbit coupling in the spinon sector; the subordinate components are  $S_x=S_y=-0.6S_z$ on site $1$, for example. At large $t'>t$, the subordinate components are small compared to the dominant components $S_x=S_y\approx -0.2S_z$. The ratio between the subordinate components and the dominant components is analyzed in Fig. \ref{fig:spin_texture_ratio}. This shows that the peculiar spin texture substantially develops by increasing the ratio $t'/t$.

As mentioned earlier, the analogy between the edge spin physics in the AQSH phase and the spin texture in the bulk in the intermediate interaction regime can be fleshed out using the argument of Laughlin \cite{laughlin}, the $U(1)$ pump of a system on a cylinder with 2 edges, in which `charge' transport on the edges would be induced under insertion of flux of such topological system on cylinder. However, the `charge' in this anisotropic spin-orbit coupling model is the `spin charge'. Fig. \ref{fig:gauge_field_configuration} illustrates how the spin physics in the two different contexts, edges versus bulk, are related. The sites around the monopole core are analogous to one edge and the infinity to another, the spin texture on different sites are then `spin charge' transported around under the insertion of a fluctuating flux. The anisotropy factor in the context of edge states of the AQSH effect is related to the type of links to which the boundary is parallel, and in the context of spin texture in the bulk, it is the type of links intersected by the line linking the monopole core and the corresponding site. These anisotropy factors determine the dominant spin polarization component when $t'>t$.

The spinon response is influenced by a plasma of monopoles rather than simply the insertion or destruction of one monopoles or two. Different from the Kane-Mele-Hubbard model in which the correlation of two monopoles separated far enough could trigger a homogeneous long-range magnetic order, the spin texture in this anisotropic spin-orbit model beyond the Mott critical point entails the coordination of several spin textures distributed around the monopole plasma; the real magnetic structure in this regime has to be considered as a statistical average of these spin textures, which remains to be explored in terms of difficulties such as frustration between spin texture induced by two juxtaposed monopoles and confinement of the $U(1)$ monopole plasma, etc.

The spin texture of two adjacent monpoles and one pair of adjacent monopole-antimonopole is provided in Appendix \ref{Two_plaquettes}. The results is heuristic and the spin-texture induced by two adjacent monopole-antimonopole seems to be in good agreement with the spiral order: when the monopole-antimonopole pair is positioned along the $x$ ($y$ or $z$)  links the spin texture on the sites shared by the two plaquetttes with fluxes penetrated would be in the $YZ$ ($XZ$ or $XY$) plane. This result tends to agree with the super-exchange Hamiltonian in Eq. (\ref{J_1_J_2_magnetism}) when $J_2\gg J_1$.

It is perhaps important to underline that the emergent magnetism induced by the Mott transition will break the time-reversal symmetry and the Kramers pairs which enables the edge spin transport shall disappear.

\section{Discussion}

To summarize, following Ref. \cite{Shitade}, we have explored the Quantum Spin Hall physics in the presence of an anisotropy in the spin-orbit coupling and taking into account the interaction between electrons then resulting in a quite generic model Hamiltonian \cite{Ruegg,RachelThomale}.

At a general level, the bulk-edge correspondence still exists in the topological band insulator phase at weak (to moderate) interactions implying for example that the system is protected by a $\mathbb{Z}_2$ topological invariant. We have shown that the helical edge states are now characterized by a prevalent spin-orbital texture driven by the anisotropy in the spin-orbit coupling and therefore this phase is referred to as the Anisotropic Quantum Spin Hall phase in Fig. 1. One could observe these
features in the edge states using current technology \cite{koenig-07s766}.

By increasing the interaction strength between electrons, by analogy with the Kane-Mele-Hubbard model \cite{RachelLeHur,Hohenadler,Wu,Lee} and in agreement with a previous analysis \cite{Ruegg}, we predict a Mott transition above which the charge and the spin of an electron becomes disentangled in the bulk and the `chargeons' become localized as a result of the dominant Hubbard interaction. At the Mott transition, the Kramers pair at the edges now shall disappear and even though the single-electron gap does not close in the bulk at the transition, the electron Green's function should eventually reveal a two-peak structure above the fact reflecting such a disentangling phenomenon of charge and spin. By applying the U(1) slave-rotor theory \cite{Florens} and considering gauge fluctuations around the mean-field saddle point, we have thoroughly analyzed how the pseudospin-orbital texture at the edges now progressively proliferates into the bulk. Note that such spin textures are different in the Kane-Mele Hubbard model, where above the Mott transition, the condensation of monopoles results in  long-range XY spin ordering \cite{RachelLeHur}. Deep in the Mott phase, our results suggest a quantum phase transition between a N\' eel and a non-colinear Spiral phase. The latter which takes place for prevalent spin-orbit couplings has also been suggested in Refs. \cite{RachelThomale,Kargarian}. The ordering wave-vectors associated with the Spiral phase are $(2\pi/(3\sqrt{3}),0)$, $(-\pi/(3\sqrt{3}),\pi/3)$.

At this point, it is perhaps relevant to pinpoint that the magnetic order induced by this spin texture analysis right above the Mott transition needs to be fleshed out. Though the heuristic consideration of the adjacent monopole-antimonopole pair  --- see Appendix \ref{Two_plaquettes}) --- appears to be in good agreement with the spiral order, we cannot definitely exclude the presence of an additional phase for moderate interactions for intermediate strengths of spin-orbit couplings, as found for example in Ref. \cite{Ruegg}, since the present analysis does not incorporate very well  frustration effects. Meanwhile, the formation of N\'eel order above the Mott critical point is a subject under current debate in similar situations \cite{Hermele_2,Hermele_3,LeeLee,Nogueira, Reuther, WesselS,Wu,Sorella,Senechal,Herbut,Wubis,Subir}. The results found in this paper may have a direct relevance for the understanding of thin films of NaIr$_2$O$_3$ \cite{Jenderka} and possibly Lithium-based iridates or artificial graphene subject to gauge fields \cite{graphene,Goldman}. We also note some analogy with the interacting spinful Hofstadter problem discussed in Ref. \onlinecite{Cocks}. Monopoles have also been investigated in spin ice materials \cite{Roderich} and also in the context of polariton quantum fluids \cite{Amo}.

\section{Acknowledgements}

We thank S. Biermann, G. Bossard, R. Coldea, M. Ferrero, A. Georges, W. Hofstetter, G. Jackeli, A. Jagannathan, C. Lhuillier, N. Perkins, A. Petrescu, S. Rachel, N. Regnault, R. Thomale, R. Valenti and W. Wu for useful discussions. This work has benefitted from discussions during CIFAR meetings in Canada, program Quantum Materials. This work has also benefitted from workshops at KITP Santa-Barbara and Aspen Center for Physics.

\begin{appendix}
\begin{widetext}
\section{Edge State Solution via Transfer Matrix}\label{Transfer_matrix}

In this Appendix, we provide an analytical solution of the edge states, formally at $U=0$, following for example \onlinecite{AlexKaryn}. We can view the system as semi-infinite with layers of one-dimensional two-sublattice chains coupled together as in Fig. \ref{fig.edge_transport} \cite{Mong_Shivamoggi,Pershoguba_Yakovenko}. We note the wave function on the $n^{th}$ layer as $\psi_{A,B}^n$, then we can write down the Schr\" odinger equation of the system:

\begin{eqnarray}\label{Edge_Pauli}
\begin{split}
&\left[-it'(e^{-i\frac{\sqrt{3}}{2}k_x}\sigma_z-e^{i\frac{\sqrt{3}}{2}k_x}\sigma_y)\tau_z-\frac{t}{2}(\tau_x+i\tau_y)\right]\psi_{n+1}+E\psi_n+\left(2t'\sin\sqrt{3}k_x\sigma_x\tau_z-2t\cos\frac{\sqrt{3}}{2}k_x\tau_x\right)\psi_n\\
&+[it'(e^{i\frac{\sqrt{3}}{2}k_x}\sigma_z-e^{-i\frac{\sqrt{3}}{2}k_x}\sigma_y)]\psi_{n-1}=0
\end{split}
\end{eqnarray}
Let us write down the wave function decaying when penetrating into the bulk: $\psi_{nJ\sigma}=\sum_i\lambda_i^nu_{iJ\sigma}$ such that the wave function vanishes at the edge $\psi_0=\sum_{i}u_{iJ\sigma}=0$. Then the Schr\" odinger equation reads:
\begin{equation}\label{schrodinger_layer}
Eu_{iJ\sigma}=[c_x^i\tau_x+c_y^i\tau_y+(m_x^i\sigma_x+m_y^i\sigma_y+m_z^i\sigma_z)\tau_z]u_{iJ\sigma}=M_iu_{iJ\sigma},
\end{equation}
in which
\begin{eqnarray}
\begin{split}
&c_x^i=\frac{t}{2}\left(\lambda_i+\frac{1}{\lambda_i}\right)-2t\cos\frac{\sqrt{3}}{2}k_x \\
&c_y^i=\frac{it}{2}(\lambda_i-\frac{1}{\lambda_i})\\
&m_x^i=-2t'\sin\sqrt{3}k_x\sigma_x\\
&m_y^i=it'(\lambda_ie^{i\frac{\sqrt{3}}{2}k_x}-\frac{1}{\lambda_i}e^{-i\frac{\sqrt{3}}{2}k_x})\\
&m_z^i=-it'(\lambda_ie^{-i\frac{\sqrt{3}}{2}k_x}-\frac{1}{\lambda_i}e^{i\frac{\sqrt{3}}{2}k_x}).
\end{split}
\end{eqnarray}
We can diagonalize the matrix in Eq. \ref{schrodinger_layer} by squaring it:
\begin{equation}\label{diagonalisation_layer}
E^2=t^2+4t^2\cos^2\left(\frac{\sqrt{3}}{2}k_x\right)+4t'^2+4t'^2\sin^2\sqrt{3}k_x+4t'^2\cos \sqrt{3}k_x+2t^2\cos\frac{\sqrt{3}}{2}k_x(\lambda_i+\frac{1}{\lambda_i})-2t'^2\cos\left(\sqrt{3}k_x\right)(\lambda_i+\frac{1}{\lambda_i})^2.
\end{equation}
Eq. \ref{diagonalisation_layer} is a second-order equation of $\lambda_i+\frac{1}{\lambda_i}$ and a fourth order equation of $\lambda_i$. There are 4 roots of
$\lambda_i$ among which two of them satisfy $|\lambda_i|<1$, and if $\lambda_i$ is a root of the equation so is $\frac{1}{\lambda_i}$. Therefore, we are allowed to write the wave function as a superposition of two eigenvectors:
\begin{equation}
\psi_n=u_1\lambda_1^n+u_2\lambda_2^n.
\end{equation}
The vanishing of the wave function at the edge gives that $u_1=-u_2=u$, then the wave function shall be written as:
\begin{equation}
\psi_n=(\lambda_1^n-\lambda_2^n)u.
\end{equation}

The fact that the two matrices $E-M_i$ ($i=1,2$) in Eq. \ref{schrodinger_layer} are sharing a null-eigenvector implies that
\begin{equation}
\hbox{Det}(E-M_1)=\hbox{Det}(E-M_2)=\hbox{Det}(a_1(E-M_1)+a_2(E-M_2))=0,
\end{equation}
in which $a_1$,$a_2$ are two arbitrary constants. This is equivalent to:
\begin{eqnarray}
\begin{split}
E^2&=(c_x^1)^2+(c_y^1)^2+(m_x^1)^2+(m_y^1)^2+(m_z^1)^2=(c_x^2)^2+(c_y^2)^2+(m_x^2)^2+(m_y^2)^2+(m_z^2)^2\\
&=c_x^1c_x^2+c_y^1c_y^2+m_x^1m_x^2+m_y^1m_y^2+m_z^1m_z^2.
\end{split}
\end{eqnarray}
Then we have:
\begin{equation}
(c_x^1-c_x^2)^2+(c_y^1-c_y^2)^2+(m_x^1-m_x^2)^2+(m_y^1-m_y^2)^2+(m_z^1-m_z^2)^2=0,
\end{equation}
\begin{equation}
(\lambda_1-\lambda_2)^2\left[2t'^2\cos\sqrt{3}k_x\left(1+\left(\frac{1}{\lambda_1\lambda_2}\right)^2\right)+\frac{t^2+4t'^2}{\lambda_1\lambda_2}\right]=0.
\end{equation}

$\lambda_1=\lambda_2$ gives a trivial solution, then we can find $\lambda_1\lambda_2$ from the above equation. If we put $L=\frac{t^2+4t'^2}{2t'^2\cos\sqrt{3}k_x}$, then:
\begin{equation}
M=\lambda_1\lambda_2=\frac{-L\pm\sqrt{L^2-4}}{2}.
\end{equation}
Since we must impose $|\lambda_i|<1$ ($i=1,2$), this implies that $|\lambda_1\lambda_2|<1$. The first Brillouin zone for the one-dimensional chain is $[0,\frac{2\pi}{\sqrt{3}}]$, resulting in:
\begin{equation}
\begin{cases}
\lambda_1\lambda_2=-\frac{L-\sqrt{L^2-4}}{2} \quad k_x\in [0, \frac{\pi}{2\sqrt{3}}] \cup [\frac{3\pi}{2\sqrt{3}},\frac{2\pi}{\sqrt{3}}]\\
\lambda_1\lambda_2=-\frac{L+\sqrt{L^2-4}}{2} \quad k_x\in [\frac{\pi}{2\sqrt{3}},\frac{3\pi}{2\sqrt{3}}].
\end{cases}
\end{equation}
From Eq. (\ref{schrodinger_layer}) we find:
\begin{equation}
\lambda_1+\frac{1}{\lambda_1}+\lambda_2+\frac{1}{\lambda_2}=(1+\frac{1}{\lambda_1\lambda_2})(\lambda_1+\lambda_2)=\frac{t^2\cos\frac{\sqrt{3}}{2}k_x}{t'^2\cos\sqrt{3}k_x}.
\end{equation}

\begin{figure}[t]
\includegraphics[scale=0.6]{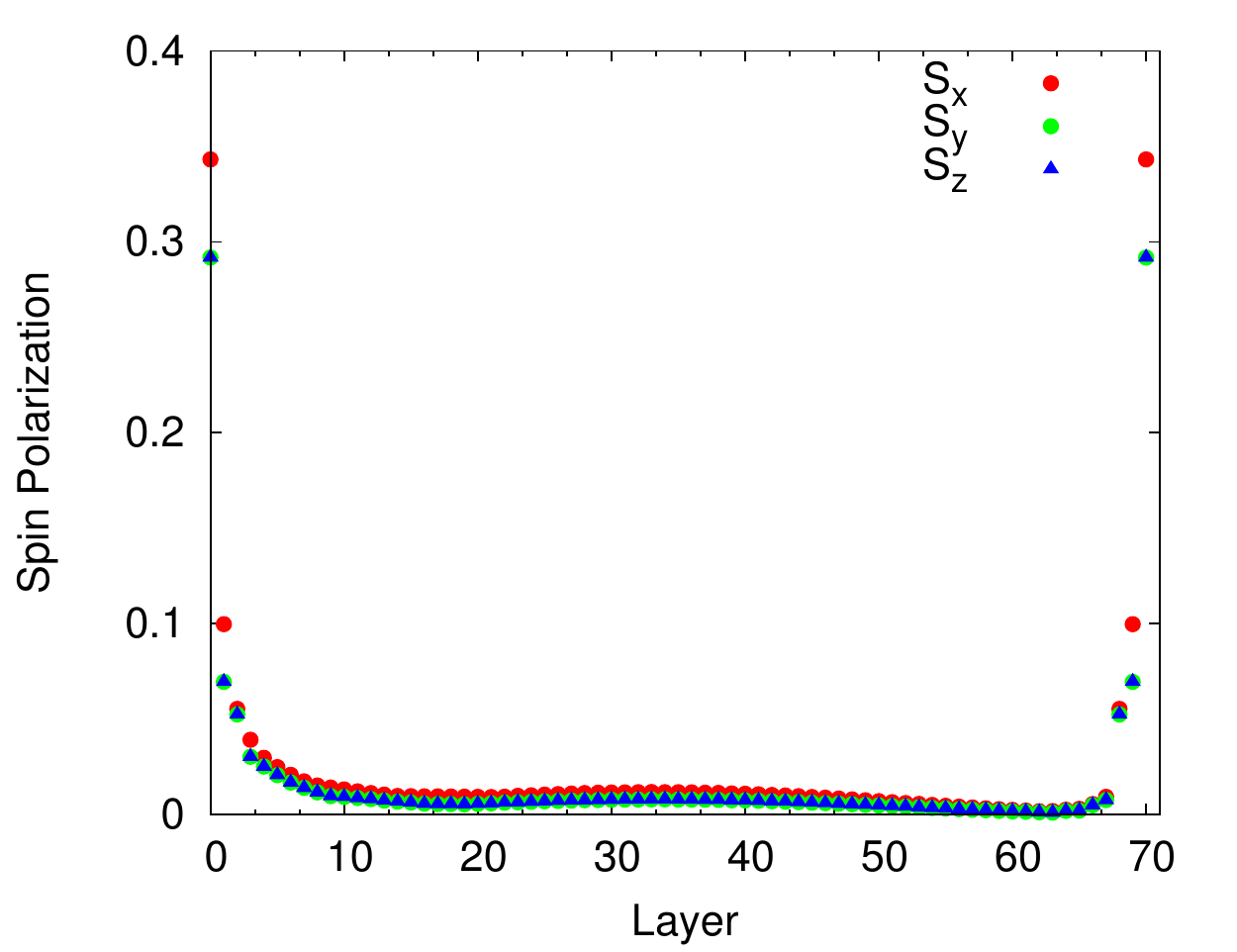}
\caption{(color online) The numerical study of spin polarization magnitude as a function of layer in the system of 70 layers of one-dimensional chains described by Eq. \ref{Edge_Pauli} at $t'=0.5t$.}\label{fig:shitade_cylinder_theory}
\end{figure}

From $\lambda_1+\lambda_2=\frac{t^2\cos\frac{\sqrt{3}}{2}k_x}{t'^2\cos\sqrt{3}k_x(1+\frac{1}{\lambda_1\lambda_2})}=N$ we can work out the two eigenvalues $\lambda_1$ and $\lambda_2$: $\lambda_{1,2}=\frac{-N\pm\sqrt{N^2-4M}}{2}$ which gives us the penetration length: $\xi_{1,2}=-\ln (\lambda_{1,2})$. From the relation:
\begin{equation}
(\lambda_1+\frac{1}{\lambda_1})(\lambda_2+\frac{1}{\lambda_2})=\frac{(\lambda_1+\lambda_2)^2}{\lambda_1\lambda_2}-2+\lambda_1\lambda_2+\frac{1}{\lambda_1\lambda_2}=\frac{E^2-(t^2+4t^2\cos^2\frac{\sqrt{3}}{2}k_x+4t'^2+4t'^2\sin^2\sqrt{3}k_x+4t'^2\cos\sqrt{3}k_x)}{2t'^2\cos\sqrt{3}k_x}
\end{equation}
we find the dispersion relation for the edge states and this fit well with the spectrum obtained numerically in Fig. \ref{fig:shitade_cylinder_theory}:
\begin{equation}
E_{\text{edge}}=\pm\sqrt{4t^2\cos^2\frac{\sqrt{3}}{2}k_x+4t'^2\sin^2\sqrt{3}k_x-\frac{4t'^4\cos^4\frac{\sqrt{3}}{2}k_x}{t^2+4t'^2-4t'^2\cos\sqrt{3}k_x}}.
\end{equation}

In order to find the wave function, we can use the projector:
\begin{equation}
P_{\pm}^i=\frac{1}{2}\left(1\pm\left(\frac{c_x^i}{E_0}\tau_x+\frac{c_y^i}{E_0}\tau_y+(\frac{m_x^i}{E_0}\sigma_x+\frac{m_y^i}{E_0}\sigma_y+\frac{m_z^i}{E_0}\sigma_z)\tau_z\right)\right)
\end{equation}
 that diagonalizes Eq. (\ref{schrodinger_layer}). The eigenvector $u$ is the intersection of the two projected spaces entailed by $P_-^{1,2}$.

\section{Lattice Gauge Field Configuration by Construction of Loop Variables}\label{lattice_gauge_configuration}

Loop variables is a tool to trace out the lattice gauge field configuration by attaching a loop variable to each plaquette. We follow the notations of Sec. III C in
the main text, except that the gauge field is renamed $\mathcal{A}$ (instead of $\mathcal{A}^c$) for simplicity.

\begin{figure}[t]
\includegraphics[width=0.45\linewidth]{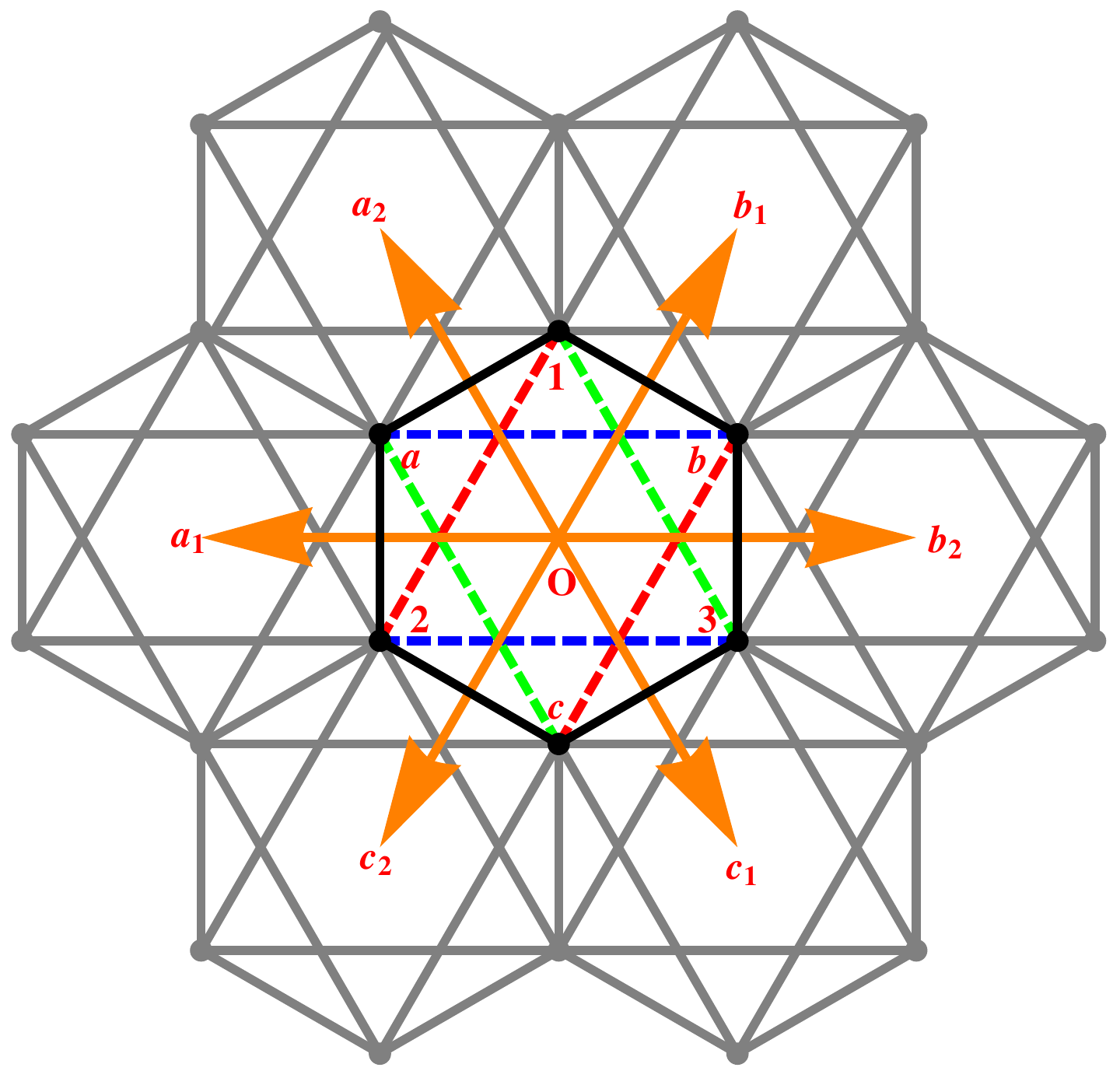}
\caption{(color online) The lattice gauge field configuration on the honeycomb lattice. On each triangular and each honeycomb plaquette, a lattice loop variable is defined. The gauge field on the counterclockwise oriented links are defined as loop variables on the left hand side minus loop variables on the right hand side of the link when travelling parallel to the link orientation. For example, $\mathcal{A}_{a2}=\phi_0-\phi_{a1}$. The loop variable construction satisfies automatically $\nabla\cdot\mathcal{A}=0$, and $\nabla\times\mathcal{A}=\Phi_m \delta_{\vec{R}_o,\vec{O}}$ is expressed by a Laplace equation in Eq. \ref{Laplacian_equation_for_loop_variables}, in which $\Phi_m$ is the magnetic flux penetrating the center of the plaquette.} \label{fig:gauge_field_configuration_1}
\end{figure}

Then the gauge field on the links as the difference of left hand side and the right hand side loop variables when one is oriented along the gauge field direction on the link or $\nabla\times\mathcal{\phi}=\mathcal{A}$ in the continuous limit. The advantage of this construction is the automatic satisfaction of $\nabla\cdot\mathcal{A}=\sum_j\mathcal{A}_{Oj}=0$ on a given site $O$. Then the equation $\nabla\times\mathcal{A}=\Phi_m$ is translated into the Laplacian equation after doing the Fourier transformation:
\begin{equation}\label{Laplacian_equation_for_loop_variables}
\nabla\times\mathcal{A}=z\phi(\vec{R}_o)-\sum_j\phi(\vec{R}_o+\vec{r}_j)=\sum_{\vec{q}}\phi(\vec{q})(z-\sum_j \exp(i\vec{q}\cdot\vec{r}_j))\exp(i\vec{q}\cdot\vec{R}_o)=\Phi_m \delta_{\vec{R}_o,\vec{O}},
\end{equation}
where $z$ is the coordinate number, $\vec{r}_j$ are vectors connecting neighbours and $\vec{R}_o$ is the center of a plaquette. Now we look at the gauge field configuration for honeycomb lattice in Fig. \ref{fig:gauge_field_configuration_1}. If we note $h(\vec{q})=\sum_j\exp(i\vec{q}\cdot\vec{r}_j)$, then making use of the fact that center of the hexagonal plaquettes form a triangular lattice which is a Bravais lattice, we can implement the Fourier transformation naturally enough.

\begin{eqnarray}\label{field_strength_premier_voisin}
\mathcal{A}_{1a}&=&\phi_o-\phi_{a_2}=\int dq \exp(i\vec{q}\cdot\vec{R}_o)\Phi_m\frac{1-\exp(i\vec{q}\cdot\vec{r}_{a_2})}{6-h(\vec{q})}\nonumber\\
\mathcal{A}_{a2}&=&\phi_o-\phi_{a_1}=\int dq \exp(i\vec{q}\cdot\vec{R}_o)\Phi_m\frac{1-\exp(i\vec{q}\cdot\vec{r}_{a_1})}{6-h(\vec{q})}\nonumber\\
\mathcal{A}_{2c}&=&\phi_o-\phi_{c_2}=\int dq \exp(i\vec{q}\cdot\vec{R}_o)\Phi_m\frac{1-\exp(i\vec{q}\cdot\vec{r}_{c_2})}{6-h(\vec{q})}\nonumber\\
\mathcal{A}_{c3}&=&\phi_o-\phi_{c_1}=\int dq \exp(i\vec{q}\cdot\vec{R}_o)\Phi_m\frac{1-\exp(i\vec{q}\cdot\vec{r}_{c_1})}{6-h(\vec{q})}\nonumber\\
\mathcal{A}_{3b}&=&\phi_o-\phi_{b_2}=\int dq \exp(i\vec{q}\cdot\vec{R}_o)\Phi_m\frac{1-\exp(i\vec{q}\cdot\vec{r}_{b_2})}{6-h(\vec{q})}\nonumber\\
\mathcal{A}_{b1}&=&\phi_o-\phi_{b_1}=\int dq \exp(i\vec{q}\cdot\vec{R}_o)\Phi_m\frac{1-\exp(i\vec{q}\cdot\vec{r}_{b_1})}{6-h(\vec{q})}.
\end{eqnarray}

The field strength on vectors connecting next-nearest-neighbors are more complicated since the loop variables defined in the center of the triangular lattice are not on a Bravais lattice, and therefore in order to obtain the right configuration we need an  extra constraint between the two sublattices to `massage' the above construction into the right Fourier transformed expression. To take the example of the sublattice of $a,b,c$ in Fig. \ref{fig:gauge_field_configuration}, we apply the following constraints derived from $\nabla\times\mathcal{A}=0$:
\begin{eqnarray}\label{premier_deuxieme_voisin}
3\phi_a&=&\phi_0+\phi_{a_1}+\phi_{a_2}\nonumber\\
3\phi_b&=&\phi_0+\phi_{b_1}+\phi_{b_2}\nonumber\\
3\phi_c&=&\phi_0+\phi_{c_1}+\phi_{c_2}.
\end{eqnarray}

Then we get eventually:

\begin{eqnarray}\label{field_strength_deuxieme_voisin}
\mathcal{A}_{12}&=&\phi_0-\phi_a=\int dq\Phi_m\frac{2-\exp(i\vec{q}\cdot\vec{r}_{a_1})-\exp(i\vec{q}\cdot\vec{r}_{a_2})}{6-h}\exp(i\vec{q}\cdot\vec{R}_o)\nonumber\\
\mathcal{A}_{23}&=&\phi_0-\phi_c=\int dq\Phi_m\frac{2-\exp(i\vec{q}\cdot\vec{r}_{c_1})-\exp(i\vec{q}\cdot\vec{r}_{c_2})}{6-h}\exp(i\vec{q}\cdot\vec{R}_o)\nonumber\\
\mathcal{A}_{31}&=&\phi_0-\phi_b=\int dq\Phi_m\frac{2-\exp(i\vec{q}\cdot\vec{r}_{b_1})-\exp(i\vec{q}\cdot\vec{r}_{b_2})}{6-h}\exp(i\vec{q}\cdot\vec{R}_o)\nonumber\\
\mathcal{A}_{ac}&=&\phi_0-\phi_2=\int dq\Phi_m\frac{2-\exp(i\vec{q}\cdot\vec{r}_{a_1})-\exp(i\vec{q}\cdot\vec{r}_{c_2})}{6-h}\exp(i\vec{q}\cdot\vec{R}_o)\nonumber\\
\mathcal{A}_{cb}&=&\phi_0-\phi_3=\int dq\Phi_m\frac{2-\exp(i\vec{q}\cdot\vec{r}_{c_1})-\exp(i\vec{q}\cdot\vec{r}_{b_2})}{6-h}\exp(i\vec{q}\cdot\vec{R}_o)\nonumber\\
\mathcal{A}_{ba}&=&\phi_0-\phi_1=\int dq\Phi_m\frac{2-\exp(i\vec{q}\cdot\vec{r}_{a_2})-\exp(i\vec{q}\cdot\vec{r}_{b_1})}{6-h}\exp(i\vec{q}\cdot\vec{R}_o).
\end{eqnarray}

\section{Spin Texture under Two Adjacent Monopoles}\label{Two_plaquettes}

\begin{figure}[htb]
\includegraphics[scale=0.45]{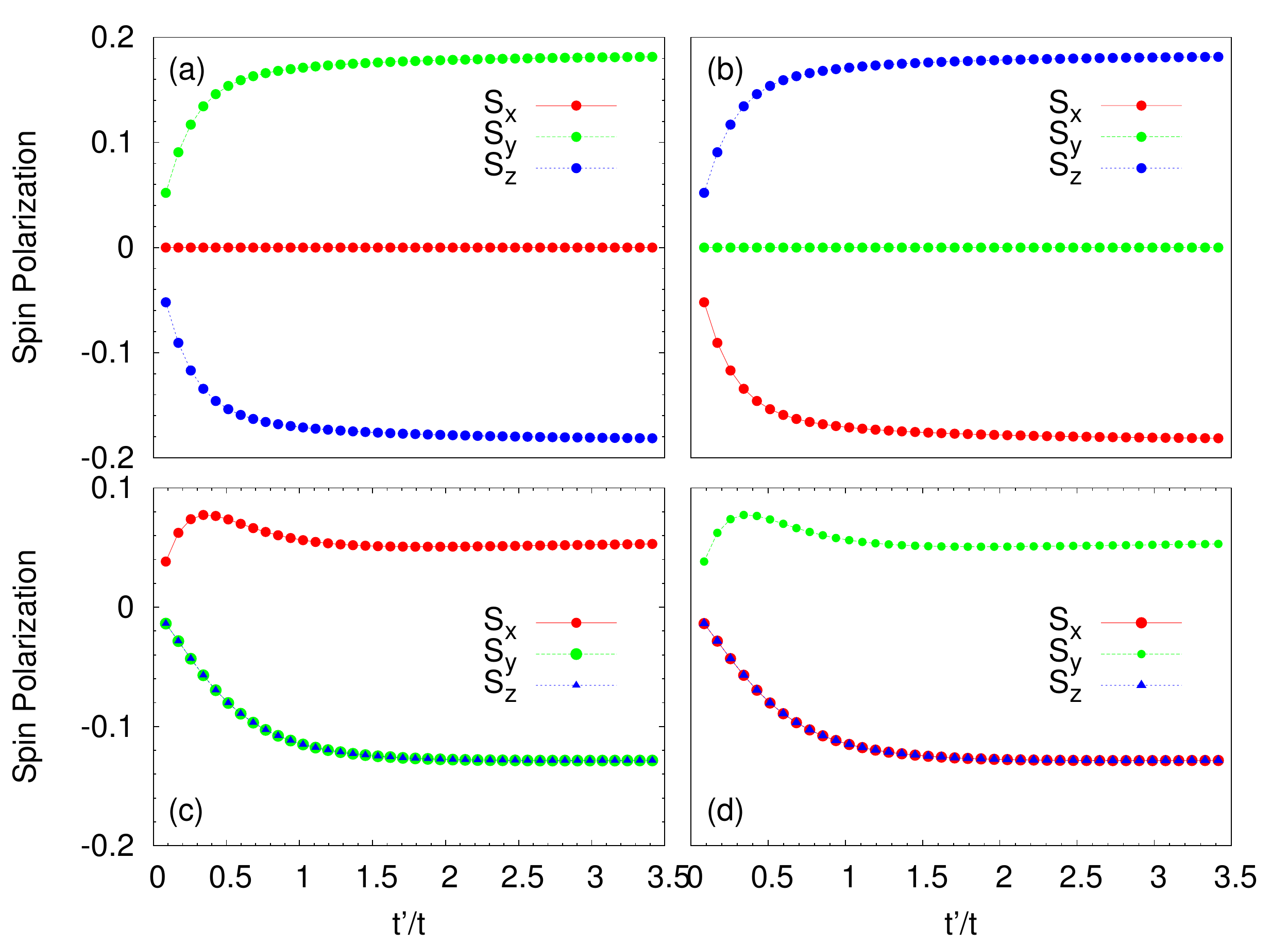}
\caption{(color online) The spin texture under the monopole-antimonople pair (two monopoles) on two adjacent plaquettes: (a) the spin texture on site 1 when the monopole-antimonopole are respectively inserted on plaquette O and b1 Fig. \ref{fig:gauge_field_configuration_1}.  (b) the spin texture on site 3 when the monopole-antimonopole are respectively inserted on plaquette O and c1. (c) the spin texture on site 1 when two monopoles are inserted on plaquette O and b1. (d) the spin texture on site 3 when two monopoles are inserted on plaquette O and c1.}\label{fig:two_plaquettes}
\end{figure}

\begin{figure}[htb]
\includegraphics[scale=0.45]{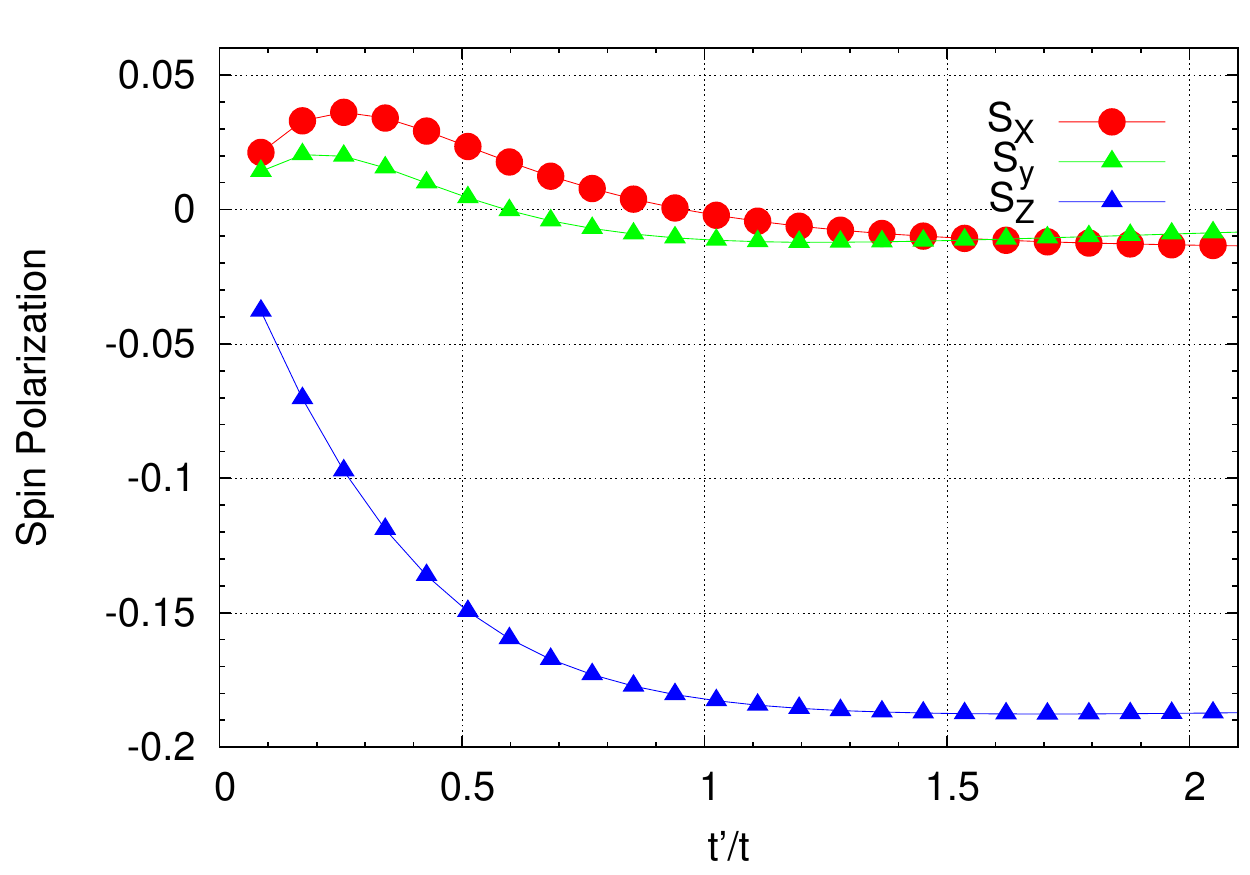}
\includegraphics[scale=0.45]{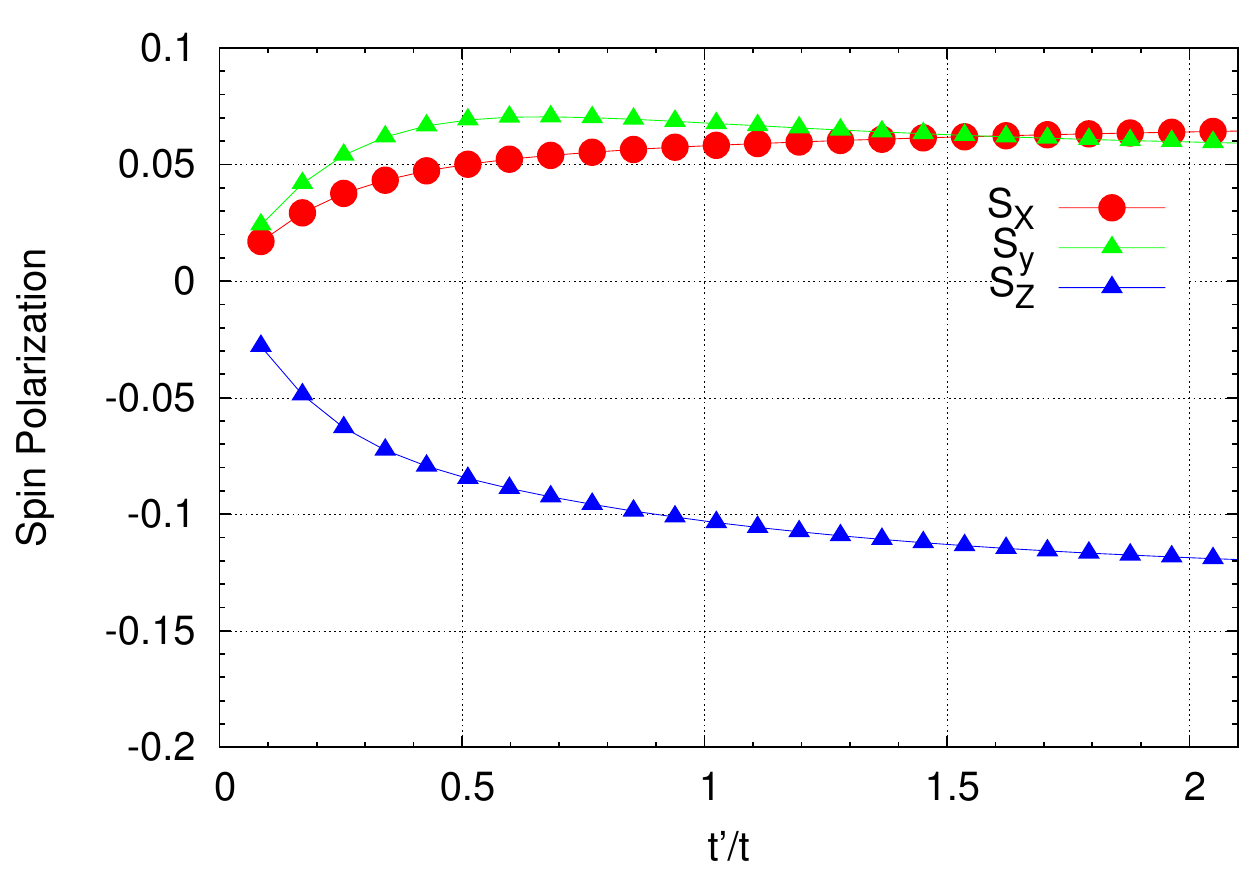}
\caption{(color online)The spin texture on site 1 when the monopole-antimonopole pair (left panel) or the two monopoles are respectively inserted on plaquette O and b2. }\label{fig:two_plaquettes_other_sites}
\end{figure}

\end{widetext}

\end{appendix}


\vskip 3cm

\begin{thebibliography}{99}

\bibitem{Laughlin_FQHE}
R. B. Laughlin, Phys. Rev. Lett. {\bf 50}, 1395 (1983).

\bibitem{Haldane1983}
F. D. M. Haldane, Phys. Lett. A {\bf 93}, 464 (1983).

\bibitem{Wen}
X. G. Wen, Phys. Rev. B. {\bf 40}, 7387 (1989); X.-G. Wen, Int. J. Mod. Phys. {\bf B4}, 239 (1990).

\bibitem{ReadMoore}
G. Moore and N. Read, NuclPhys B{\bf 360}, 362 (1991); N. Read and D. Green, Phys.Rev. B {\bf 61}, 10267 (2000).

\bibitem{Leboeuf}
P. Leboeuf, J. Kurchan, M. Feingold and D. P. Arovas, Phys. Rev. {\bf 65}, 3076 (1990).

\bibitem{Kitaev}
A. Kitaev, Annals of Physics {\bf 321}, 2-111 (2006).

\bibitem{Nayak}
C. Nayak, S. H. Simon, A. Stern, M. Freedman, and S. Das Sarma, Rev. Mod. Phys.
{\bf 80}, 1083 (2008).

\bibitem{DoucotIoffe}
B. Dou\c{c}ot and L. B. Ioffe, Rep. Prog. Phys. {\bf 75}, 072001 (2012).

\bibitem{ti-reviews}
M.~Z. Hasan and C.~L. Kane, Rev. Mod. Phys. {\bf 82},  3045  (2010);
J. Moore, Nature {\bf 464},  194  (2010); X.-L. Qi and S.-C. Zhang, Rev. Mod. Phys. {\bf 83}, 1057 (2011).

\bibitem{Majoranas}
A. Y. Kitaev Phys.-Usp. {\bf 44} 131 (2001);
Y. Oreg, G. Refael, and F. von Oppen, Phys. Rev. Lett.
{\bf 105}, 177002 (2010); R. M. Lutchyn, J. D. Sau, and S. Das Sarma, Phys. Rev.
Lett. {\bf 105}, 077001 (2010); J. Alicea, Reports on Progress in Physics
{\bf 75}, 076501 (2012); Jason Alicea, Yuval Oreg, Gil Refael, Felix von Oppen, Matthew P. A. Fisher Nature Physics {\bf 7}, 412-417 (2011);
V. Mourik, K. Zuo, S. M. Frolov, S. R. Plissard, E. P.
A. M. Bakkers, and L. P. Kouwenhoven, Science {\bf 336}, 1003 (2012); L. P. Rokhinson, X. Liu, and J. K. Furdyna, Nat Phys
{\bf 8}, 795 (2012); J. R. Williams, A. J. Bestwick, P. Gallagher, S. S.
Hong, Y. Cui, A. S. Bleich, J. G. Analytis, I. R. Fisher, and D. Goldhaber-Gordon, Phys. Rev. Lett. {\bf 109}, 056803 (2012); A. Das {\it et al.} Nature Phys. {\bf 8}(12) 887 (2012); 
S. Gangadharaiah, B. Braunecker, P. Simon and D. Loss Phys. Rev. Lett. {\bf 107}, 036801 (2011); E.M. Stoudenmire, J. Alicea, O. A. Starykh, M. P.A. Fisher Phys. Rev. B {\bf 84}, 014503 (2011); D. Sticlet, C. Bena and P. Simon
Phys. Rev. Lett. {\bf 108}, 096802 (2012).

\bibitem{spinl}
C. Waldtmann {\it et al.} Eur. Phys. Jour. B {\bf 2}, 501 (1998); P. Lecheminant, B. Bernu, C. Lhuillier, L. Pierre, P. Sindzingre, Phys. Rev. B {\bf 56}, 2521-2529 (1997); L. Messio, B. Bernu and C. Lhuillier, Phys. Rev. Lett. {\bf 108}, 207204 (2012); B. Fak {\it et al.} Phys. Rev. Lett. {\bf 109}, 037208 (2012); S. Yan, D. A. Huse and S. R. White, Science {\bf 332}, 1173-1176 (2011); S. Depenbrock, I. P. McCulloch and U. Schollwoeck, Phys. Rev. Lett. {\bf 109}, 067201 (2012); H. C. Jiang, Z. Wang and L. Balents, Nature Phys. {\bf 8}, 902 (2012); Y. Iqbal, F. Becca, S. Sorella and D. Poilblanc,  Phys. Rev. B {\bf 87}, 060405(R) (2013);
S. S. Gong, D. N. Sheng, O. I. Motrunich and M. P. A Fisher, arXiv:1306.6067; L. Balents, M. P. A. Fisher and S. M. Girvin, Phys. Rev. B {\bf 65}, 224412 (2002).

\bibitem{Misguich}
G. Misguich and C. Lhuillier, in {\it frustrated spin models}, edited by H. T. Diep (World Scientist, New Jersey, 2004).

\bibitem{Mong}
R. Mong {\it et al.} arXiv:1307.4403.

\bibitem{TKNN}
D.J. Thouless, M. Kohmoto, M. P. Nightingale and M. den Nijs, Phys. Rev. Lett. {\bf 49}, 405 (1982).

\bibitem{Haldane}
F. D. Haldane, Phys. Rev. {\bf 61}, 2015 (1988).

\bibitem{kane-mele}
C.~L. Kane and E.~J. Mele, Phys. Rev. Lett. {\bf 95},  146802  (2005);
{\it ibid}.  {\bf 95}, 226801  (2005).

\bibitem{fu-kane}
L. Fu and C. L. Kane, Phys. Rev. B {\bf 76}, 045302 (2007).

\bibitem{moore-07prb121306}
J.~E. Moore and L. Balents, Phys. Rev. B {\bf 75},  121306(R)  (2007).

\bibitem{disorder}
A.~P. Schnyder, S. Ryu, A. Furusaki, and A.~W.~W. Ludwig,
Phys. Rev. B {\bf 78}, 195125 (2008); E. Prodan, T.~L. Hughes, and B.~A. Bernevig, Phys. Rev. Lett. {\bf 105}, 115501 (2010).

\bibitem{wu-06prl106401}
C. Wu, B.~A. Bernevig, and S.-C. Zhang, Phys. Rev. Lett. {\bf 96},  106401 (2006).

\bibitem{ti+int}
C. Xu and J.~E. Moore, Phys. Rev. B {\bf 73},  045322  (2006); M. Levin and A. Stern, Phys. Rev. Lett. {\bf 103}, 196803 (2009);
Z. Wang, X.-L. Qi, and S.-C. Zhang, {\it ibid}. {\bf 105},  256803 (2010); V. Gurarie, Phys. Rev. B {\bf 83},  085426  (2011); T. Neupert, L. Santos, C. Chamon and  C. Mudry, Phys. Rev. B {\bf 86}, 165133 (2012).

\bibitem{pesin-10np376}
D.~A. Pesin and L. Balents, Nature Phys. {\bf 6},  376  (2010).

\bibitem{RachelLeHur}
S. Rachel and K. Le Hur, Phys. Rev. B {\bf 82}, 075106 (2010).

\bibitem{Kallin}
M. W. Young, S.-S. Lee and C. Kallin, Phys. Rev. B {\bf 78}, 125316 (2008).

\bibitem{Hohenadler}
M. Hohenadler,  T. C. Lang, F. F. Assaad Phys. Rev. Lett. {\bf 106}, 100403 (2011); M. Hohenadler {\it et al.} Phys. Rev. B {\bf 85}, 115132 (2012); M. Hohenadler and F. F. Assaad  J. Phys.: Condens. Matter {\bf 25}, 143201 (2013).

\bibitem{Wu}
W. Wu, S. Rachel, W.-M. Liu and K. Le Hur, Phys. Rev. B {\bf 85}, 205102 (2012); Y. Yamaji and M. Imada, Phys. Rev. B {\bf 83}, 205122 (2011); S. L. Yu, X. C. Xie and J. X. Li, Phys. Rev. Lett. {\bf 107}, 010401 (2011); D. Zheng, G.-M Zhang and C. Wu, Phys. Rev. B {\bf 84}, 205121 (2011).

\bibitem{koenig-07s766}
M. K{\"o}nig, S. Wiedmann, C. Br{\"u}ne, A. Roth, H. Buhmann, L.~W. Molenkamp, X.-L. Qi, and S.-C. Zhang,
 Science {\bf 318},  766  (2007). C. Br\" une {\it et al.} Nature Physics {\bf 8}, 486 (2012); K. C. Nowack {\it et al.} arXiv:1212.2203; Y. Ma {\it et al.} arXiv:1212.6441.

\bibitem{bernevig-06s1757}
B.~A. Bernevig, T.~L. Hughes, and S.-C. Zhang, Science {\bf 314},  1757
  (2006).

\bibitem{Hsieh_1}
D. Hsieh {\it et al.} Nature {\bf 452}, 970-974 (2008).

\bibitem{Hsieh_2}
D. Hsieh {\it et al.} Science {\bf 323}, 919-922 (2009).

\bibitem{Zhang_H}
H. Zhang {\it et al.} Nature Phys. {\bf 5}, 438 (2009).

\bibitem{BenoitKamran}
B. Fauqu\' e {\it et al.} Phys. Rev. B {\bf 87}, 035133 (2013).

\bibitem{Marsi}
M. Hajlaoui {\it et al.} Nano Lett. {\bf 12}, 3532 (2012).

\bibitem{Brune}
C. Br\" une {\it et al.}, Phys. Rev. Lett. {\bf 106}, 126803 (2011).

\bibitem{Crauste}
O. Crauste {\it et al.}, arXiv:1307.2008.

\bibitem{Chen}
Y. L. Chen {\it et al.} Science {\bf 325}, 178-181 (2009).

\bibitem{Schmidt}
T. L. Schmidt, S. Rachel, F. von Oppen and L. Glazman, Phys. Rev. Lett. {\bf 108}, 156402 (2012).

\bibitem{Adroguer}
P. Adroguer, C. Grenier, D. Carpentier, J. Cayssol, P. Degiovanni and E. Orignac, Phys. Rev. B {\bf 82}, 081303(R) (2010).

\bibitem{IonKaryn}
I. Garate and K. Le Hur, Phys. Rev. B {\bf 85}, 195465 (2012).

\bibitem{Raghu}
S. Raghu, Xiao-Liang Qi, C. Honerkamp and Shou-Cheng Zhang  Phys.Rev.Lett. {\bf 100}, 156401 (2008);  J. Wen,  A. Ruegg, C.-C. J. Wang and G. A. Fiete, Phys. Rev. B {\bf 82}, 075125 (2010).

\bibitem{atoms}
A. L. Fetter, Rev. Mod. Phys. {\bf 81}, 647 (2009);
J. Dalibard, F. Gerbier, G. Juzeli\"{u}nas and P. Ohberg,  Rev. Mod. Phys. {\bf 83}, 1523 (2011);
D. Jaksch and P. Zoller, Annals of Physics {\bf 315}, 52-79 (2005); K. Osterloh {\it et al.}, Phys. Rev. Lett. {\bf 95}, 010403 (2005);
Yu-Ju Lin {\it et al.}, Nature {\bf 462}, 628 (2009); N. Goldman {\it et al.}, Phys. Rev. Lett. {\bf 105}, 255302 (2010); M. Aidelsburger \textit{et al.}, Phys. Rev. Lett. \textbf{107}, 255301 (2011); J. Heinze et al., Phys. Rev. Lett. \textbf{107}, 135303 (2011); J. Struck \textit{et al.}, Phys. Rev. Lett. \textbf{108}, 225304 (2012); N. R. Cooper and J. Dalibard, Phys. Rev. Lett. {\bf 110}, 185301 (2013); A. Petrescu and K. Le Hur, arXiv:1306.5986.

\bibitem{photons}
M. C. Rechtsman {\it et al.}, Nature {\bf 496}, 196-200 (2013).

\bibitem{photons1}
T. Kitagawa {\it et al.} Nature Comm. {\bf 3}, 882 (2012);
M. Bellec, U. Kuhl, G. Montambaux and F. Montessagne, Phys. Rev. Lett. {\bf 110}, 033902 (2012);
F. D. M. Haldane and S. Raghu, Phys. Rev. Lett. {\bf 100}, 013904 (2008); S. Raghu and F. D. M. Haldane, Phys. Rev. A {\bf 78}, 033834 (2008); Z. Wang, Y. D. Chong, J. D. Joannopoulos and M. Soljacic, Phys. Rev. Lett. {\bf 100}, 013905 (2008); Z. Wang, Y. Chong, J. Joannopoulos and M. Soljacic, Nature {\bf 461}, 772 (2009); M. Hafezi, E. Demler, M. Lukin and J. Taylor, Nature Phys. {\bf 7}, 907 (2011); M. Hafezi, J. Fan, A. Migdall and J. Taylor, arXiv:1302.2153; J. Koch {\it et al.} Phys. Rev.  A {\bf 82} 043811 (2010);  A. B. Khanikaev {\it et al.}, Nature Mat. {\bf 12}, 233 (2013); I. Carusotto and C. Ciuti, Rev. Mod. Phys. {\bf 85}, 299 (2013).

\bibitem{AlexKaryn}
A. Petrescu, A. A. Houck and K. Le Hur, Phys. Rev. A {\bf 86}, 053804 (2012).

\bibitem{Floquet}
T. Kitagawa, E. Berg, M. Rudner and E. Demler, Phys. Rev. B {\bf 82}, 235114 (2010); N. H. Lindner, G. Refael and V. Galitski, Nature Physics {\bf 7}, 490-495 (2011); P. Delplace, A. Gomez-Leon, G. Platero, arXiv:1304.6272; J. Cayssol, B. D\' ora, F. Simon and R. Moessner, Phys. Status Solidi RRL {\bf 7}, 101 (2013).

\bibitem{Taillefumier}
M. Taillefumier, V. K. Dugaev, B. Canals, C. Lacroix and P. Bruno, Phys. Rev. B {\bf 84}, 085427 (2011).

\bibitem{Su}
W. P. Su, J. R. Schrieffer, and A. J. Heeger, Phys. Rev. Lett. {\bf 42} , 1698 (1979).

\bibitem{Tsui}
D. C. Tsui, H. L. Stormer, and A. C. Gossard, Phys. Rev.
Lett. {\bf 48}, 1559 (1982).

\bibitem{Jain}
J. K. Jain, Phys. Rev. Lett. {\bf 63}, 199 (1989).

\bibitem{FCI}
E. Tang, J.-W. Mei and X.-G. Wen, Phys. Rev. Lett. {\bf 106}, 236802 (2011); T. Neupert, L. Santos, C. Chamon and C. Mudry, Phys. Rev. Lett. 106, 236804 (2011);
N. Regnault and B. A. Bernevig, Phys. Rev. X {\bf 1}, 021014 (2011); T. Liu, C. Repellin, B. A. Bernevig and N. Regnault, Phys. Rev. B {\bf 87}, 205136 (2013);
Y.-F. Wang {\it et al.} Phys. Rev. Lett. 108, 126805 (2012);  M. O. Goerbig, Eur. Phys. J B {\bf 85}(1), 15 (2012); N. Y. Yao {\it et al.}  Phys. Rev. Lett. {\bf 110}, 185302 (2013).

\bibitem{Krempa}
W. W. Krempa, G. Chen, Y.-B. Kim and L. Balents, arXiv:1305.2193.

\bibitem{Ari}
A. M. Turner and A. Vishwanath, arXiv:1301.0330, review article

\bibitem{Nakatsuji}
Y. Machida, S. Nakatsuji, S. Onoda, T. Tayama and T. Sakakibara, Nature {\bf 463}, 210-213 (2010).

\bibitem{Krempa0}
W. W. Krempa, T. P. Choy and Y. B. Kim, Phys. Rev. B {\bf 82} 165122 (2010).

\bibitem{Ruegg}
A. R\" uegg and G. A. Fiete, Phys. Rev. Lett. {\bf 108}, 046401 (2012).

\bibitem{Takagi1}
Y. Okamoto, M. Nohara, H. Aruga-Katori, and H. Takagi, Phys. Rev. Lett. {\bf 99}, 137207 (2007);

\bibitem{BJKim}
B. J. Kim et al., Phys. Rev. Lett. {\bf 101}, 076402 (2008).

\bibitem{Takagi2}
B. J. Kim, H. Ohsumi, T. Komesu, S. Sakai, T. Morita, H. Takagi, and T. Arima, Science {\bf 323}, 1329 (2009).

\bibitem{Martins}
C. Martins, M. Aichhorn, L. Vaugier and S. Biermann, Phys. Rev. Lett. {\bf 26}, 266404 (2011).

\bibitem{Alaska}
A. Subedi, Phys. Rev. B {\bf 85}, 020408(R) (2012).

\bibitem{Shitade}
A. Shitade, H. Katsura, J. Kunes, X.L. Qi, S.C. Zhang and N. Nagaosa, Phys. Rev. Lett {\bf 102}, 256403 (2009).

\bibitem{Jackeli_Khaliullin}
G.Jackeli and G.Khaliullin Phys. Rev. Lett. {\bf 102}, 017205 (2009).

\bibitem{Chaloupka}
J. Chaloupka, G. Jackeli and G. Khaliullin Phys. Rev. Lett. {\bf 105}, 027204 (2010).

\bibitem{Jiang}
H. C. Jiang, Z. C.Gu, X. L.Qi, and S. Trebst Phys. Rev. B {\bf 83}, 245104 (2011).

\bibitem{Natalia}
C. Price and N. B. Perkins, Phys. Rev. Lett. {\bf 109}, 187201 (2012).

\bibitem{Maria}
I. Rousochatzakis, U. K. R\" ossler, J. van den Brink and M. Daghofer arXiv:1209.5895

\bibitem{RachelThomale}
J. Reuther, R.Thomale and S. Rachel, Phys. Rev. B {\bf 86}, 155127 (2012).

\bibitem{Kargarian}
M. Kargarian, A. Langari and G. A. Fiete, Phys. Rev. B {\bf 86}, 205124 (2012).

\bibitem{Bhatt}
S. Bhattacharjee, S.-S. Lee, and Y. B. Kim, New Journal of Physics {\bf 14}, 073015 (2012).

\bibitem{You}
Y. Z.You, I. Kimchi and A.Vishwanath, Phys. Rev. B {\bf 86}, 085145 (2012).

\bibitem{Vishwanath}
I. Kimichi and A.Vishwanath, arxiv 1303.3290.

\bibitem{Kimichi_2}
I. Kimichi and Y. Z.You, Phys. Rev. B {\bf 84}, 180407 (2011).

\bibitem{Roser}
I. I. Mazin {\it et al.} arXiv:1304.2258.

\bibitem{Gegenwart1}
X. Liu {\it et al.} Phys. Rev. B {\bf 83}, 220403(R) (2011).

\bibitem{Gegenwart2}
S. K. Choi {\it et al.} Phys. Rev. Lett. 108, 127204 (2012).

\bibitem{Damascelli}
R. Comin {\it et al.} Phys. Rev. Lett. {\bf 109}, 266406 (2012).

\bibitem{Gretarsson}
H. Gretarsson {\it et et al.} Phys. Rev. Lett. {\bf 110}, 076402 (2013).

\bibitem{Takagi}
K. Matsuhira {\it et al.} J. Phys. Soc. Jpn. {\bf 82} 023706 (2013).

\bibitem{Valenti}
K. Foyevtsova {\it et al.} arXiv:1303.2105.

\bibitem{Trebst}
Y. Singh {\it et al.} Phys. Rev. Lett. {\bf 108}, 127203 (2012); F.Ye et al. Phys. Rev. B {\bf 85}, 180403 (2012); Z. Nussinov and J. van den Brink, arXiv:1303.5922.

\bibitem{Jackeli}
J. Chaloupka {\it et al.}, Phys. Rev. Lett. {\bf 110}, 097204 (2013).

\bibitem{Liu}
X.Liu et al, Phys. Rev. B {\bf 83}(R), 220403 (2011).

\bibitem{Singh}
Y.Singh and Y.Gegenwart, Phys. Rev. B {\bf 82}, 064412 (2010).

\bibitem{Cao}
G. Cao, T. F. Qi, L. Li, J. Terzic, S. J. Yuan, M. Tovar, G. Murthy and R. K. Kaul, arXiv:1307.2212.

\bibitem{Jenderka}
M. Jenderka {\it et al}, arxiv 1303.5245.

\bibitem{Lhuillier}
J. B. Fouet, P. Sindzingre and C. Lhuillier, Eur. Phys J. B, {\bf 20}, 241-254 (2001); P. Lecheminant, B. Bernu, C. Lhuillier and L. Pierre,  Phys. Rev. B {\bf 52}, 6647 (1995);

\bibitem{Florens}
S. Florens and A. Georges, Phys. Rev. B {\bf 70}, 035114 (2004).

\bibitem{Paramekanti}
E. Zhao and A. Paramekanti, Phys. Rev. B {\bf 76}, 195101 (2007).

\bibitem{LeeLee}
S.-S. Lee and P. A. Lee, Phys. Rev. Lett {\bf 95}, 036403 (2005).

\bibitem{Miniatura}
G. Wang, M. O. Goerbig, C. Miniatura and B. Gr\' emaud, EuroPhys. Lett. {\bf 95}, 47013 (2011).

\bibitem{KrempaKim}
W. Witczak-Krempa and Y. B. Kim, Phys. Rev. B {\bf 85}, 045124 (2012).

\bibitem{GurarieWessel}
T. C. Lang, A. M. Essin, V. Gurarie and S. Wessel, Phys. Rev. B {\bf 87}, 205101 (2013).

\bibitem{Budich}
J. C. Budich, B. Trauzettel and G. Sangiovanni, Phys. Rev. B {\bf 87}, 235104 (2013).

\bibitem{Wang}
Z. Wang, X.-L. Qi and S.-C. Zhang, Phys. Rev. B {\bf 85}, 165126 (2012).

\bibitem{Wang_SCZhang}
Z. Wang and S.-C. Zhang, Phys. Rev. X {\bf 2}, 031008 (2012).


\bibitem{Biermann}
L. de' Medici, A. Georges and S. Biermann Phys. Rev. B {\bf 72}, 205124 (2005).

\bibitem{Nand}
R. Nandkishore, M. A. Metlitski and T. Senthil, Phys. Rev. B {\bf 86}, 045128 (2012).

\bibitem{laughlin}
R.B. Laughlin, Phys. Rev. B {\bf 23}, 5632 (1981).

\bibitem{J1J2}
P. Chandra and B. Dou\c{c}ot, Phys. Rev. B {\bf 38}, 9335 (1988); N. Read and S. Sachdev, Phys. Rev. Lett. {\bf 66}, 1773 (1991); B. K. Clark, D. A. Abanin and S. L. Sondhi, Phys. Rev. Lett. {\bf 107}, 087204 (2011); H. C. Jiang, H. Yao and L. Balents Phys. Rev. B {\bf 86}, 024424 (2012); Yuan-Ming Liu and Ying Ran, Phys. Rev. B {\bf 84}, 024420 (2011); W. Wu, M. Scherer, C. Honerkamp and K. Le Hur, Phys. Rev. B {\bf 87}, 094521 (2013); L. Wang, D. Poilblanc, Z.-C. Gu, X.-G. Wen and F. Verstraete, arXiv:1301.4492.

\bibitem{Bishop}
P. H. Y. Li {\it et al.} Phys. Rev. B {\bf 86} 144404 (2012).

\bibitem{polyakov}
A. M. Polyakov, {Gauge Fields and Strings}, Contemporary Concepts in Physics, Taylor and Francis, 1987.

\bibitem{IoffeLarkin}
L. B. Ioffe and A. Larkin, Phys. Rev. B {\bf 39}, 8988 (1988).

\bibitem{HerbutSubir}
I. Herbut, B. H. Seradjeh, S. Sachdev and G. Murthy, Phys. Rev. B {\bf 68}, 195110 (2003).

\bibitem{pepin}
C. P\' epin, Phys. Rev. B {\bf 77}, 245129 (2008).

\bibitem{fu-kane_2}
L. Fu and C. L. Kane, Phys. Rev. B {\bf 74}, 195312 (2006).

\bibitem{Lee}
D.-H. Lee, Phys. Rev. Lett. {\bf 107}, 166806 (2011).

\bibitem{RueggSigrist}
A. R\"uegg, S. D. Huber and M. Sigrist, Phys. Rev. B {\bf 81}, 155118 (2010).

\bibitem{SenthilFisher1}
T. Senthil and M. P. A. Fisher, Phys. Rev. B {\bf 63}, 134521 (2001).

\bibitem{SenthilFisher2}
T.Senthil and M. P. A. Fisher, Phys. Rev. B {\bf 64}, 214511 (2001).

\bibitem{SubirZ2}
Matthias Punk, Debanjan Chowdhury, Subir Sachdev,  arXiv:1308.2222.

\bibitem{HermeleRanLeeWen}
M. Hermele, Y. Ran, P. A. Lee, and X.G. Wen, Phys. Rev. B {\bf 77}, 224413 (2008).

\bibitem{Hermele_2}
M. Hermele, Phys. Rev. B {\bf 76}, 035125 (2007).

\bibitem{Hermele_3}
M. Hermele, T. Senthil, M. P. A. Fisher, P.A.Lee, N.Nagaosa and X.G.Wen, Phys. Rev. B {\bf 70}, 214437 (2004).

\bibitem{Nogueira}
F. S. Nogueira, H. Kleinert, {\bf 95}, Phys. Rev. Lett. 176406 (2005).

\bibitem{Mong_Shivamoggi}
R. S. K. Mong, V. Shivamoggi, Phys. Rev. B {\bf 83}, 125109 (2011).

\bibitem{Pershoguba_Yakovenko}
S. S.Pershoguba and V.Yakovenko, Phys. Rev. B {\bf 86}, 075403 (2012).

\bibitem{Reuther}
J. Reuther, D. Abanin and R. Thomale, Phys. Rev. B {\bf 84}, 014417 (2011).

\bibitem{WesselS}
Z. Y. Meng {\it et al.} Nature {\bf 464}, 847 (2010).

\bibitem{Sorella}
S. Sorella, Y. Otsuka and S. Yunoki, Scientific Reports {\bf 2}, 992 (2012).

\bibitem{Senechal}
S. R. Hassan and D. S\' en\' echal, Phys. Rev. Lett. {\bf 110}, 096402 (2013).

\bibitem{Herbut}
F. Assaad and I. Herbut, arXiv:1304.6340.

\bibitem{Wubis}
A. Liebsch and W. Wu, Phys. Rev. B {\bf 87}, 205127 (2013).

\bibitem{Subir}
C. Xu and S. Sachdev, Phys. Rev. Lett. {\bf 105} 057201 (2010).

\bibitem{graphene}
Marco Polini, Francisco Guinea, Maciej Lewenstein, Hari C. Manoharan and Vittorio Pellegrini, arXiv:1304.0750.

\bibitem{Goldman}
N. Goldman {\it et al.} New Journal of Physics {\bf 15} 013025 (2013).

\bibitem{Cocks}
D. Cocks, P. P. Orth, S. Rachel, M. Buchhold, K. Le Hur and W. Hofstetter, Phys. Rev. Lett. {\bf 109}, 205303 (2012); N. R. Cooper and J. Dalibard, Phys. Rev. Lett. {\bf 110}, 185301 (2013); P. P. Orth, D. Cocks, S. Rachel, M. Buchhold, K. Le Hur and W. Hofstetter, J. Phys. B At. Mol. Opt. Phys. {\bf 46}, 134004 (2013).

\bibitem{Roderich}
C. Castelnovo, R. Moessner and S. L. Sondhi, Nature {\bf 451}, 42 (2008);
S. T. Bramwell {\it et al.} Nature 461, 956-959 (2009); D. J. P. Morris {\it et al.} Science {\bf 326}, 411 (2009);
K. Kimura {\it et al.} Nature Communications {\bf 4} (2013); N. Shannon, O. Sikora, F. Pollmann, K. Penc, P. Fulde Phys. Rev. Lett. {\bf 108}, 067204 (2012).

\bibitem{Amo}
R. Hivet {\it et al.} Nature Phys. {\bf 8}, 724 (2012).

\end{thebibliography}
\end{document}